\documentclass[useAMS,usenatbib]{mn2e}
\usepackage{mathptmx}
\usepackage{color}
\usepackage{graphics,graphicx,amssymb,amsmath,natbib}
\usepackage{deluxetable}
\usepackage{float}
\usepackage{color}
\usepackage{lscape,graphicx}
\usepackage{amsmath}
\usepackage{amssymb}
\usepackage{multirow}
\usepackage{afterpage}
\usepackage{threeparttable}
\usepackage{subfigure}
\usepackage{rotating}
\usepackage{lscape}
\usepackage{caption}

\newcommand{\gca}{Geochim.~Cosmochim.~Acta}
\newcommand{\aap}{A$\&$A}
\newcommand{\apj}{ApJ}
\newcommand{\mnras}{MNRAS}
\newcommand{\apjs}{ApJS}
\newcommand{\apjl}{ApJL}
\newcommand{\araa}{ARA$\&$A}
\newcommand{\aj}{AJ}

\title[A1095 and A1926]{Baryon content and dynamic state of galaxy clusters: {\sl XMM-Newton} observations of A1095 and A1926}
\author
[Chong Ge et al.]{Chong Ge$^{1,2,3,4}$, Q.Daniel Wang$^{1,4}$\thanks{E-mail: wqd@astro.umass.edu; qsgu@nju.edu.cn}, Todd M. Tripp$^4$, Zhiyuan Li$^{1,2,3}$, Qiusheng Gu$^{1,2,3}$\footnotemark[1]
\newauthor
and Li Ji$^{3,5,6}$\\
$^{1}$School of Astronomy and Space Science, Nanjing University, Nanjing 210093, China\\
$^{2}$Key Laboratory of Modern Astronomy and Astrophysics (Nanjing University), Ministry of Education, Nanjing 210093, China\\
$^{3}$Collaborative Innovation Center of Modern Astronomy and Space Exploration, Nanjing 210093, China\\
$^{4}$Department of Astronomy, University of Massachusetts, Amherst, MA 01003, USA\\
$^{5}$Purple Mountain Observatory, Chinese Academy of Sciences, Nanjing 210008, China\\
$^{6}$Key Laboratory of Dark Matter and Space Astronomy, PMO, CAS, Nanjing 210008, China\\
}
\begin{document}
\date{Accepted. Received; in original form}

\pubyear{2016}

\maketitle

\label{firstpage}

\begin{abstract}
We have initiated a program to study the baryon content and dynamic state of galaxy clusters. Here we present results primarily from {\it XMM-Newton} observations of two optically-selected galaxy clusters, A1095 ($z \simeq 0.210$) and A1926 ($z \simeq 0.136$). We find that both of them are actually cluster pairs at similar redshifts. We characterize the temperatures of these individual clusters through X-ray spectral fits and then estimate their gravitational masses. 
We show a rich set of substructures, including large position offsets between the diffuse X-ray centroids and the brightest galaxies of the clusters, which 
suggests that they are dynamically young.
For both A1095 and A1926, we find that the mass required for the cluster pairs to be bound is smaller than the total gravitational mass. Thus both cluster pairs appear to be ongoing major mergers.
Incorporating SDSS and NVSS/FIRST data, we further examine the large-scale structure environment and radio emission of the clusters to probe their origins, which also leads to the discovery of two additional X-ray-emitting clusters ($z \simeq 0.097$ and $z \simeq 0.147$) in the field of A1926.
We estimate the hot gas and stellar masses of each cluster, which compared with the expected cosmological baryonic mass fraction, leave ample room for warm gas.

\end{abstract}


\section{Introduction}
\label{s:int}
One of the most enduring problems in our understanding of cosmological structure formation is the missing baryon problem --- after summing up the known mass in the form of stars, interstellar material, hot (X-ray emitting) intragalactic medium in clusters, and observed intergalactic gas, a large portion of the expected baryonic mass is still missing (Fukugita, Hogan, \& Peebles 1998; Bregman 2007). Theoretically, these missing baryons are believed to reside primarily in the so-called warm-hot intergalactic medium (WHIM) at moderate overdensity (Cen \& Ostriker 1999, 2006), heated to temperatures of $10^5-10^7$ K by adiabatic compression and shocks during formation of clusters/galaxies via a hierarchical sequence of mergers and accretion of smaller systems (e.g., Vazza et al. 2009). Observationally, the baryon fraction of the total gravitational mass in the inner region of a typical cluster ($r \leq r_{500}$, within which the mean mass density is 500 times the critical density of the Universe), for example, has been shown to be substantially smaller than the universal value as expected from the standard cosmology (Andreon 2010; Gonzalez et al. 2013); 
this missing baryon problem is most convincing observed in relatively low- and intermediate-mass clusters ($M_{500} \lesssim 10^{14} M_{\odot}$; Lagan{\'a} et al. 2013). Similar conclusions have also been drawn for the baryon fraction in individual galaxies (McGaugh et al. 2010; Werk et al. 2014).

On one hand, X-ray emission is routinely observed in the hot intracluster medium (ICM), mostly from inner regions ($r \lesssim r_{500}$), and recently from outer regions of rich clusters ($r \sim r_{500} - r_{200}$) as well (Bonamente et al. 2012; Eckert et al. 2012; Walker et al. 2013; Wang \& Walker 2014). 
These observations characterize the gravitational mass and dynamic state, as well as the hot ICM properties of an individual cluster, by measuring its X-ray temperature, morphology and structure.
The non-homogeneous or multi-temperature (or even multiphase) nature of the ICM is inferred, especially for the outer regions, and is indeed expected from simulations and theoretical models of the structure formation (Roncarelli et al. 2006, Molnar et al. 2009). These simulations and models
show a complicated shock heating/cooling history of the ICM in the outer regions and beyond (up to $\sim 2r_{200}$ , where the strong external shock of the accretion flow is located, while the virial radius is typically located at $r \sim r_{200}$; Molnar et al. 2009). In addition, the cluster environment can strongly affect the gaseous halos of individual galaxies, via processes such as ram-pressure stripping and pressure compression (e.g., Lu \& Wang 2011).

On the other hand, observations of UV absorption lines from warm gas along the sight-lines toward background QSOs have shown a significant reservoir of baryonic matter in the outskirts of galaxies on scales of their virial radii. For example, Ly-$\alpha$ and/or O VI absorbers are often found to be around individual galaxies of luminosities $L > 0.5L^{\star}$ to $r\sim150$ kpc with a high covering factor of $\approx 90\%$ for blue, star-forming galaxies and a substantially lower covering factor for red, passive galaxies (Prochaska et al. 2011; Tumlinson et al. 2011a, 2013). Occasionally, such absorbers are also observed to much greater distances ($r \gtrsim 1$ Mpc) away from field galaxies. The exact nature of such absorbers (e.g, collisional ionization vs. photoionization for O VI-bearing gas) remains uncertain (e.g., Oppenheimer \& Schaye 2013). So far only a handful of sight-lines have been probed for the ICM or intragroup medium via far-UV (FUV) absorption lines. Nevertheless, sight-lines to several nearby groups of galaxies clearly show absorption lines in the intragroup medium, with definite signals of a multiphase plasma as lines arising from species of very different ionization states (such as C III and O VI are present for same absorbers; Shull, Tumlinson \& Giroux 2003; Pisano et al. 2004; Tripp et al. 2011; Tumlinson et al. 2011b; Muzahid et al. 2015).

The fortuitous alignment of UV-bright background QSOs behind X-ray emitting foreground clusters provides a unique opportunity to probe the mass content and physical conditions of the multiphase ICM. 
With suitable cluster/QSO pairs, the hot cluster gas can be sensitively studied with X-ray emission diagnostics while the ``warm-hot" ICM can be probed with ultraviolet absorption lines imprinted on the spectrum of the background QSO. Such pairs provide a laboratory to examine the physical processes of the ICM, freshly accreted from the intergalactic medium and/or stripped out of individual galaxies, as well as the gaseous halos of individual cluster galaxies.
Clearly, the understanding of these phenomena and physical processes is important not only for studying the environmental impact on galaxy evolution and for determining the baryon content and physical/dynamic state of the clusters, but for their utility as cosmology probes as well (e.g., via the observed Sunyaev-Zel'dovich effect signal; Carlstrom, Holder \& Reese 2002).

We have started a program to use cluster/QSO pairs to examine the multiphase intracluster medium in and around cluster galaxies using {\it XMM-Newton}/{\it Chandra} and the Hubble Space Telescope ({\sl HST}). 
An important factor in the selection of the background QSOs is that the FUV and near-UV (NUV) fluxes have been accurately measured. Although most QSOs by nature are UV bright in the rest frame, the probability of intersecting a Lyman limit system (log$\rm{N_{HI}} >$ 17.2), which effectively blocks all radiation below (1 + $z_c$)$\times$912 \AA, increases with redshift. Hence, not all QSOs with $z_Q > z_c$ are suitable for this work. We cross-reference the Sloan Digital Sky Survey (SDSS; York et al. 2000) DR7 quasar catalog with the {\it GALEX} all-sky survey data to select targets that have reliable FUV and NUV flux measurements. 
Similarly, not all clusters are suited for absorption studies with {\sl HST}/Cosmic Origins Spectrograph (COS). The O VI doublet (1032 and 1038 \AA), in particular, is not redshifted into an observable region below $z_c < 0.10$.
Below these redshifts, the sensitivity of COS declines, and the absorption from H$_2$ in the
Milky Way can impede detection and measurement of weak O VI lines. Thus we
cross-correlate our list of UV bright QSOs with the SDSS GMBCG catalog (Hao et al. 2010), consisting of over 55,000 clusters, to search for proper z$_c >$ 0.10 targets. For an
optimal use of the {\it XMM-Newton} and {\sl HST} observing time, we select only those relatively
rich clusters with the temperatures $\gtrsim$ 2 keV and 0.10 $< z_c <$ 0.25 paired with
background QSOs of $m_{FUV} <$ 18.3 and projected distances $\lesssim 1.5 \times r_{200}$. The projected distances all lie well within the external accretion shock radii expected for the clusters (e.g., Molnar et al. 2009).
In addition, we check to see if the O VI lines would lie near lines from the Milky Way
interstellar medium (ISM) to ensure that there is no blending with ISM lines.
In total, we have identified 10 potential targets. Two of these targets, A1095 and A1926, have been observed with {\it XMM-Newton} and {\sl HST} for our program.

Here, we present a study of these two clusters based primarily on our {\it XMM-Newton} observations. 
The rest of the present paper is organized as follows: Section 2 describes
the {\it XMM-Newton} observations and our data reduction and analysis procedures; Section 3 presents the results based on the X-ray observations; 
Section 4 compares multi-wavelength observations to further the exploration of the clusters, including their dynamic state, substructure, large-scale environment, and baryon content, as well as implication of our findings; Section 5  summarizes our results and conclusions.
We use the standard cold dark matter cosmology with $H_{0}=\rm{70\ km\ s^{-1}\ Mpc^{-1}}$, $\Omega_m$=0.3 and $\Omega_\Lambda$=0.7. 

\section{X-ray Observations and data analysis}
\label{s:obs}
Table~\ref{t:obs} provides a log of the {\it XMM-Newton} observations employed in this paper. This study uses data collected from the European Photon Imaging Cameras (EPIC), which were set in the full-frame mode, using the Thin1 filter.
We processed the data using the Extended Source Analysis Software (ESAS; Snowden et al. 2008; Kuntz \& Snowden 2008),
as part of the {\it XMM-Newton} Science Analysis System (SAS, version 13.5.0.), with the associated Current Calibration Files (CCF)\footnote{ftp://xmm.esac.esa.int/pub/ccf/constituents/extras/esas\_caldb}.

\begin{table*}
\caption{\textit{XMM-Newton} Observations of our sample galaxy clusters}
\tabcolsep=0.11cm
\centering
\begin{tabular}{lcccccrrrcrrr}
\hline\hline\noalign{\smallskip}
\multicolumn{1}{l}{Target}            & 
\multicolumn{1}{c}{R.A.}                & 
\multicolumn{1}{c}{Dec.}                & 
\multicolumn{1}{c}{Obs.\ ID}            & 
\multicolumn{1}{c}{Obs.\ data}   & 
\multicolumn{3}{c}{$T_{obs}^{a}$(ks)} & 
\multicolumn{1}{c}{}        &
\multicolumn{3}{c}{$T_{filt}^{b}$(ks)}   \\  
\\
\cline{6-8}
\cline{10-12}
\multicolumn{1}{c}{}        &
\multicolumn{2}{c}{(J2000)} &
\multicolumn{1}{c}{}        &
\multicolumn{1}{c}{(yyyy-mm-dd)}     &
\multicolumn{1}{c}{MOS1}      &
\multicolumn{1}{c}{MOS2}    &
\multicolumn{1}{c}{pn}    &
\multicolumn{1}{c}{}        &
\multicolumn{1}{c}{MOS1}      &
\multicolumn{1}{c}{MOS2}    &
\multicolumn{1}{c}{pn}    \\
\hline
\noalign{\smallskip}
A1095 & 10:47:29.0 & $+$15:14:02 & 0721880101  & 2013-12-17 & 29.5 & 29.4 & 28.2 & & 23.6 & 25.9 & 14.8 \\
A1926 & 14:30:33.1 & $+$24:38:43 & 0728170101  & 2014-01-05 & 38.4 & 38.4 & 44.5 & & 19.0 & 18.6 & 14.0 \\	
\hline
\end{tabular}
\begin{tablenotes}
      \item $^a$: The total duration of the observation from EPIC MOS1, MOS2 and pn.\\
      $^b$: The length of the good time intervals for each instrument after using the XMM-ESAS routines to filter flaring.
    \end{tablenotes} 
\label{t:obs}
\end{table*}
\subsection{Imaging analysis}
\label{ss:obs-ima}
We use the XMM-ESAS routines, {\it emchain} and {\it epchain}, to create the raw event files of the Metal Oxide Semiconductor (MOS) and pn CCDs. When {\it XMM-Newton} orbits the Earth, solar protons with energies less than a few hundred eV are funneled towards the detectors. These soft protons create a time-variable (unpredictable flaring) instrumental background component inside the open fields of view (FOV) of the detectors. The affected time intervals are filtered out with the routines {\it mos-filter} and {\it pn-filter}, which fit a Gaussian to the peak of the distribution of counts collected from the FOV in the 2.5-12 keV and filter time intervals with the count rates deviating more than 1.5 $\sigma$  from the peak. The useful exposure times after this filtering are listed in Table~\ref{t:obs}. We use the routine {\it cheese} to perform source detection on the filtered data, using a flux threshold of $\rm{3.0 \times 10^{-15}\ erg\ cm^{-2\ }s^{-1}}$, which is estimated from the relation (Watson et al. 2001) of detection limit with exposure time. We apply the routines {\it mos-spectra} and {\it pn-spectra} to create spectra and images.
The instrumental or quiescent particle background is modeled with routines {\it mos$\_$back} and {\it pn$\_$back}, which use data from the
corners (unexposed pixels) of the detectors, and filter-wheel closed data sets with hardness ratios and count rates similar to those measured during our observations.
Data from the CCDs that were in anomalous state or damaged are excluded from subsequent analysis. The MOS1, MOS2 and pn images are combined with the routine {\it comb}. We use the routine {\it adapt} to create the background subtracted, exposure corrected EPIC images, which are binned by a factor of 2 and adaptively smoothed with a minimum number of 50 counts per bin. Our final combined images of the clusters are shown in Figs.~\ref{fig1} and \ref{fig2}.

\subsection{X-ray source detection}
\label{ss:obs-sou}
We search for the X-ray sources in the S (0.5-2 keV), H (2-7 keV) and B (0.5-7 keV) broad bands.
The task {\it cheese} uses two algorithms to detect sources: {\it eboxdetect} and {\it emldetect}. The former runs twice, while the latter once for each observation. In the first pass, {\it eboxdetect} uses a `local' mode, estimating the background from a region local to each source detection box. Detected source candidates from this `local' box method are masked out from the image and a smooth global background map is produced by fitting a spline surface to the remaining data. The second {\it eboxdetect} pass then applies this map to detect sources with greater sensitivity. This new source list forms the input to {\it emldetect} for the final Maximum-Likelihood PSF fitting, which then exports the results of source detection. 

We excise all detected sources within
regions where the surface brightness of the point source
is expected to be $ > 25\%$ of the surrounding background to 
construct diffuse X-ray maps of the clusters. 

\subsection{Spectral analysis}
\label{ss:obs-spe}

We extract the on-cluster spectra from the regions outlined by the bold cyan circles shown in Figs.~\ref{fig1}b and \ref{fig2}b.
The spectra are analysed using the X-ray spectral fitting package XSPEC (version 12.8.0).
We model an on-cluster {\it XMM-Newton} spectrum as a linear combination of various components:

\noindent \textbf{The source:} the cluster emission is modeled with an absorbed thermal component (\emph{wabs*apec}).
The temperature and normalization are free parameters with metallicity fixed at 0.3 $Z_{\odot}$ (Anders \& Grevesse 1989), and the redshift (see Table~\ref{t:clusters} for details) is set to the value of the brightest cluster galaxy (BCG) in each cluster.

\noindent \textbf{The quiescent particle background}: the continuum can be subtracted as background spectra in XSPEC, while the instrumental lines due to the interaction of the particle background with the detectors, which may vary from observation to observation, cannot be completely removed by subtracting a spectrum constructed from the filter wheel closed data. These lines are fit as \emph{gaussians} with line energies and widths allowed to vary.

\noindent \textbf{The cosmic X-ray background}: this contribution can be modeled with three components:
an unabsorbed thermal one (\emph{apec}) for emission from the Local Hot Bubble or heliosphere;
an absorbed thermal one (\emph{wabs*apec}) for emission from the Galactic halo and/or intergalactic medium;
and an absorbed power-law (\emph{wabs*powerlaw}) for emission from an unresolved background of cosmological sources.
The ROSAT all-sky survey (RASS) spectrum from the HEASARC X-ray Background Tool\footnote{http://heasarc.gsfc.nasa.gov/cgi-bin/Tools/xraybg/xraybg.pl} is added to constrain the contribution of the cosmic background. The RASS spectrum is extracted from a $1^\circ-2^\circ$ annulus surrounding the cluster with an appropriate response file also provided by the tool.

\noindent \textbf{The soft protons}: residual soft proton contamination may remain in some {\it XMM-Newton} observations even after light curve screening and  is modeled as an additional unfolded \emph{powerlaw} component, not affected by detector response.

\noindent \textbf{The solar wind charge exchange:} this process (Snowden, Collier \& Kuntz 2004) can also create additional emission lines in the observed spectra. Therefore, an additional \emph{gaussian} component is included to model this possible emission.

The parameters of the above models are set and linked as suggested by Snowden et al. (2008) and the XMM-ESAS Users Guide\footnote{http://heasarc.gsfc.nasa.gov/docs/xmm/esas/cookbook/xmm-esas.html}.

We also conduct both morphological and spectral analyses of many of the prominent X-ray sources/features to determine their nature. Their spectra are extracted from regions shown in Figs.~\ref{fig1}a and \ref{fig2}a with the background contributions estimated locally.

\section{Results}
\label{s:res}
We present the results from the source detection in Tables~\ref{t:s-a1095}
and \ref{t:s-a1926}. 
A total of 47 and 78 sources are detected in the A1095 and A1926 fields, respectively. 
While most of these sources are likely  background
AGNs, some may represent peaks of the ICM emission. We will discuss such cases  in the context of ICM substructures (see \S~\ref{ss:dis-sub}).
Here we focus on the global diffuse X-ray properties of the clusters.

\subsection{Global X-ray morphological properties}
\label{ss:x-morph}
\begin{figure*}
 	\begin{center}
\includegraphics[width=0.495\textwidth,keepaspectratio=true,clip=true]{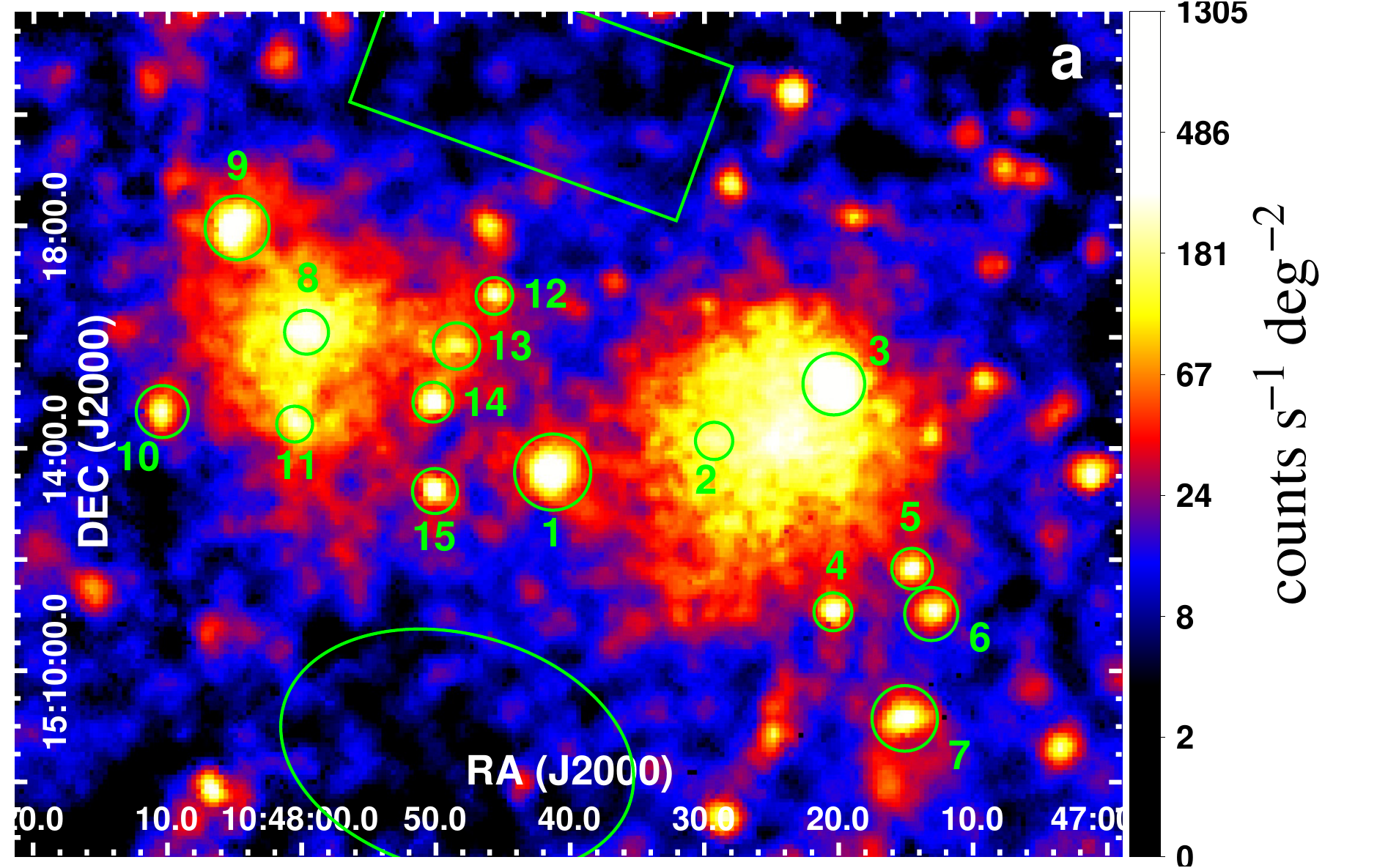}
\includegraphics[width=0.495\textwidth,keepaspectratio=true,clip=true]{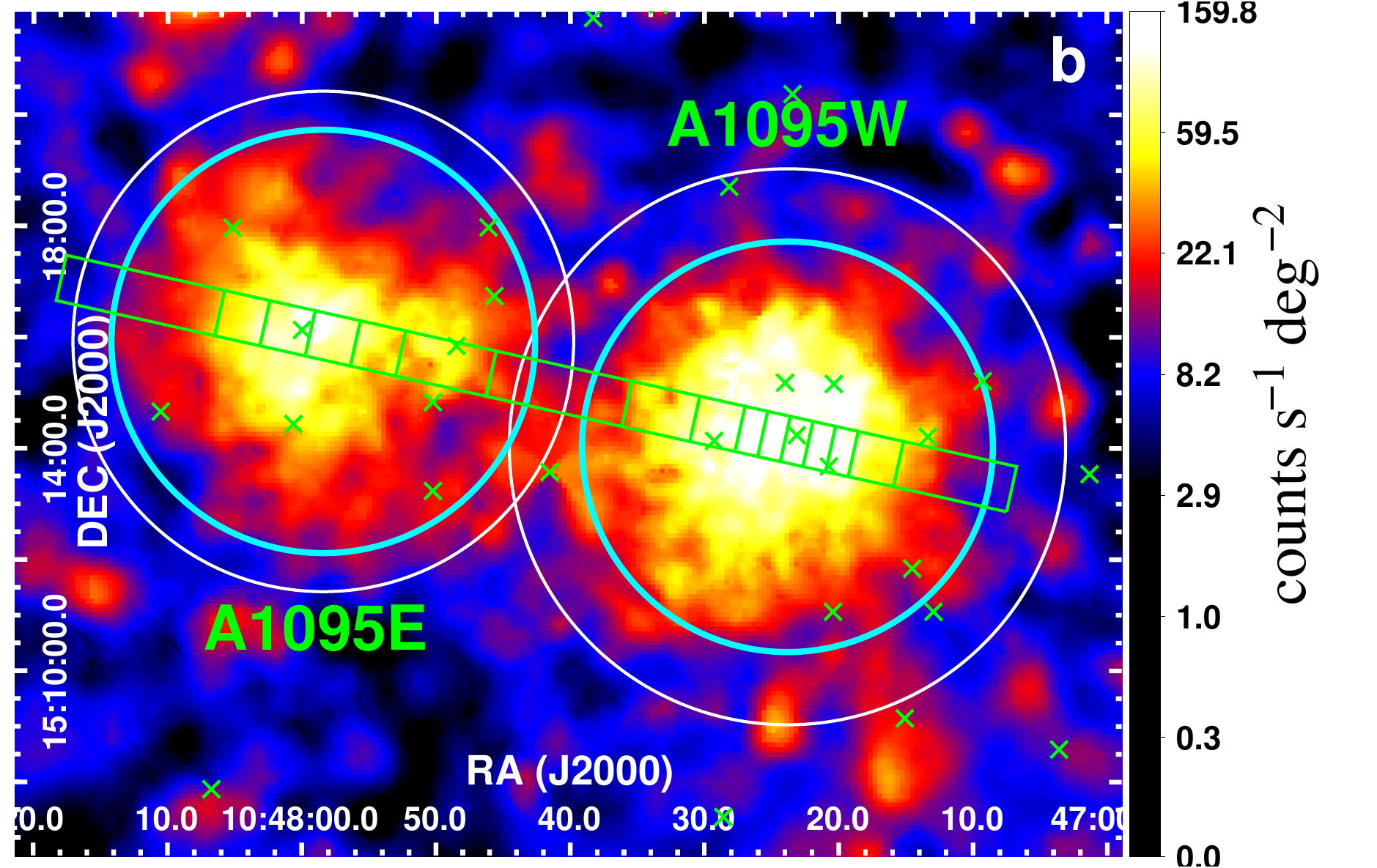}
\includegraphics[width=0.495\textwidth,keepaspectratio=true,clip=true]{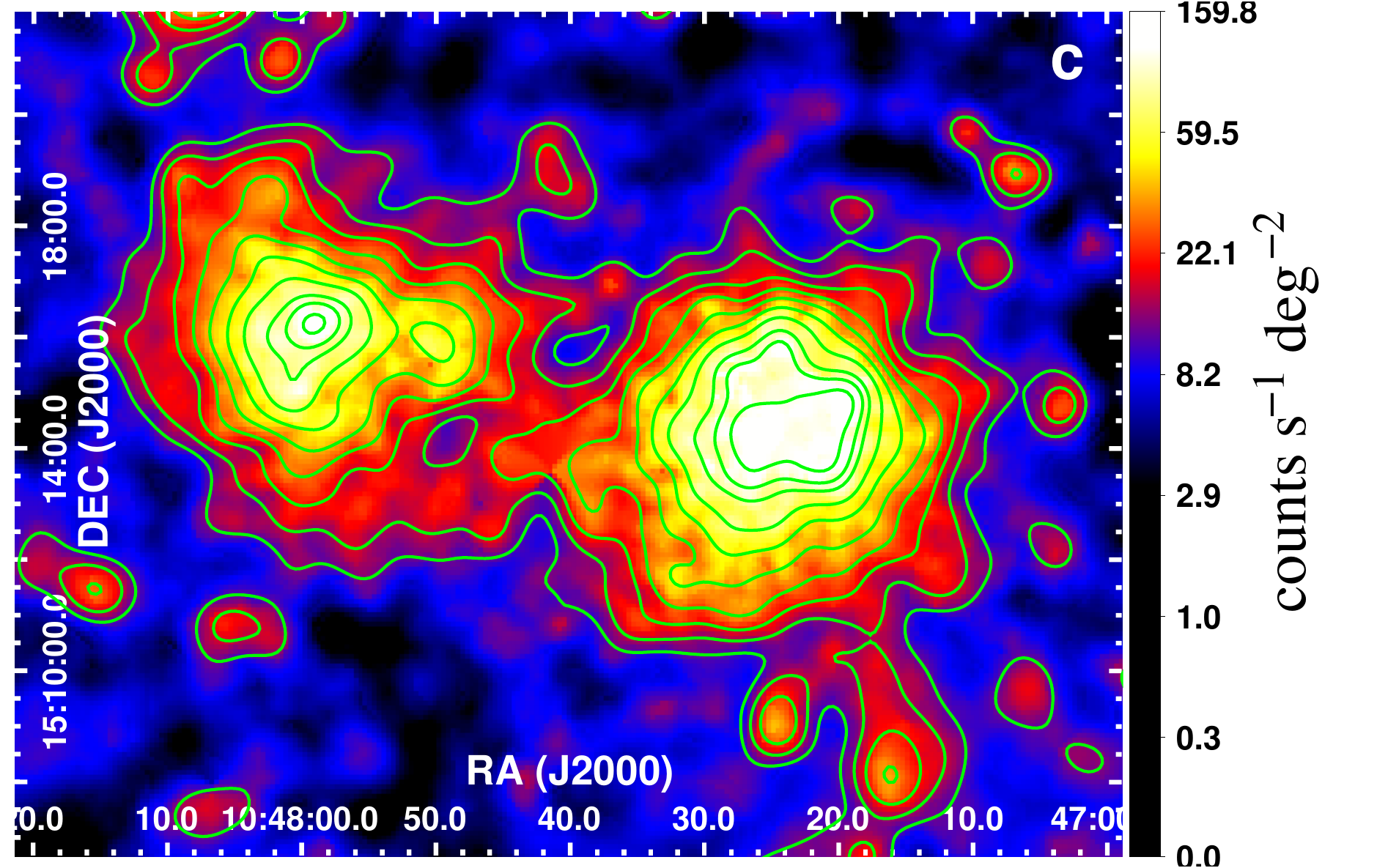}
\includegraphics[width=0.495\textwidth,keepaspectratio=true,clip=true]{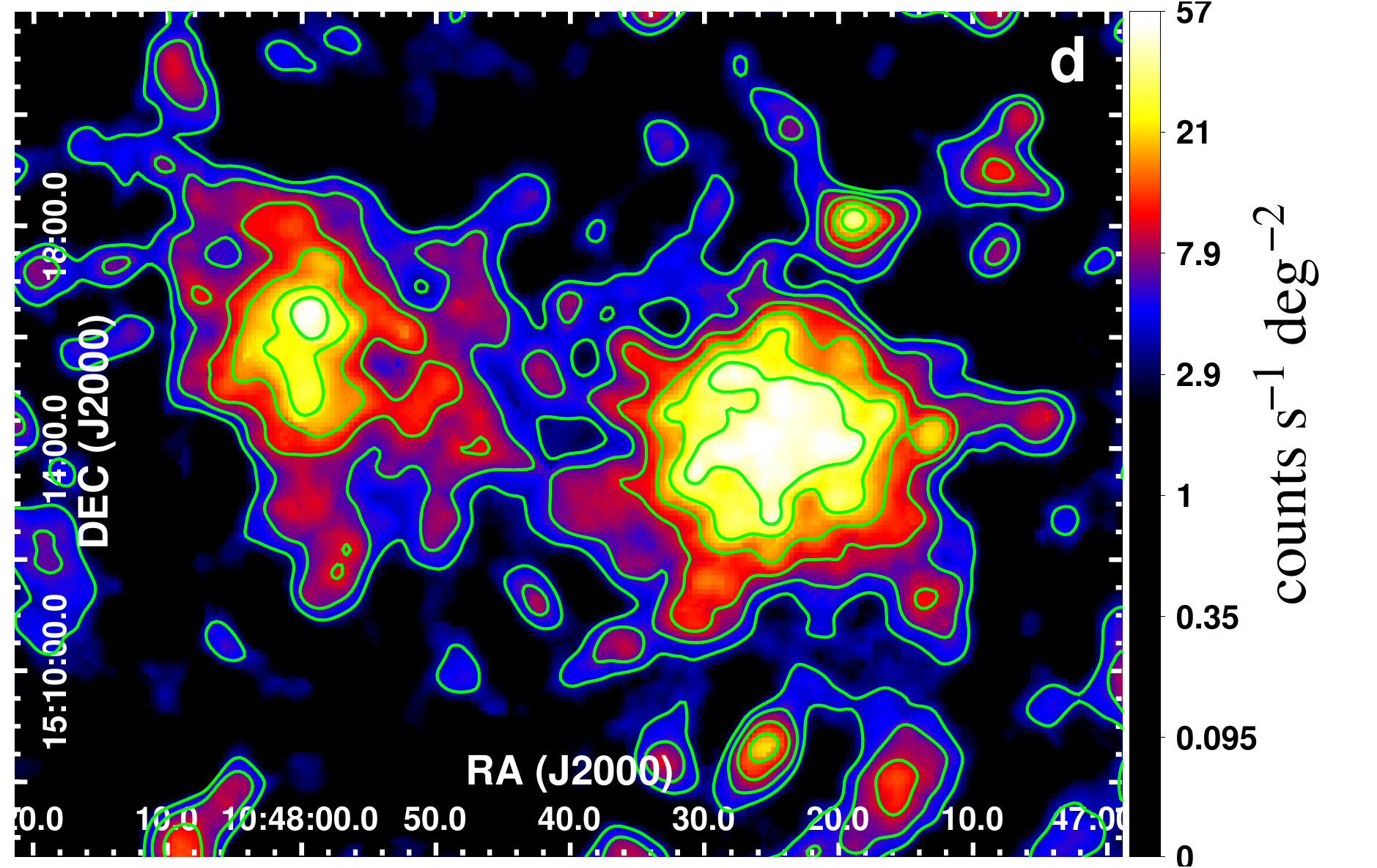}
 	\end{center}
	\caption{X-ray images of A1095W and A1095E.
(a) 0.5-7 keV intensity with the outlined regions for various spectral extractions:  luminous X-ray sources (circles) and  the local background regions (large ellipse plus rectangle). 
(b) Diffuse 0.5-2 keV intensity with point sources (marked by crosses) removed; rectangles outline the regions for constructing the temperature profile  along the major axis of the cluster pair, while the bold cyan circles represent the extraction regions for the on-cluster spectra. The larger white circles outline the regions for the statistical estimate of the photometric redshifts of galaxies.
(c) Diffuse 0.5-2 keV image, together with the contours (in units of $\rm{counts\ s^{-1}\ deg^{-2}}$) at 11.4, 15.2, 20.3, 27.1, 36.0, 47.9, 63.6, 84.4, 97.3, and 112.1.
(d) The same as (c), but for the 2-7 keV band and the contours at 3.9, 6.1, 9.5, 14.8, 22.9, and 35.4.}
	 \label{fig1}
\end{figure*}

\begin{figure*}
 	\begin{center}
\includegraphics[width=0.495\textwidth,keepaspectratio=true,clip=true]{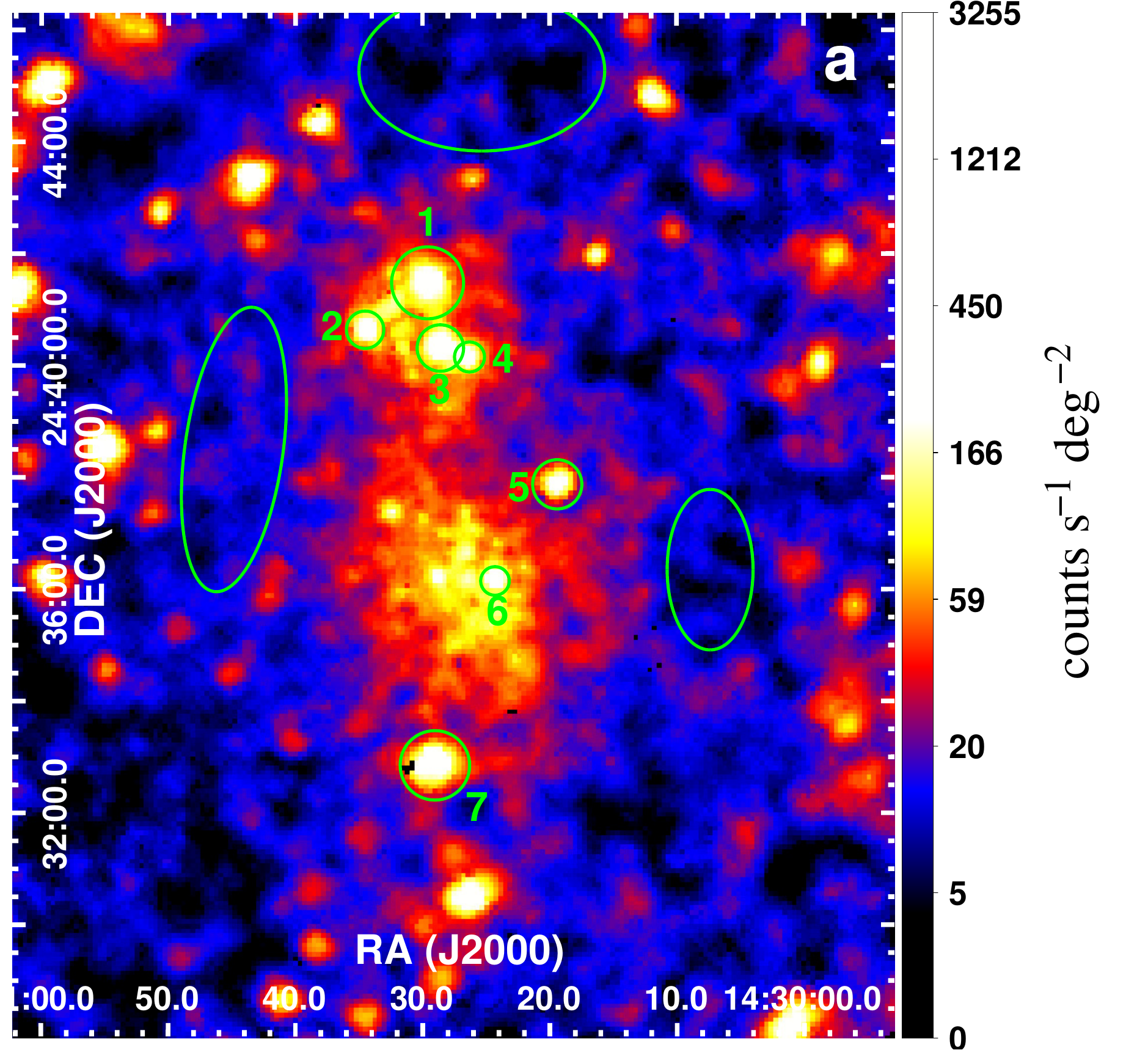}
\includegraphics[width=0.495\textwidth,keepaspectratio=true,clip=true]{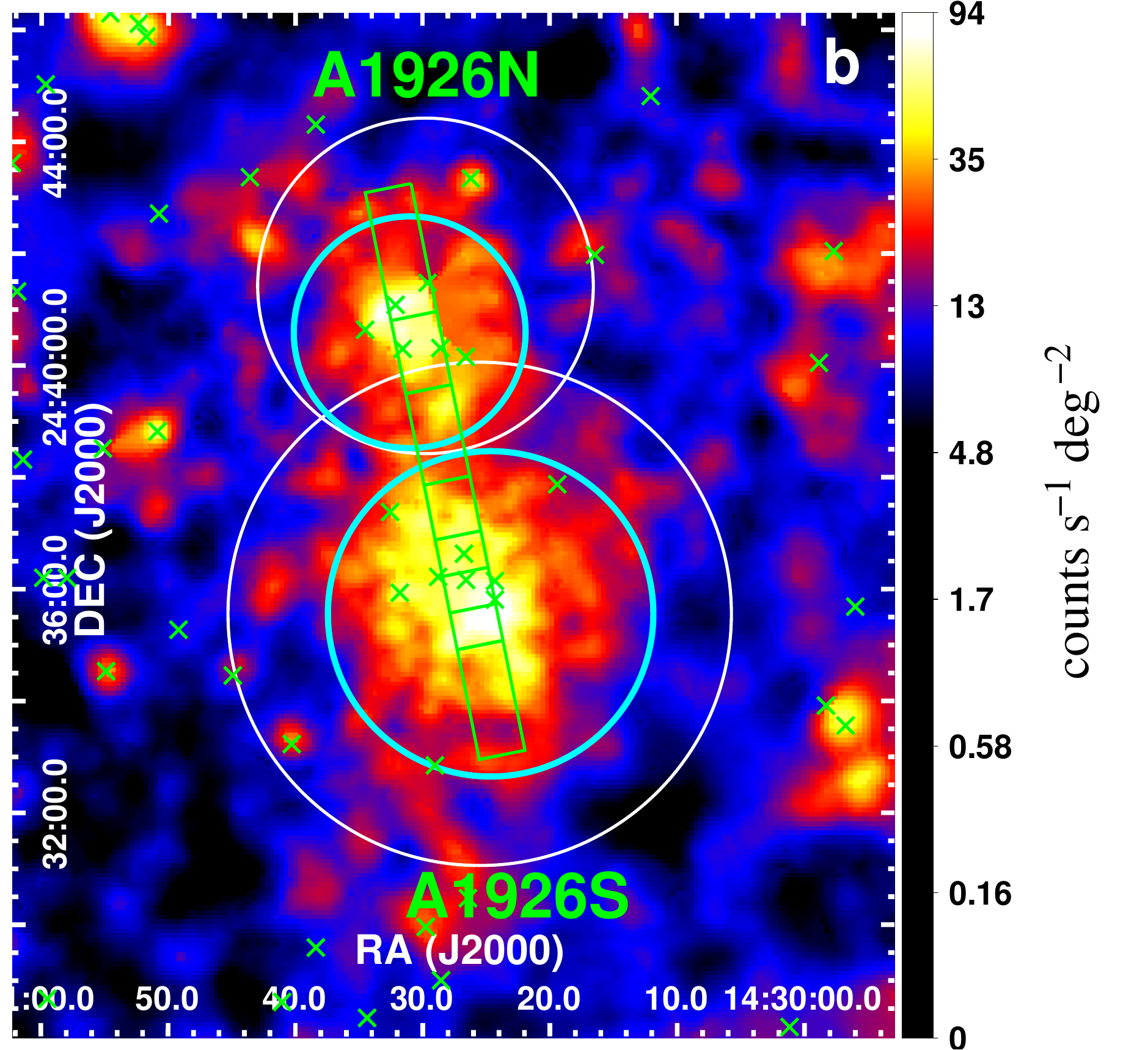}
\includegraphics[width=0.495\textwidth,keepaspectratio=true,clip=true]{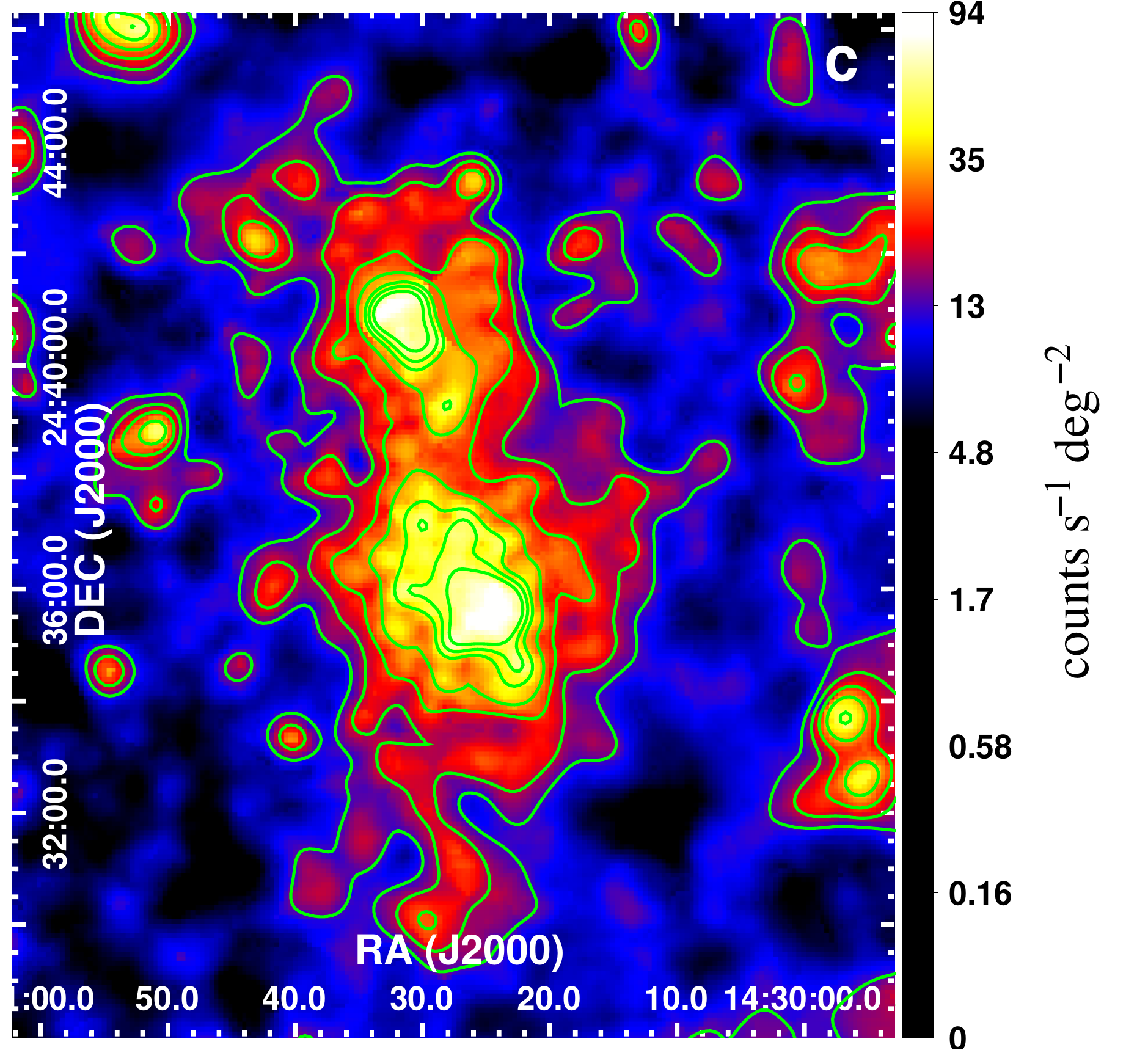}
\includegraphics[width=0.495\textwidth,keepaspectratio=true,clip=true]{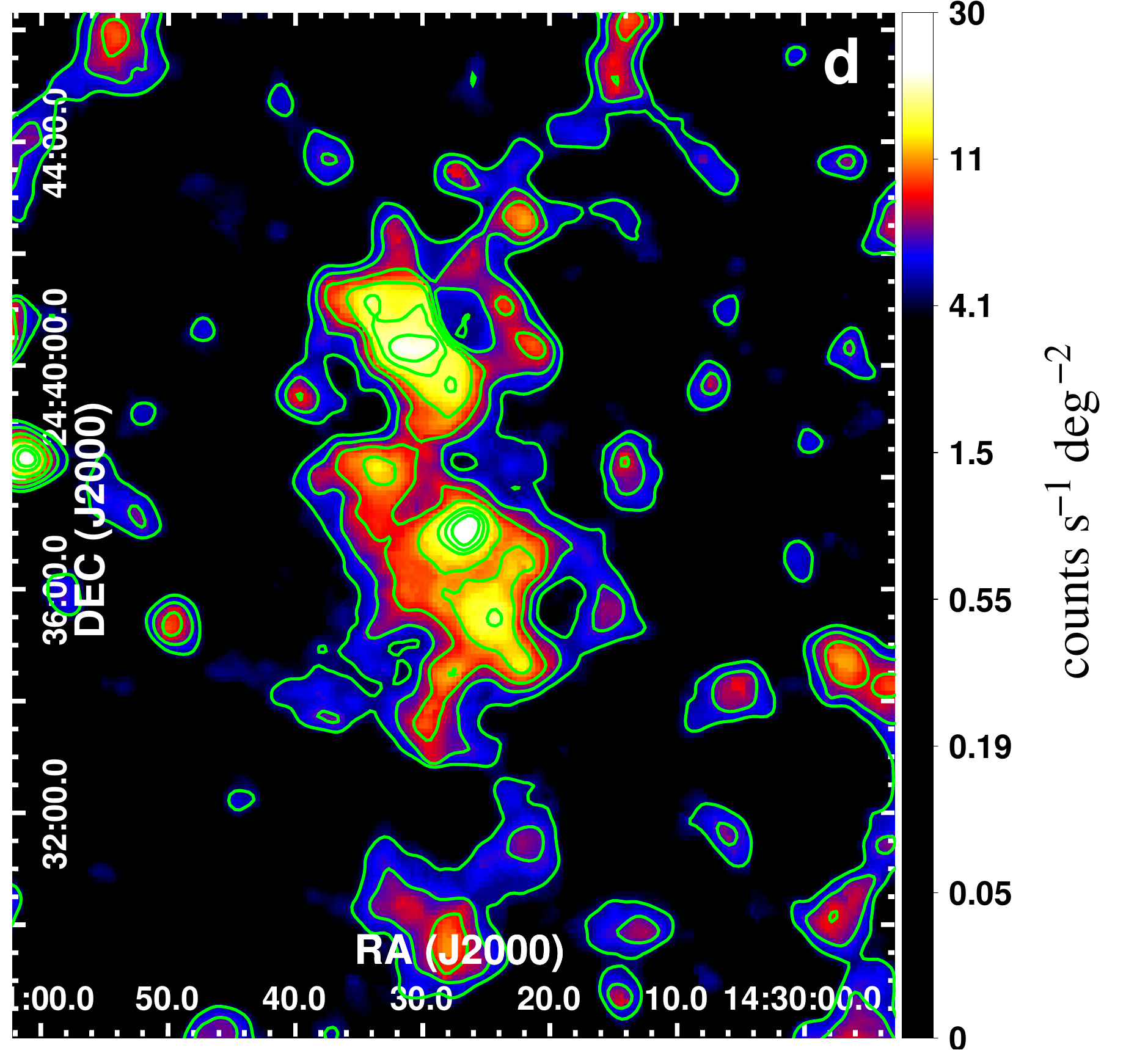}
 	\end{center}
    \caption{%
The same as in Fig.~\ref{fig1}, but for A1926S and A1926N and
the contours at 13.2, 17.6, 23.5, 31.2, 41.5, 47.9, and 55.2 (c) and contours at 4.7, 6.3, 8.5, 11.4, 15.2, 17.6, and 20.3 (d).
     }%
   \label{fig2}
\end{figure*}

Figs.~\ref{fig1} and \ref{fig2} show the X-ray images of the A1095 and A1926 fields. Each contains two apparent clusters with comparable size and brightness. In fact, they have been separately identified  in  some optical catalogues of clusters (Table~\ref{t:clusters}; see \S~\ref{ss:dis-com} for further discussion). For ease of 
reference here, we label GMBCG J161.87087+15.23391 (Hao et al. 2010) 
as A1095W and its companion, GMBCG J162.00202+15.26823, as A1095E in 
Fig.~\ref{fig1}b,
and we label SDSS J143021.94+243429.1 as A1926S and GMBCG J217.61927+24.67202 
as A1926N in Fig.~\ref{fig2}b. A1095W and A1095E are relatively well separated. 
The X-ray morphology of A1095W itself is quite round globally (though elongated in the inner region), whereas A1095E appears irregular, showing roughly a triangle shape. A1926S and A1926N are not
clearly separated in projection. They together appear to have a ``peanut" shaped 
morphology; both are elongated in the direction of the line connecting their centers 
and contain substantial substructures.

\begin{figure*}
 	\begin{center}
\includegraphics[width=0.24\textwidth]{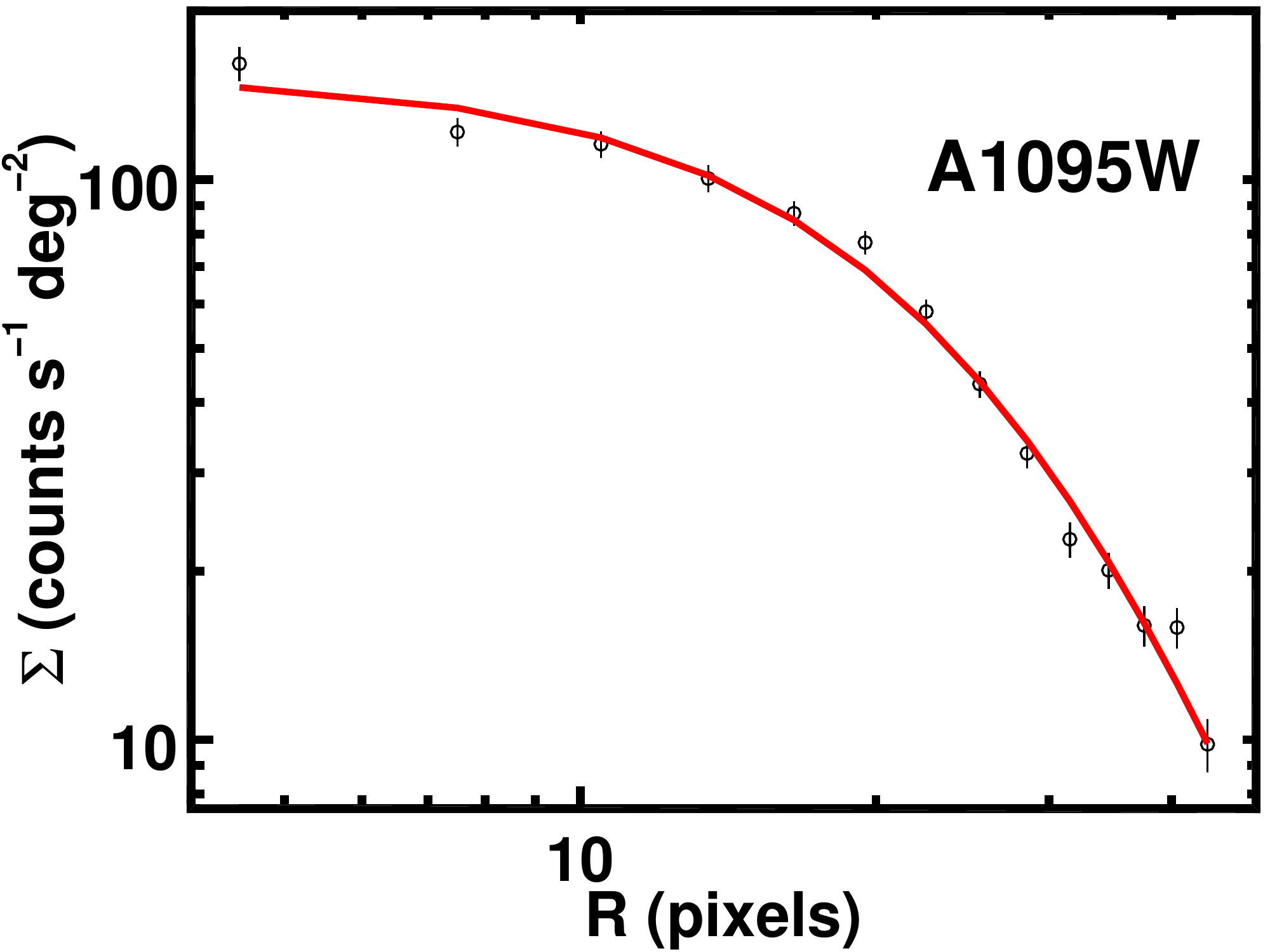}
\includegraphics[width=0.24\textwidth]{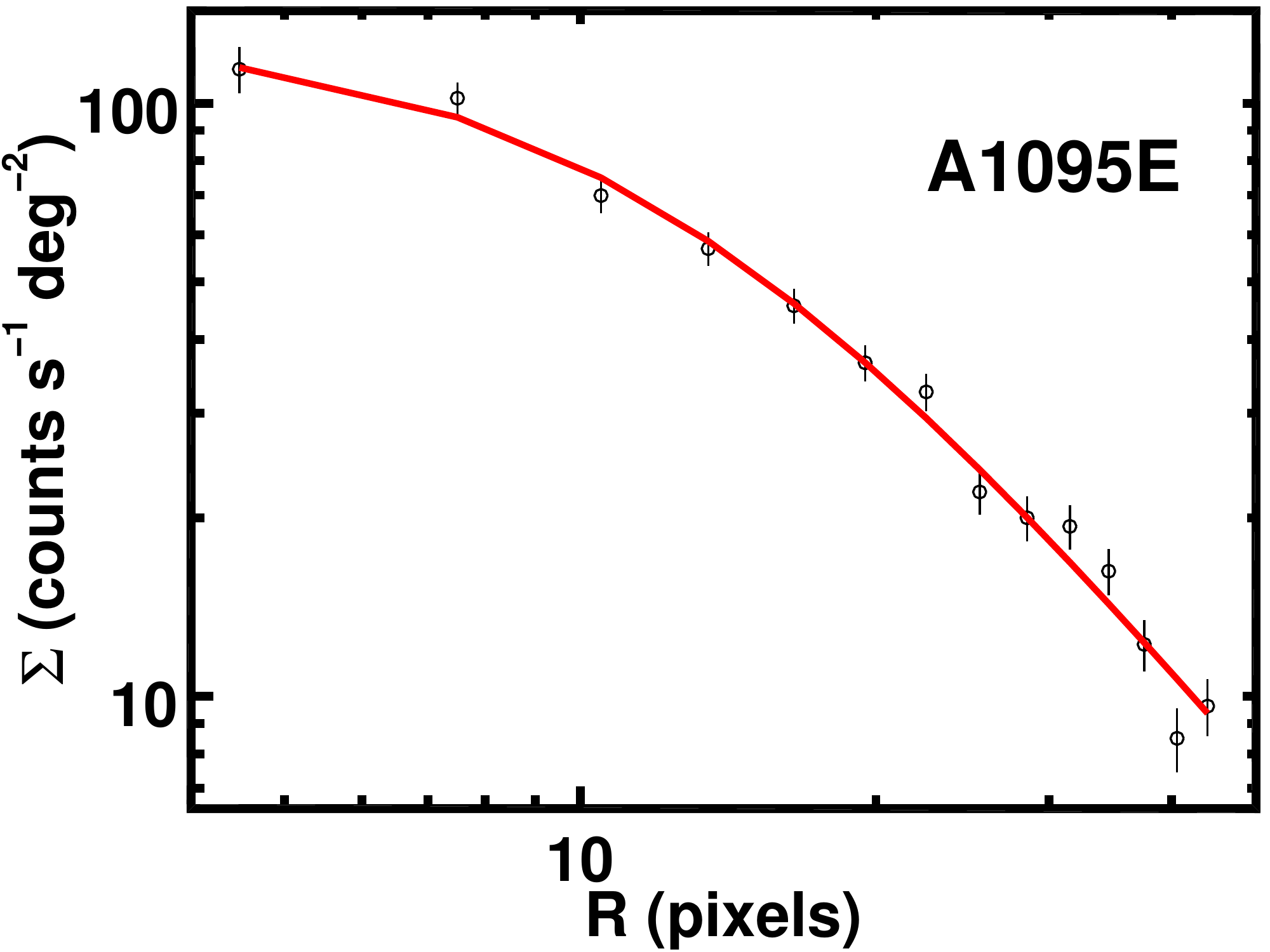}
\includegraphics[width=0.24\textwidth]{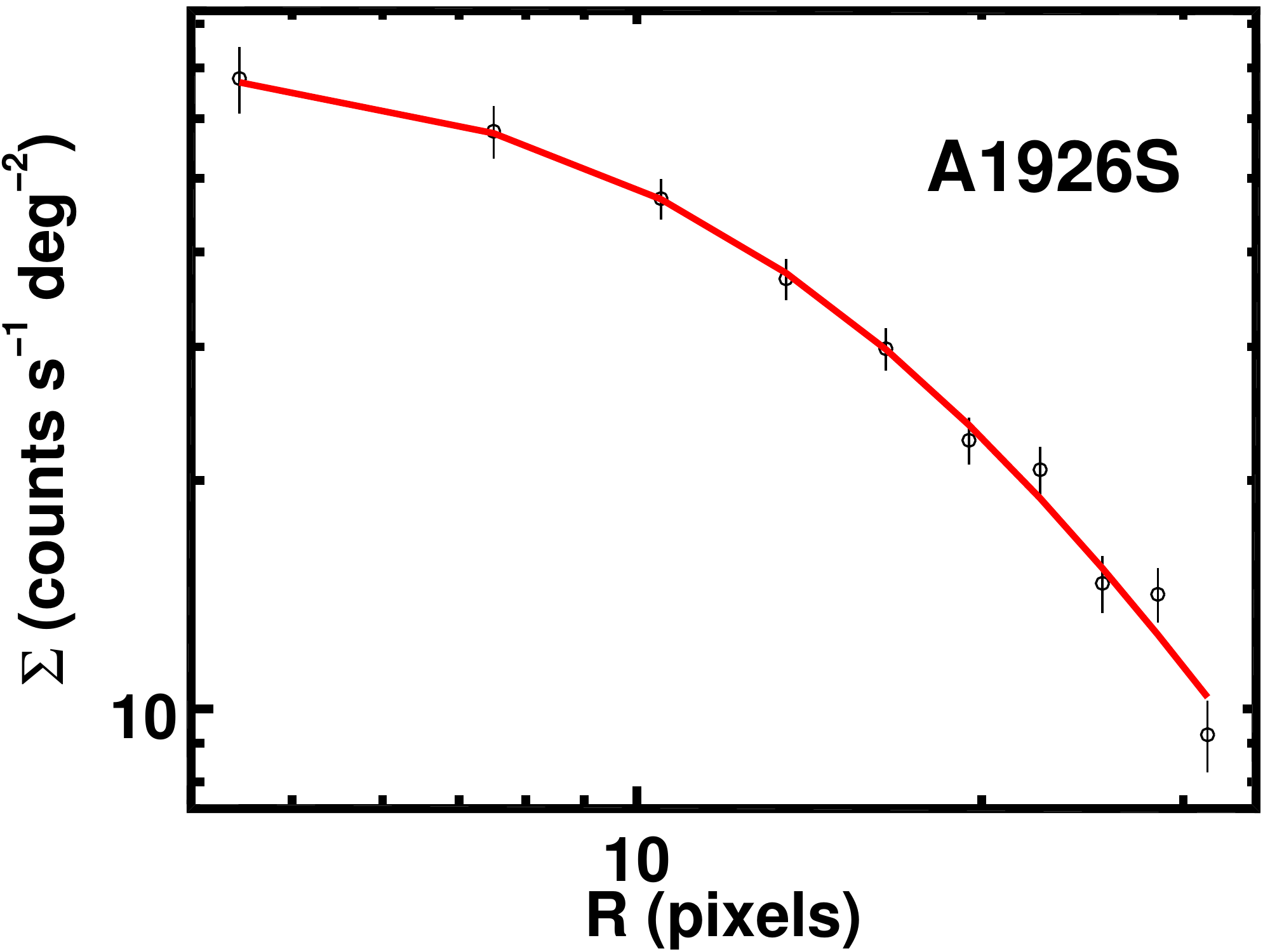}
\includegraphics[width=0.24\textwidth]{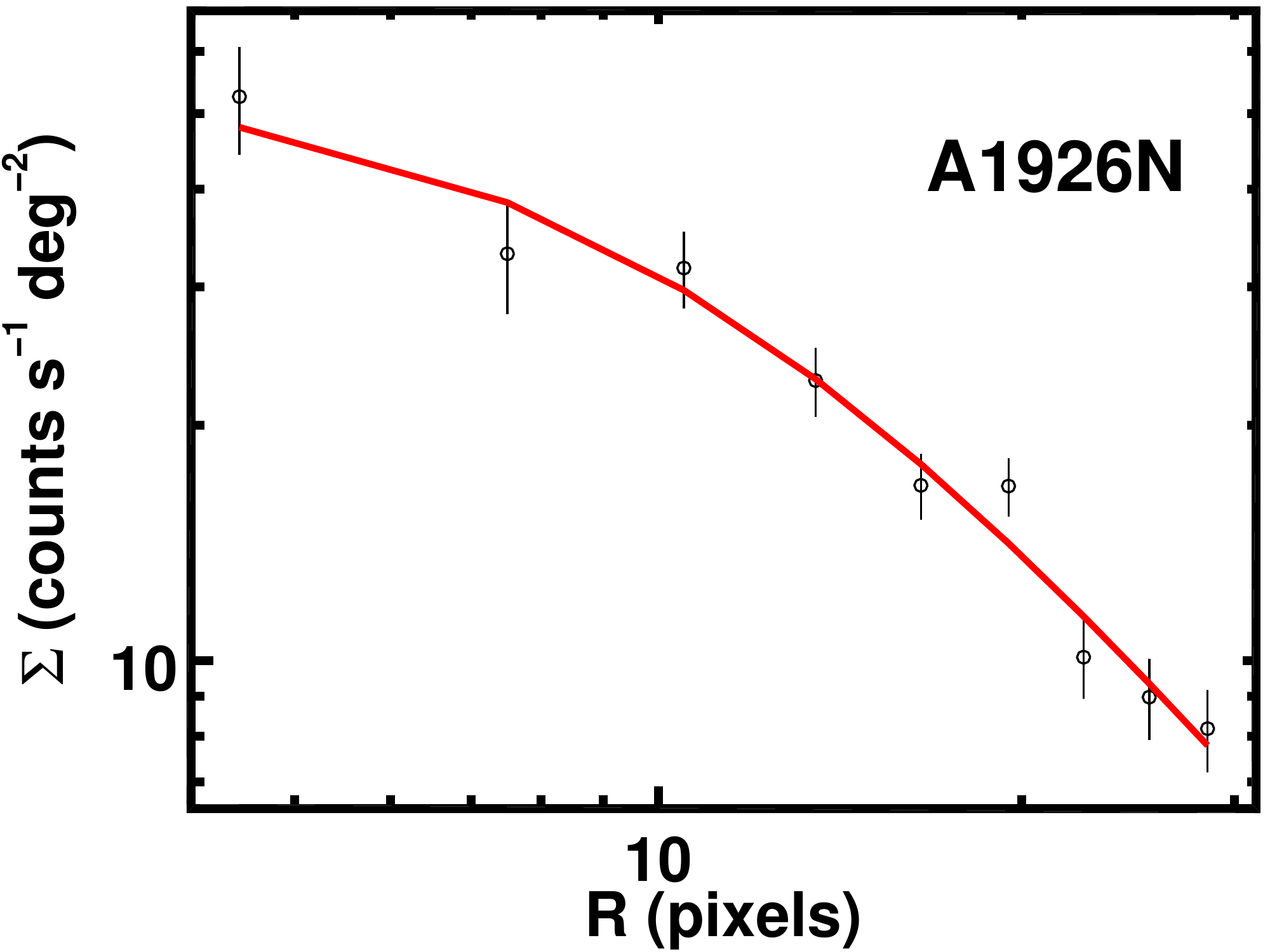}
\includegraphics[width=0.24\textwidth]{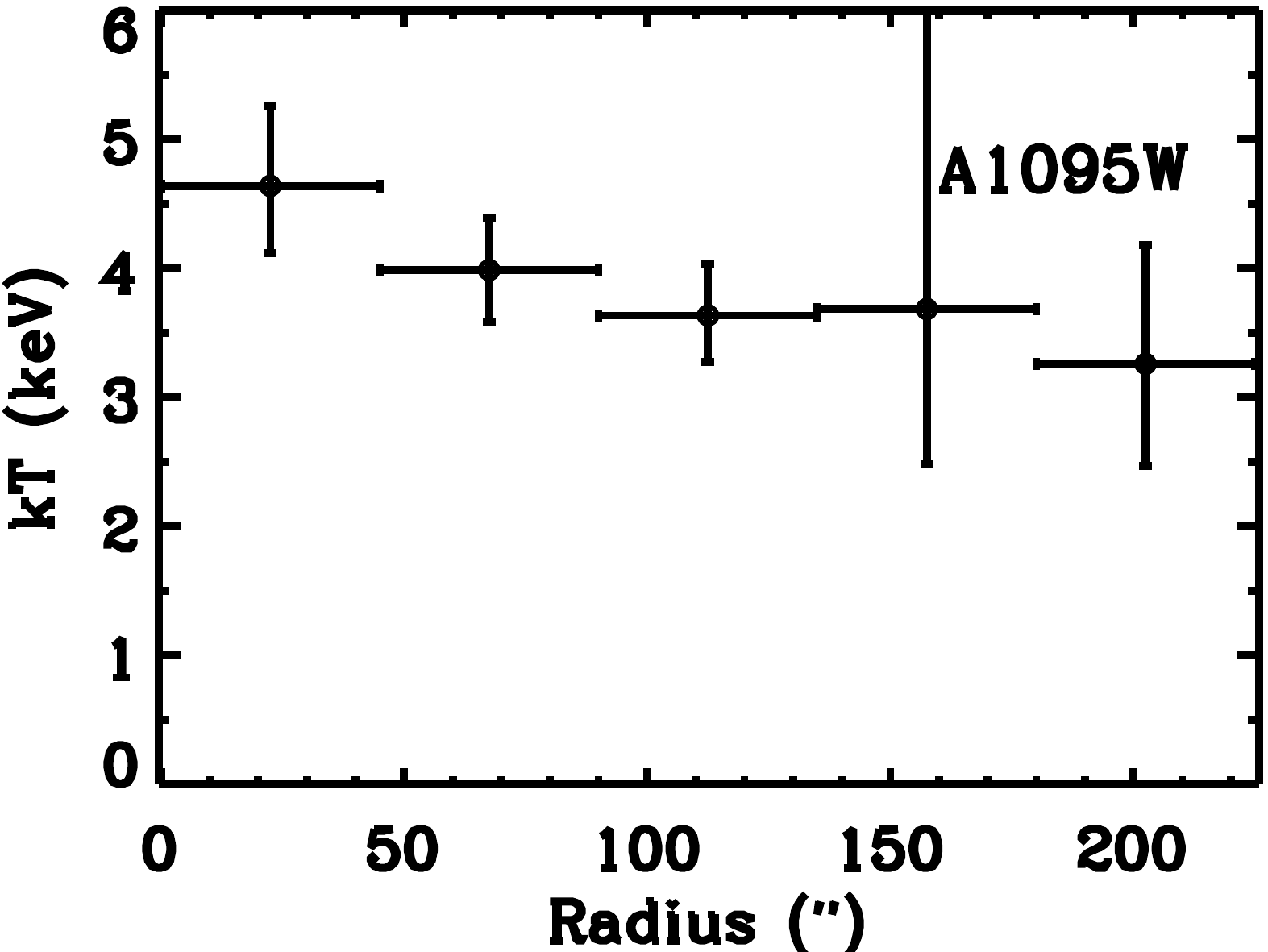}
\includegraphics[width=0.24\textwidth]{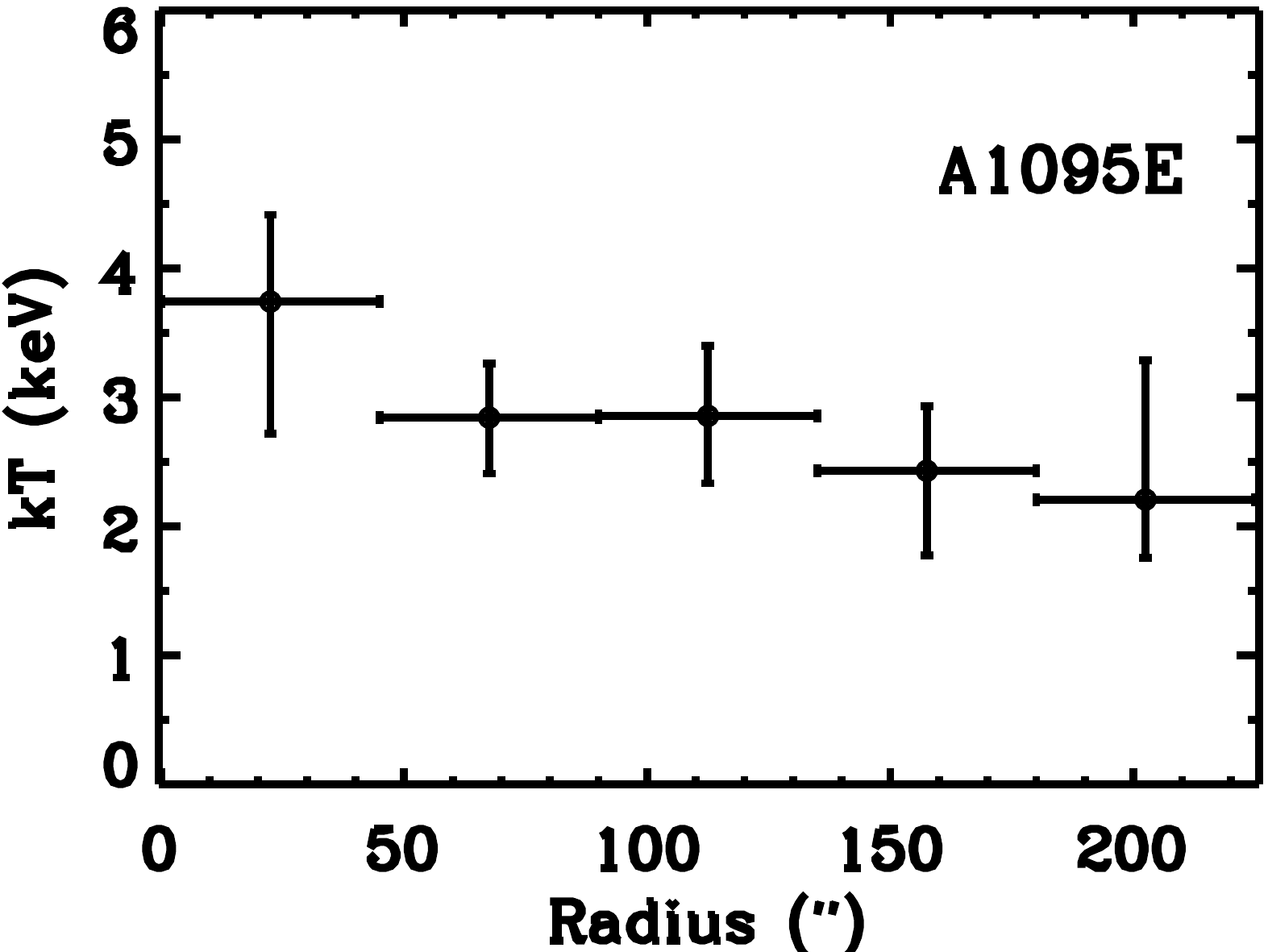}
\includegraphics[width=0.24\textwidth]{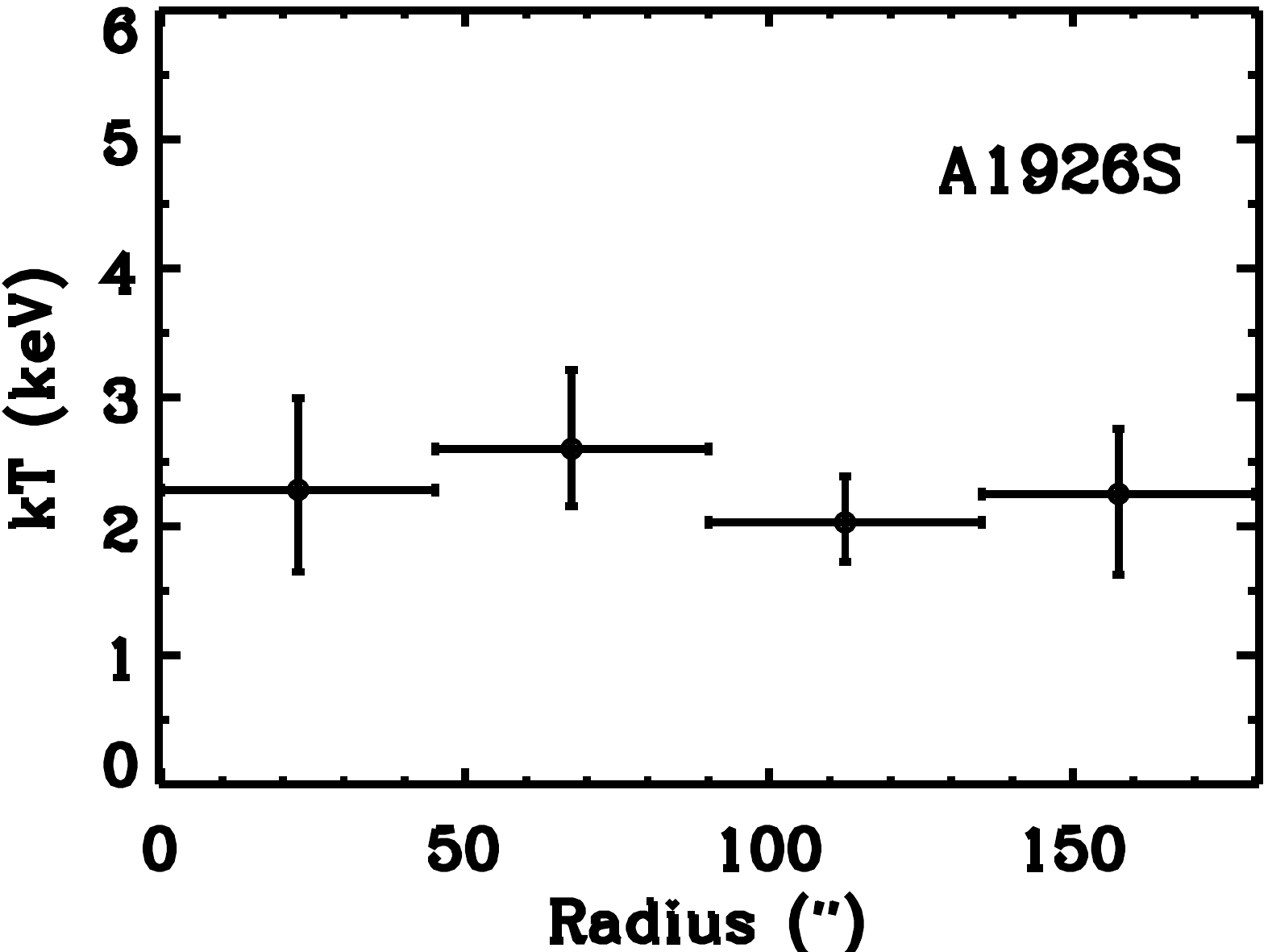}
\includegraphics[width=0.24\textwidth]{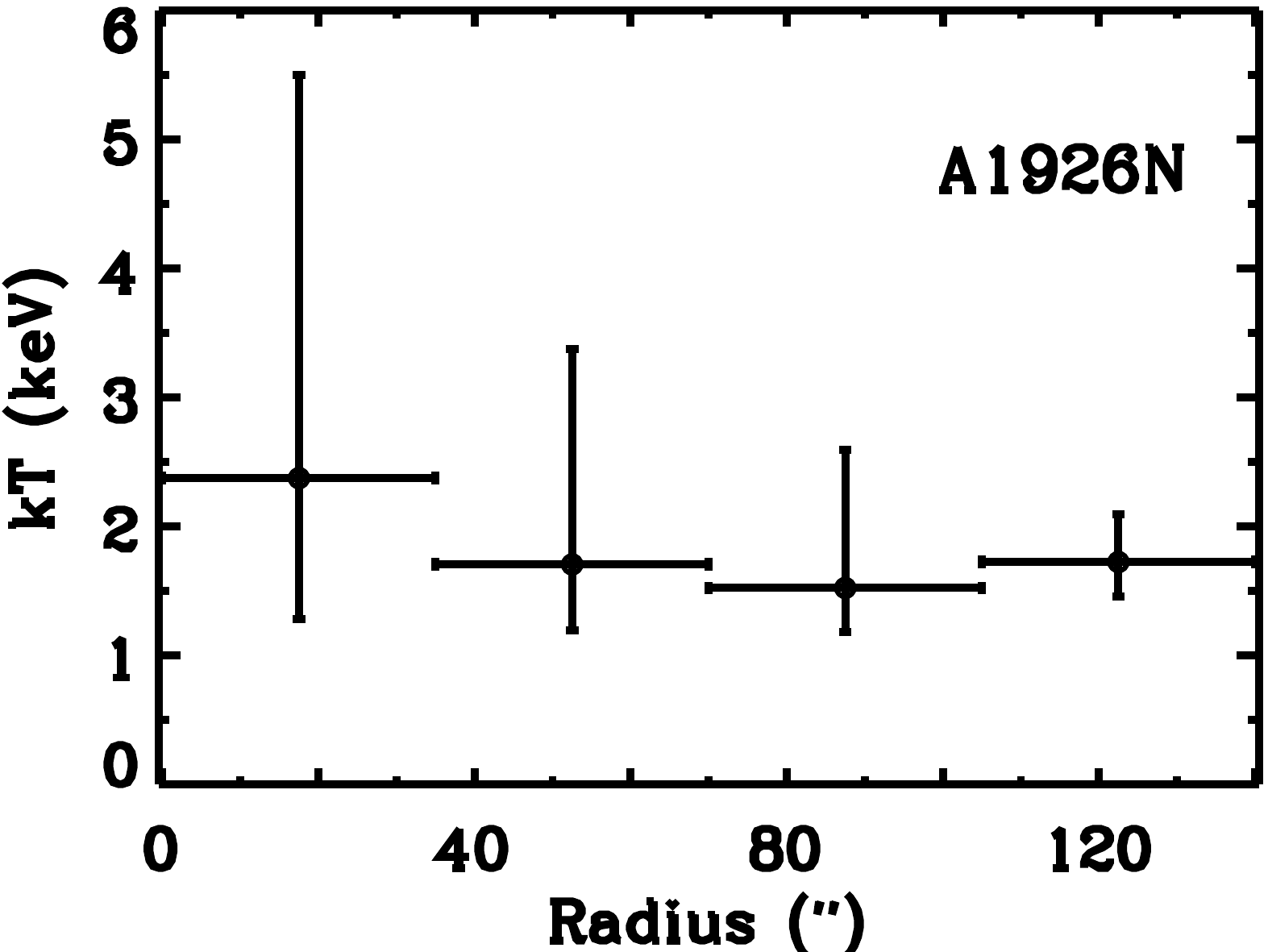}
 	\end{center}
    \caption{%
Top panels: X-ray surface brightness profiles of the diffuse 0.5-2 keV emission for each cluster, together with the best-fitting $\beta$-model. The pixel size is 5$^{\prime\prime}$, the interval between data points is 15$^{\prime\prime}$. These radial intensity profiles probe the regions within $r_{500}$.
Bottom panels: radial temperature profiles of the individually labelled clusters.
            }%
   \label{fig3}
\end{figure*}

We present the radial 0.5-2 keV intensity profiles of each cluster as a function of the projected off-center radius ($R$) in the top panels of Fig.~\ref{fig3}.
Constructed by accumulating counts in annuli of width 15$^{\prime\prime}$ around each cluster X-ray centroid, the profiles are fitted in SHERPA with the standard $\beta$-model (Cavaliere \& Fusco-Femiano 1976) of the form: 
\begin{equation}
I=I_{0} (1+x^{2})^{1/2-3\beta},
\end{equation}
where $x=R/r_{c}$, while $I_{0}$, $r_c$ and $\beta$ are the free parameters.
Their best-fitting values are included in Table~\ref{t:clusters}.

\begin{figure*}
 	\begin{center}
\includegraphics[width=0.45\textwidth]{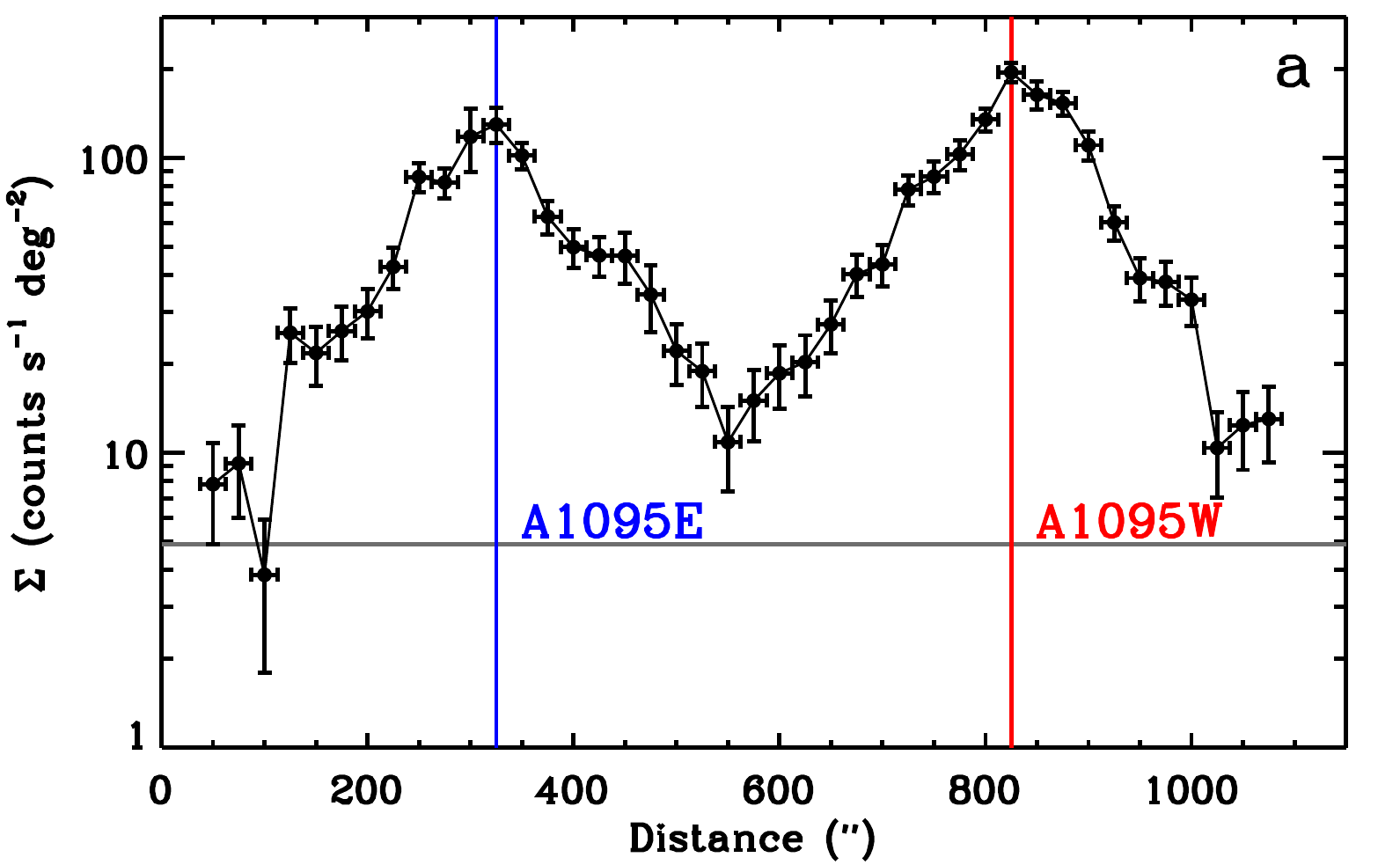}
\includegraphics[width=0.45\textwidth]{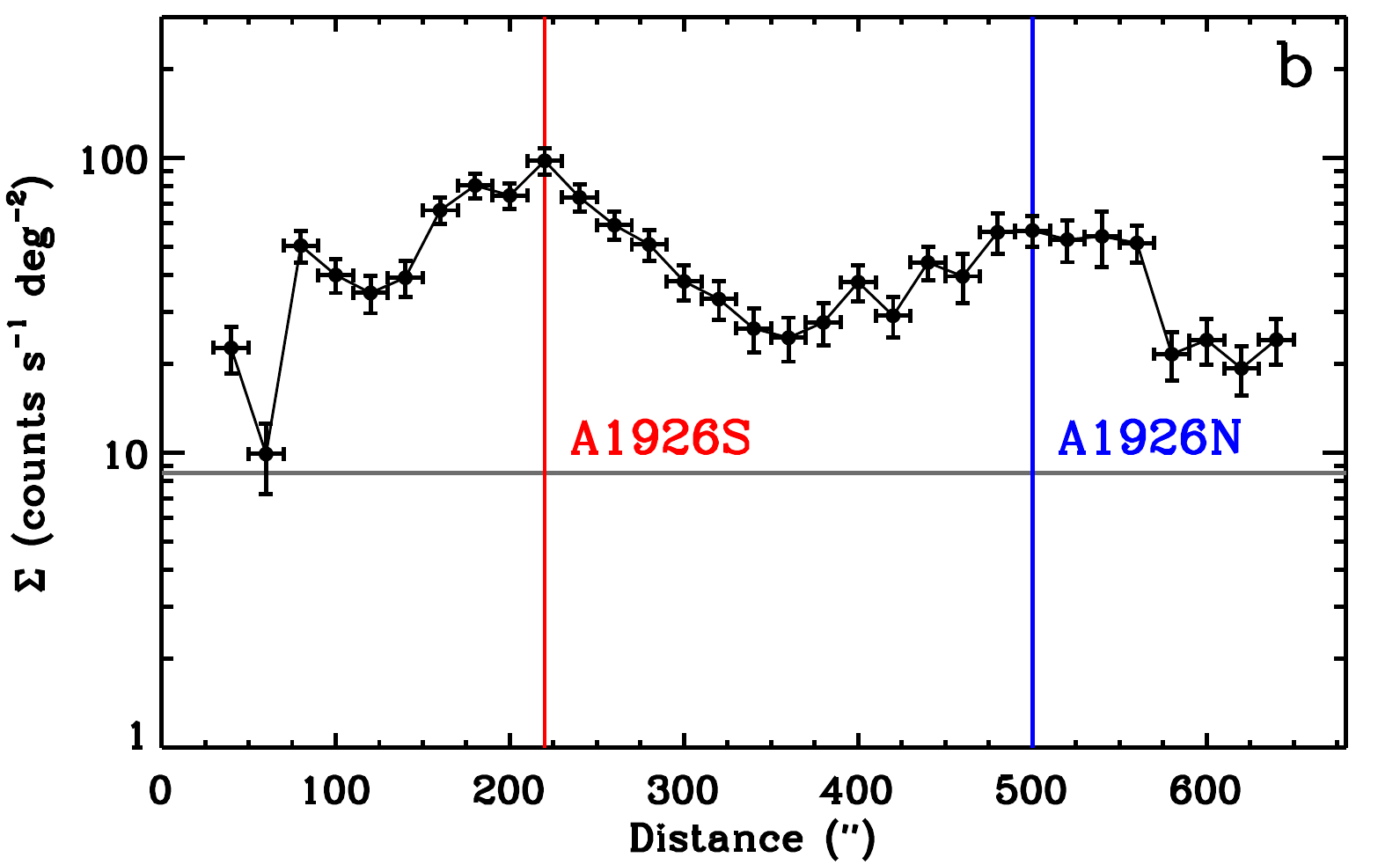}
\includegraphics[width=0.45\textwidth]{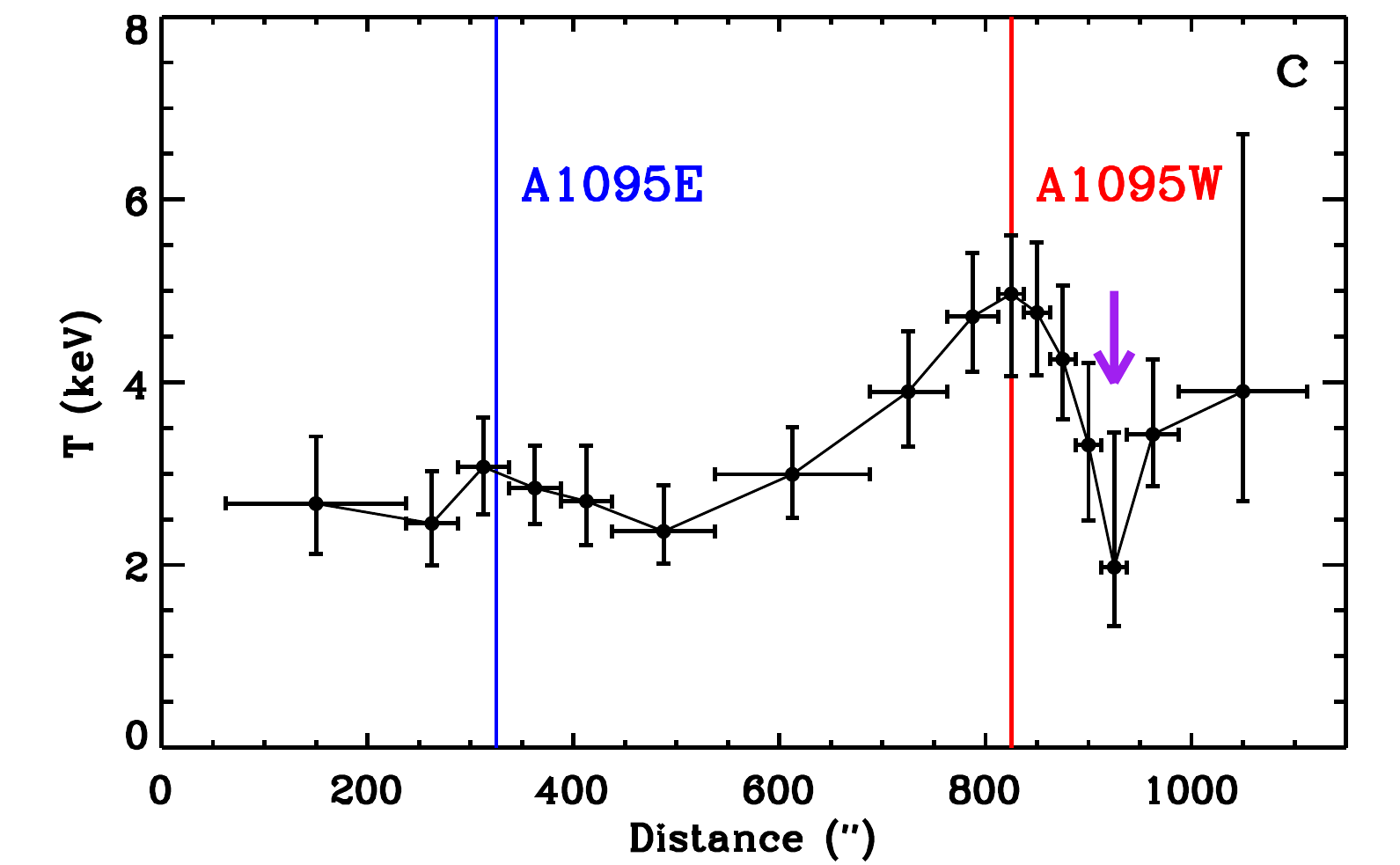}
\includegraphics[width=0.45\textwidth]{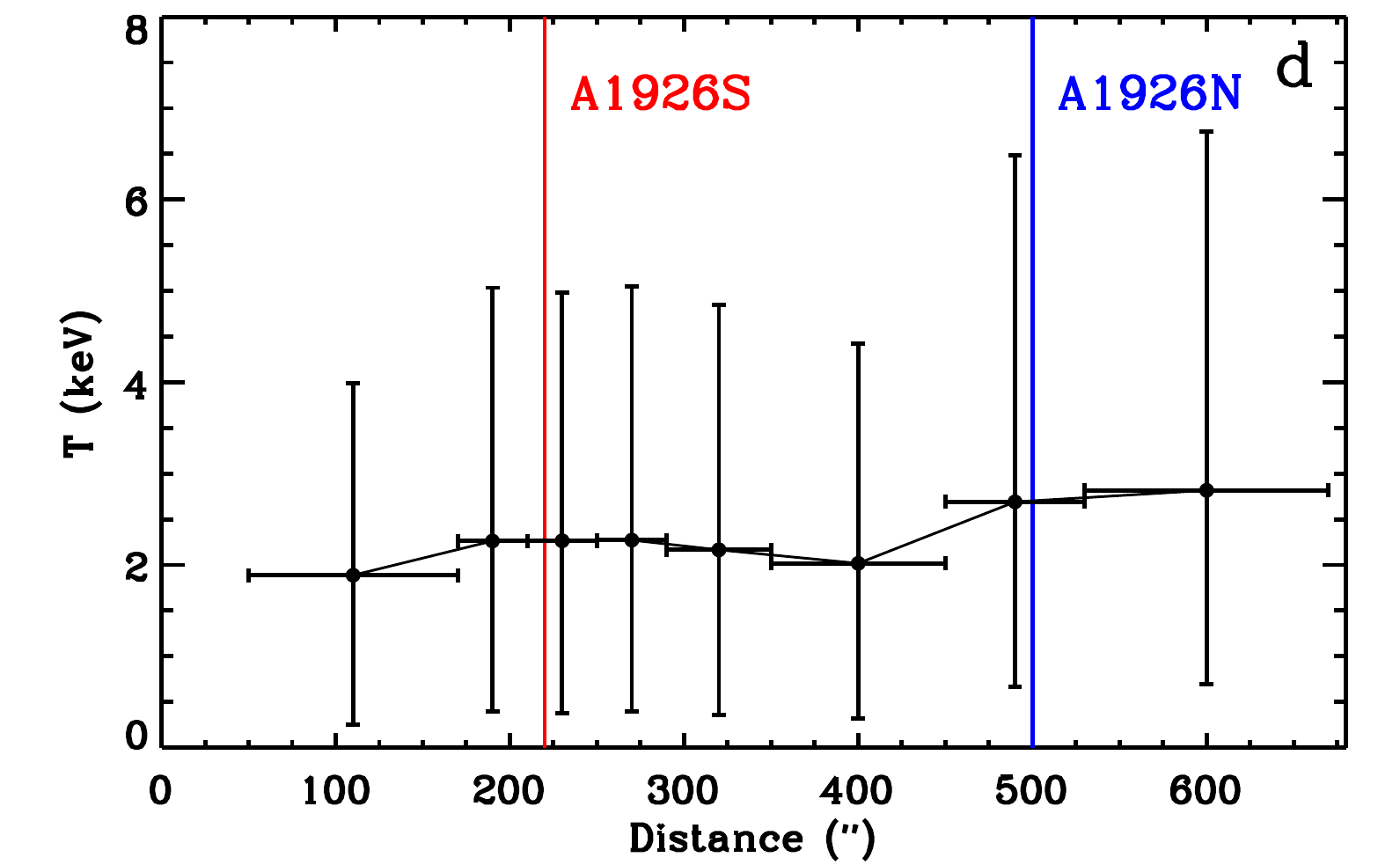}
 	\end{center}
    \caption{%
Diffuse 0.5-2 keV intensity profiles (panels a and b) and temperature profiles (c and d)  along the major axis of each cluster pair. The red and blue lines mark the positions of the diffuse X-ray centroids of individual clusters, while the gray lines in panels a and b represent the background levels. The purple arrow in the panel c indicates a temperature drop associated with a substructure on the western side of A1095W.}%
   \label{fig4}
\end{figure*}

We also extract the intensity profiles along the major axes of the cluster pairs, by accumulating counts in slices of $25^{\prime\prime} \times 50^{\prime\prime}$ and $20^{\prime\prime} \times 50^{\prime\prime}$ for A1095 and A1926 pairs, respectively  (outlined as the long rectangles in Figs.~\ref{fig1}b and \ref{fig2}b). The profiles (Figs.~\ref{fig4}a,b) show various shoulders (even local peaks), as well as asymmetries (relative to the cluster centers), indicating the presence of substantial
substructures in the ICM density distributions of the clusters.

\subsection{X-ray spectral properties}
\label{ss:x-spec}
We model the X-ray spectra of the clusters as optically-thin thermal plasmas in collisional ionization equilibrium (\emph{apec}; \S~\ref{ss:obs-spe}), multiplied by a \emph{wabs} foreground absorption with the Galactic hydrogen column density fixed at $N_H=2.76 \times 10^{20}\ \rm{and}\  2.59 \times 10^{20}\ cm^{-2}$ (Dickey \& Lockman 1990) for A1095 and A1926, respectively.
The results for individual clusters are listed in Table~\ref{t:clusters}. 
The bottom panels of Fig.~\ref{fig3} presents the radial temperature profiles, which are extracted from a set of 45$^{\prime\prime}$-wide annuli  around the X-ray centroid of 
each cluster (except for 35$^{\prime\prime}$ 
in the A1926N case). 

In order to examine evidence for possible shocks, 
we also obtain the temperature profiles (Figs.~\ref{fig4}c,d) 
along the major axes (Figs.~\ref{fig1}b and 
\ref{fig2}b) of the cluster pairs. In general, the counting statistics from individual
spectral extraction regions are only moderate, and the angular resolution is not sufficient, which results in uncertain data that cannot reveal sharp temperature jumps.
But one can readily see a steep temperature drop 
(indicated by the arrow in Fig.~\ref{fig4}c), associated with 
an intensity substructure on the western side of A1095W (marked as Source 3 in 
Fig.~\ref{fig1}a).

\begin{table*}
\caption{Properties of galaxy clusters }
\tabcolsep=0.10cm
\footnotesize
\begin{tabular}{lcccc}
\hline\hline
Name & A1095W & A1095E & A1926S & A1926N\\
Redshift & 0.210 & 0.213 & 0.136 & 0.136\\
X-ray centroid position & $10^h47^m24.8^s, +15^\circ14^{\prime}13^{\prime\prime}$ & $10^h47^m59.6^s, +15^\circ16^{\prime}05^{\prime\prime}$  & $14^h30^m26.5^s, +24^\circ36^{\prime}09^{\prime\prime}$ & $14^h30^m30.5^s, +24^\circ40^{\prime}44^{\prime\prime}$\\
BCG position & $10^h47^m29.0^s, +15^\circ14^{\prime}02^{\prime\prime}$ & $10^h48^m00.5^s, +15^\circ16^{\prime}06^{\prime\prime}$  & $14^h30^m21.9^s, +24^\circ34^{\prime}29^{\prime\prime}$ & $14^h30^m28.6^s, +24^\circ40^{\prime}19^{\prime\prime}$\\
Offset ($^\prime$/Mpc) & 1.0/0.21 & 0.21/0.04 & 2.0/0.29 & 0.6/0.09 \\
$CR (\rm{count\ s^{-1}})/\eta (10^{-4})$ & 0.67/16.5 & 0.32/9.6 & 0.20/4.7 & 0.10/2.2 \\
Temperature (keV) & $3.6^{+0.2}_{-0.2}$ & $3.3^{+0.2}_{-0.4}$ & $2.2^{+0.2}_{-0.2}$ & $2.0^{+0.9}_{-0.4}$\\
$\chi_{T}^{2}/$d.o.f. & 859/979 & 600/601 & 148/126 & 104/85\\
$M_{500}/M_{200} (10^{14} M_\odot)$ & $1.9^{+0.2}_{-0.2}/2.7^{+0.3}_{-0.3}$  &$1.7^{+0.2}_{-0.4}/2.3^{+0.3}_{-0.6}$ & $0.9^{+0.2}_{-0.1}/1.3^{+0.2}_{-0.2}$  &$0.8^{+0.6}_{-0.3}/1.1^{+0.8}_{-0.4}$\\
$r_{500}/r_{200}$ (Mpc) & $0.82^{+0.03}_{-0.03}/1.24^{+0.04}_{-0.04}$ & $0.78^{+0.03}_{-0.06}/1.19^{+0.04}_{-0.09}$ & $0.65^{+0.04}_{-0.03}/0.99^{+0.06}_{-0.04}$ & $0.62^{+0.16}_{-0.08}/0.94^{+0.24}_{-0.12}$\\
$I_0 \rm {(count\ s^{-1}\ deg^{-2})}$ & $153.5^{+10.1}_{-9.4}$ & $130.5^{+20.4}_{-16.4}$ & $73.8^{+12.5}_{-9.7}$ & $55.9^{+28.5}_{-14.4}$\\
$r_c$ ($^\prime$/Mpc) & $2.9^{+0.7}_{-0.5}/0.59^{+0.14}_{-0.10}$ & $1.0^{+0.3}_{-0.2}/0.21^{+0.06}_{-0.04}$ & $1.3^{+0.6}_{-0.4}/0.19^{+0.09}_{-0.06}$ & $0.9^{+0.9}_{-0.5}/0.13^{+0.13}_{-0.07}$ \\
$\beta$ & $1.13^{+0.38}_{-0.22}$ & $0.50^{+0.06}_{-0.04}$ & $0.57^{+0.22}_{-0.10}$ & $0.48^{+0.28}_{-0.10}$ \\
$\chi_{\beta}^{2}/$d.o.f. & 23.9/11 & 14.1/11 & 5.4/7 & 6.8/6 \\
$n_0$ (10$^{-3}$ cm$^{-3}$) & $0.84$ & $1.03$ & $0.76$ & $0.70$ \\
$M_{gas}/f_{gas} (10^{13} M_\odot/\%)$ & 1.7/9.0 & 1.3/7.4 & 0.50/5.5 & 0.34/4.4 \\
$M_{star}/f_{star} (10^{13} M_\odot/\%)$ & 0.35(0.45)/1.8(2.4) & 0.35(0.42)/2.0(2.5) & 0.33(0.30)/3.6(3.3) & 0.26(0.28)/3.3(3.6) \\
QSO projected distance ($^\prime$) & 4.2 & 5.0 & 14.8 & 12.7 \\
$N_p$ ($10^{20}$ cm$^{-2}$) & $6.9$ & $14.9$ & $3.2$ & $4.9$ \\ 
\hline 
\end{tabular}
\begin{tablenotes}
      \small
      \item $Note.$ Errors are at the 90\% confidence level. The redshifts of three GMBCG clusters are determined from the values of BCGs, while the redshift of A1926S is estimated from average redshifts of 7 galaxies with spectroscopic redshifts near the X-ray centroid. The offset is the projected distance between the X-ray centroid and the BCG. The count rate/normalization ($CR/\eta$) as well as
the temperature are from the best-fitting {\it apec} model of each cluster. The cluster mass ($M_{500}/M_{200}$) and radius ($r_{500}/r_{200}$) are estimated from the temperature (Arnaud et al. 2005). The $\beta$-model parameters of Eq. (1) are from the best fit to the 0.5-2 keV intensity profile. The inferred parameters include the central proton density ($n_0$), the hot gas mass ($M_{gas}$) and stellar mass ($M_{star}$) within $r_{500}$, and the ratio of $M_{gas}$ or $M_{star}$ to $M_{500}$ ($f_{gas}$ or $f_{star}$), the stellar mass in the parentheses is estimated with the relation of Gonzalez et al. (2013). The proton column density ($N_p$) is at the projected distance of the corresponding UV-bright background QSO (SDSS J104741.75+151332.2 for the A1095W/A1095E pair; SDSS J143125.88+244220.6 for the A1926S/A1926N pair).
    \end{tablenotes} 
\label{t:clusters}    
\end{table*}

Table~\ref{t:spec} presents the spectral fit results of luminous compact X-ray sources/features (marked in Figs.~\ref{fig1}a and \ref{fig2}a)
in the field of each cluster. 
These sources are also detected by the XMM-ESAS source detection routine, marked with the same numbers (in Figs.~\ref{fig1}a and \ref{fig2}a) as Column (2) of Tables~\ref{t:s-a1095} and \ref{t:s-a1926}.
These spectra are typically well fitted with an absorbed power-law model, except for Sources 2, 3, 8, 11 and 13 in the A1095 field and Source 5 in the A1926 field, which are better fitted with a thermal plasma model.
This modeling is mainly intended to characterize the overall spectral shapes of the sources, but may also provide hints about their nature.

\section{Discussion}
\label{s:dis}
In this section we discuss the relationship between cluster pairs (Section 4.1), 
their dynamical states and substructure properties (Section 4.2),
their large scale environment (Section 4.3),
the complementarity of the optical and X-ray detections of clusters (Section 4.4),
and their baryon content (Section 4.5).
This discussion is largely based on a comparison of our X-ray analysis results with existing multi-wavelength data, mostly in optical and radio bands.

\subsection{Relationship between cluster pairs}
\label{ss:dis-rel}
\begin{figure*}
 	\begin{center}
\includegraphics[width=0.45\textwidth]{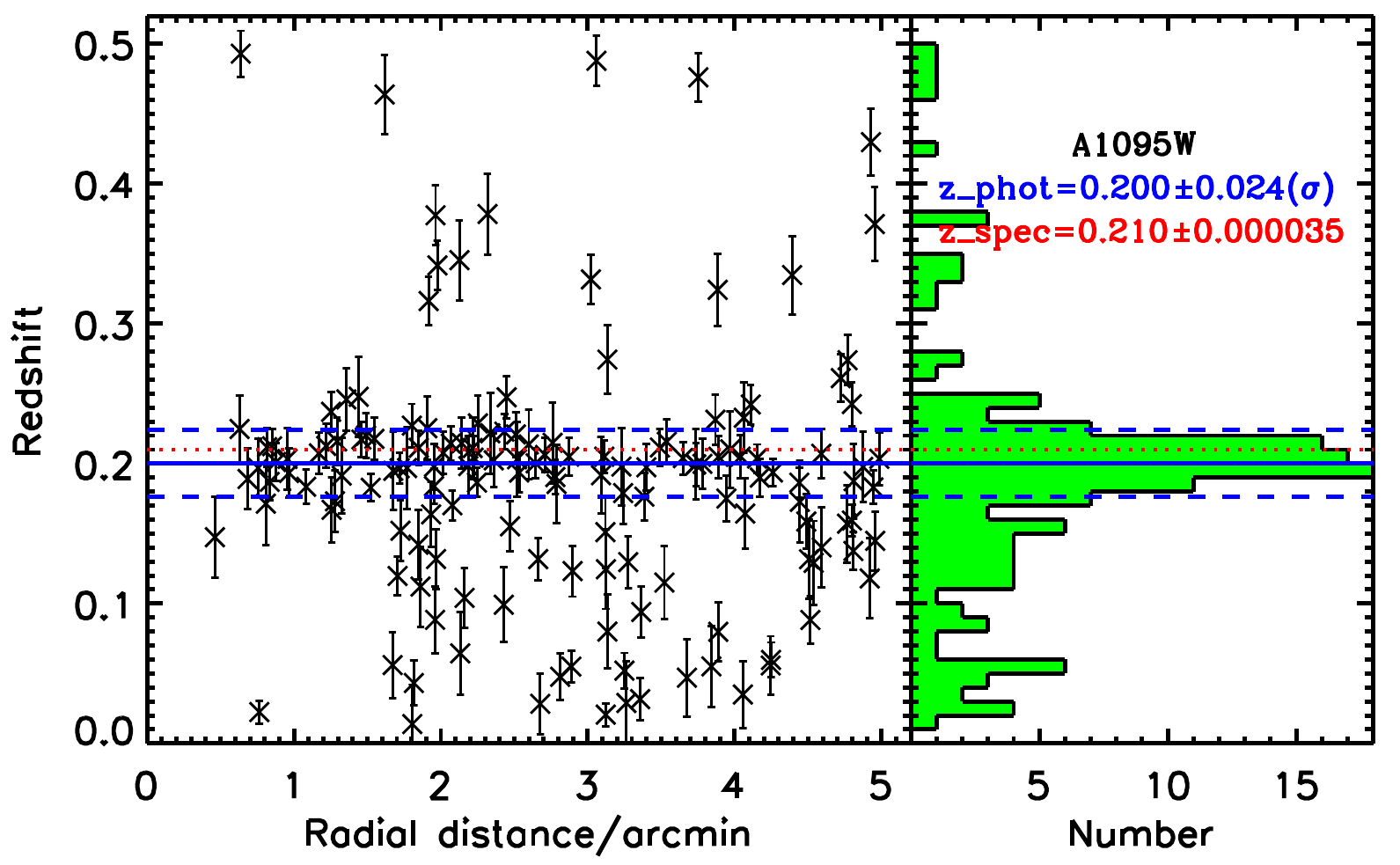}
\includegraphics[width=0.45\textwidth]{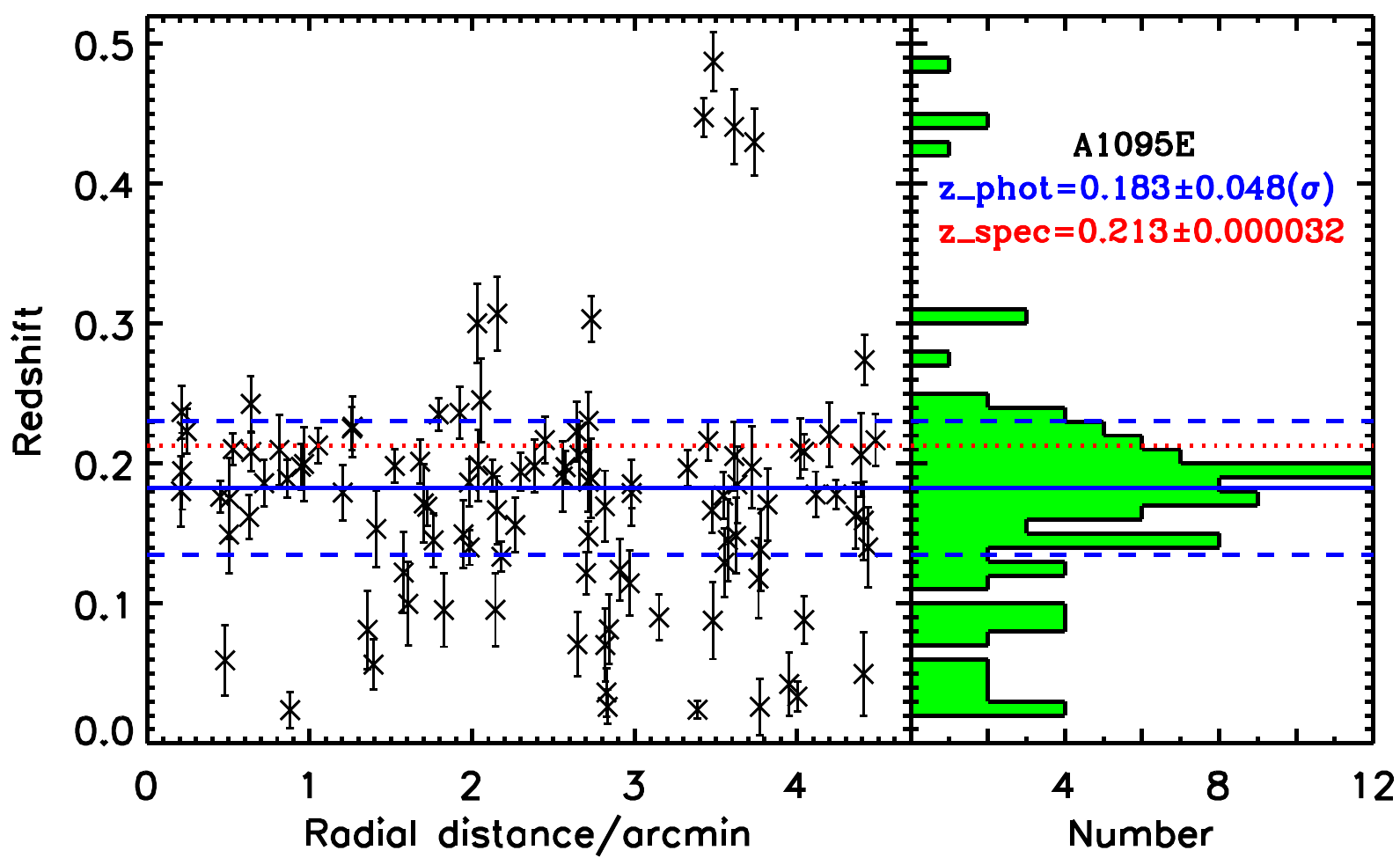}
\includegraphics[width=0.45\textwidth]{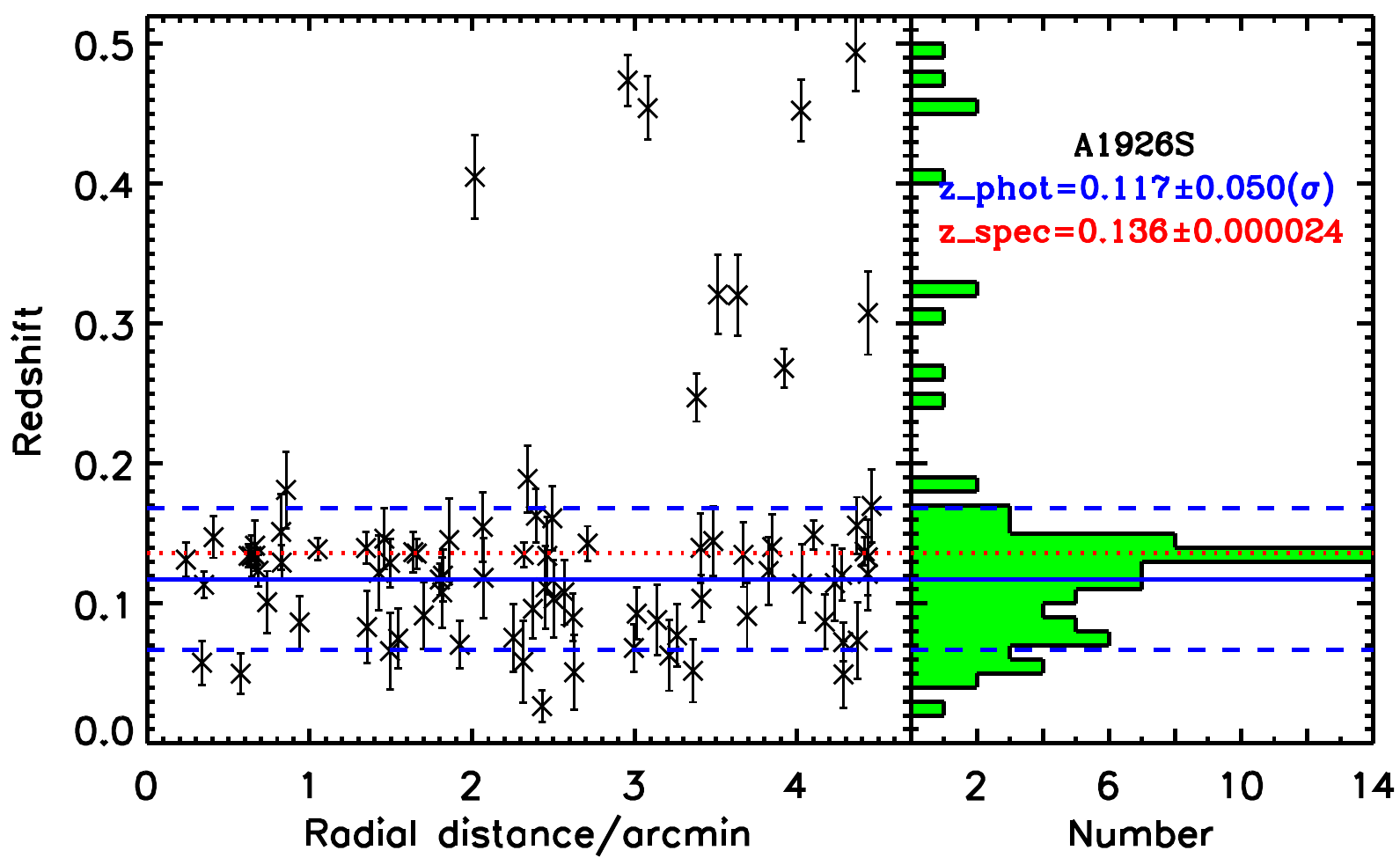}
\includegraphics[width=0.45\textwidth]{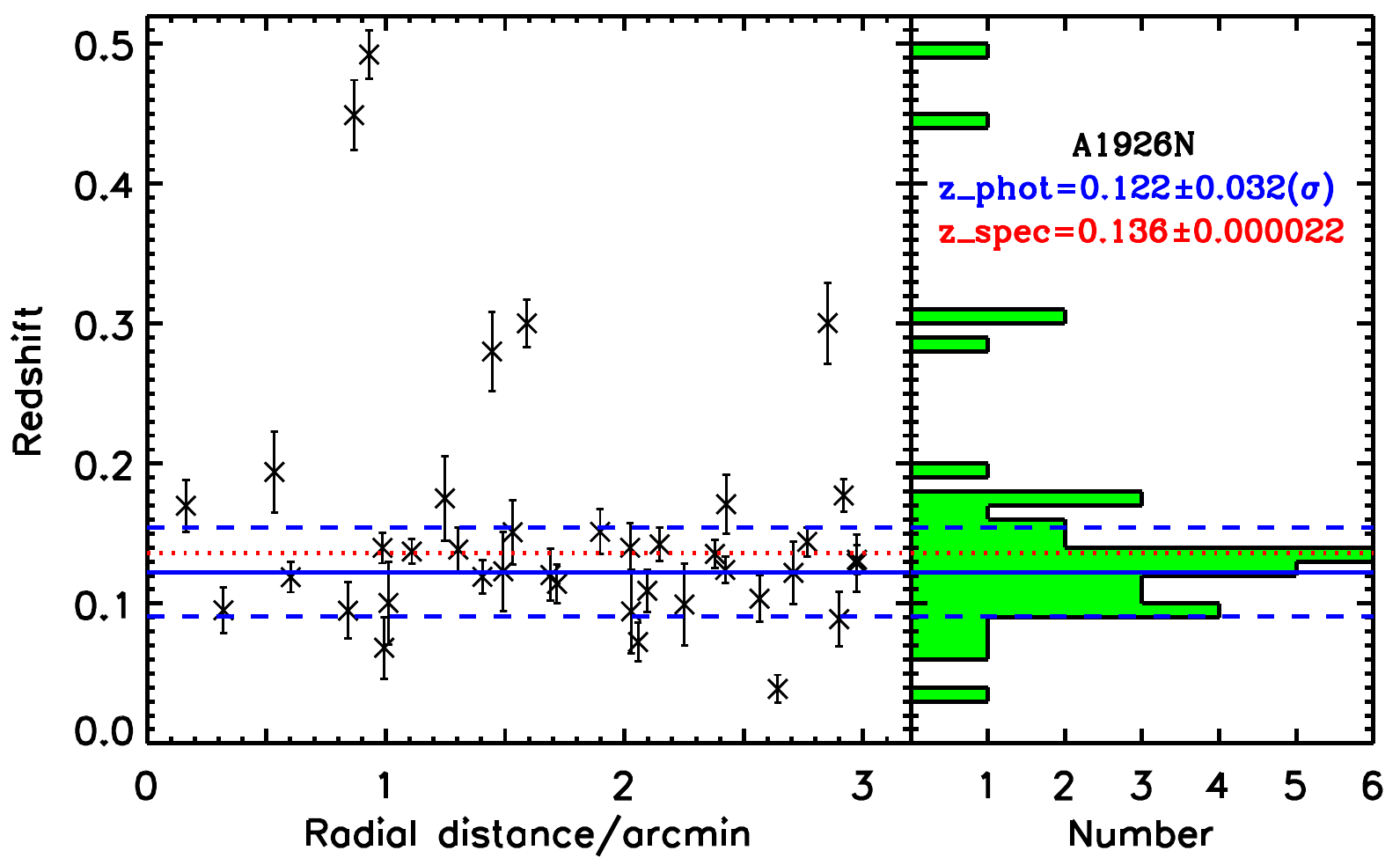}
 	\end{center}
    \caption{%
Distribution of galaxy photometric redshifts as a function of the radial projected distance away from the X-ray centroid of each cluster, and the redshift histograms. 
In each panel, the solid blue line represents the Gaussian-fitted mean redshift (z\_phot) of the galaxies. This redshift is somewhat biased toward low redshift, because of  the increasing incompleteness of the galaxy sample with increasing redshift. The dashed blue lines mark the $\pm \sigma$ redshift boundaries of the galaxies selected for the stellar mass estimate of the cluster.
The dotted red line represents the spectroscopic redshift (z\_spec) of the BCG (except for A1926S, for which the average spectroscopic redshift of the 7 galaxies near the cluster center is adopted).
            }%
   \label{fig5}
\end{figure*}

\begin{figure*}
\centering
\includegraphics[width=0.8\textwidth , angle=0]{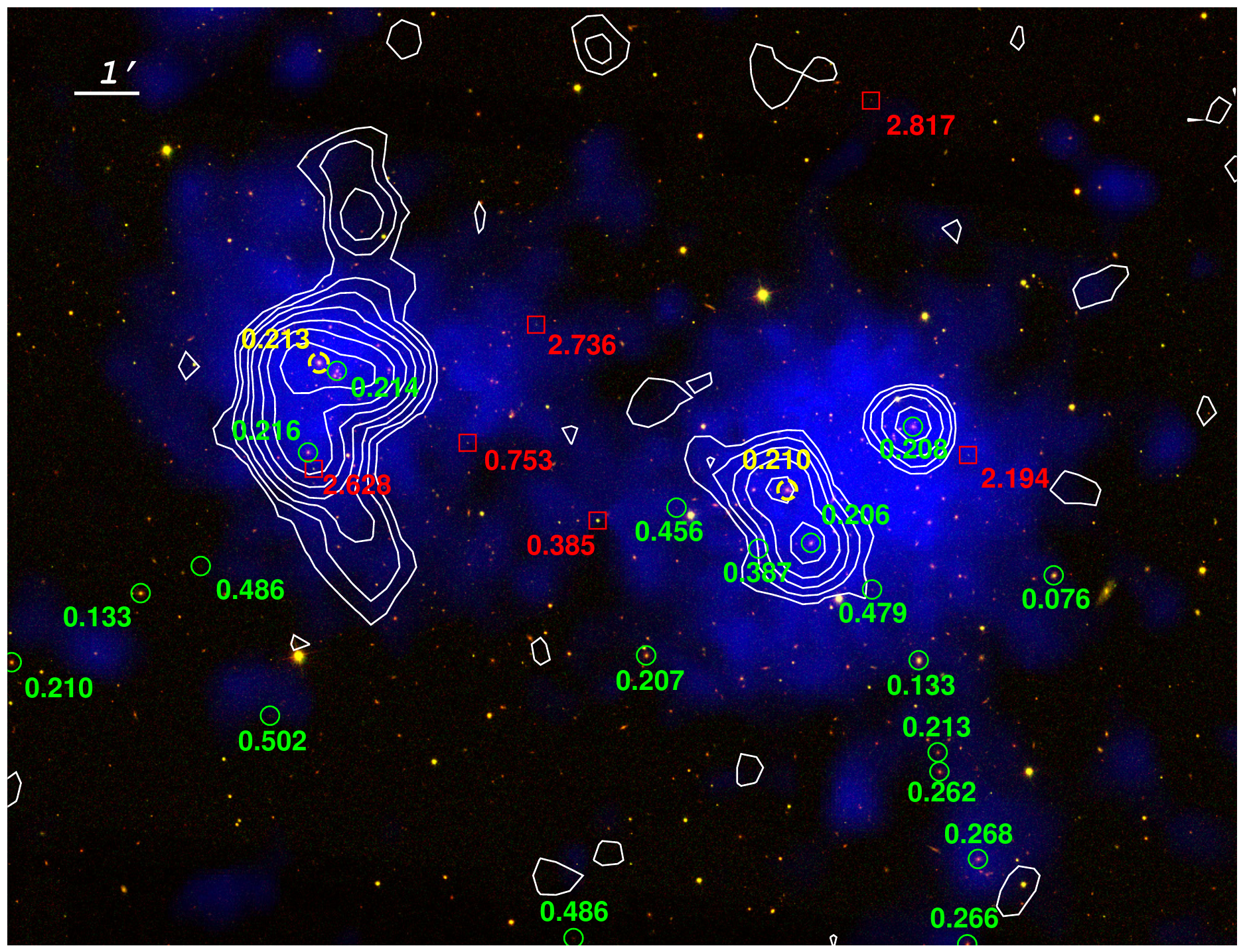}
\caption{Multi-wavelength montage of the A1095W and A1095E field: SDSS r-band (red), g-band (green) and  diffuse 0.5-2 keV X-ray emission (blue), marked with positions of galaxies (solid green circles), BCGs (dashed yellow circles) and QSOs (red boxes), together with the numbers of SDSS spectroscopic redshifts. Overlaid contours are from the NVSS map of the 1.4 GHz continuum intensity at 1, 2, 4, 8, 16, and 32 $\rm{mJy\ beam^{-1}}$.}
\label{fig6}
\end{figure*}

\begin{figure*}
\centering
\includegraphics[width=0.8\textwidth , angle=0]{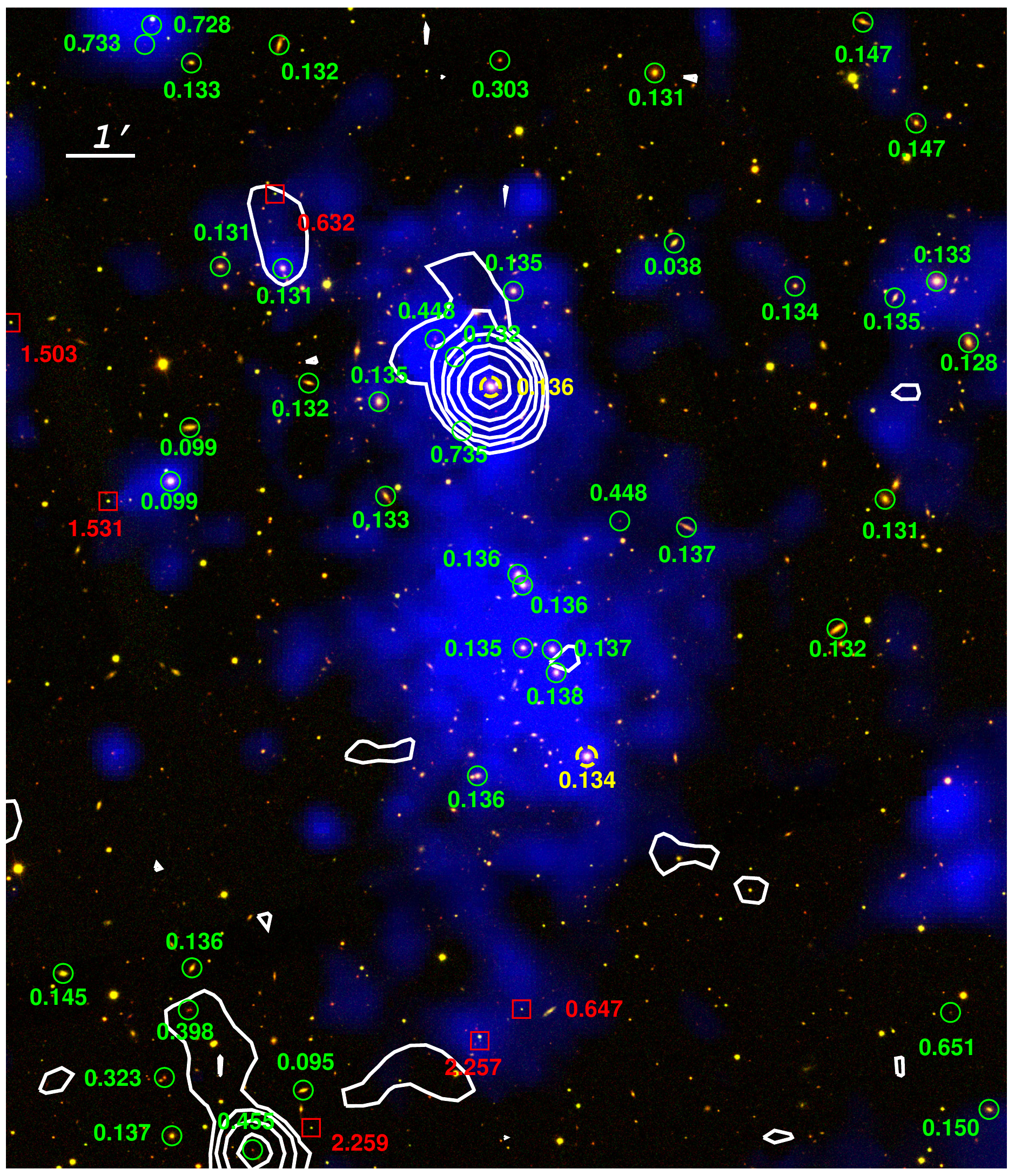}
\caption{Same as Fig~\ref{fig6} but for the A1926S and A1926N field.}
\label{fig7}
\end{figure*}

Given the projected proximity of the 
clusters in each pair, the question is whether or not they are physically associated. We address this question by comparing the galaxy redshift distributions of the clusters. Fig.~\ref{fig5} shows the SDSS photometric redshift distribution of galaxies along the projected radial distance from the X-ray centroid of each cluster and the corresponding redshift histograms. Galaxies with SDSS spectroscopic redshifts are shown in Figs.~\ref{fig6}-\ref{fig7}. The BCGs are marked with yellow circles, whose redshifts are adopted as those of the clusters, except for A1926S, whose BCG (z=0.134) is spatially far off the X-ray centroid. Instead, the average spectroscopic redshift (0.136) of the 7 galaxies near the centroid (Fig.~\ref{fig7}) is adopted for A1926S.
The blue and red lines in Fig.~\ref{fig5} represent the peak photometric redshifts of galaxies and the spectroscopic redshifts of the BCGs in the clusters.
Accounting for uncertainties in these redshift estimates and their biases, the two clusters in each pair appear to have consistent redshifts.

With the redshift of each cluster, we can now use the above X-ray temperature
measurements to estimate the gravitational mass ($M_{200}$) and 
virial radius ($r_{200}$). We use the scaling relations of Arnaud, Pointecouteau \& Pratt (2005): $h(z)M_{200} = 5.3 \times 10^{14} M_{\odot} (kT/5 \rm{keV})^{1.72}$, and $h(z)r_{200} = 1.7 \times 10^{3}  \rm{kpc}(\emph{kT}/5 \rm{keV})^{0.57}$, where $h^2(z)=\Omega_m(1+z)^3+\Omega_{\Lambda}$. We also estimate the gravitational mass $M_{500}$ within $r_{500}$, to which our X-ray data are sensitive,
and list the results in Table~\ref{t:clusters}. The clusters in each pair have comparable mass and size because they have a similar temperature. Although we do not have clear spectral evidence for inter-cluster shocks, 
the morphological irregularity and orientation of the paired clusters, as shown
in \S~\ref{ss:x-morph}, do strongly indicate that they are
interacting with each other. We explore this issue further in the following
subsection.

\subsection{Dynamic states and substructures of the clusters}
\label{ss:dis-sub}

We use the two-body model (e.g., Beers, Geller \& Huchra 1982; Mohr \& Wegner 1997) to estimate the total mass needed for a cluster pair to be bound.
The system is bound if it has a negative total energy:
\begin{equation}
\frac{V^2}{2} \leq \frac{GM_{tot}}{R},
\label{e:v}
\end{equation}
where the relative line-of-sight velocity $V_{r}$ and the projected separation $R_p$ of the two clusters are related to the intrinsic parameters in the form of:
\begin{equation}
V_r=V\ sin\alpha,\ R_p=R\ cos\alpha,
\end{equation}
depending on the angle ($\alpha$) between the line joining the clusters and the plane of the sky.
Eq.~\ref{e:v} can be re-arranged as
\begin{equation}
V^2_rR_p \leq 2GM_{tot}\ sin^2\alpha\ cos\alpha,
\label{e:m}
\end{equation}
so that its left side contains  only the observable parameters.
We estimate $V_{r}$ as $c\Delta z/(1+z)$, where $z$ is the spectroscopic redshift of a BCG, which tends to be a good tracer of the cluster gravitational mass center (e.g., Oguri et al. 2010; George et al. 2012; Zitrin et al. 2012; Cui et al. 2016). Because the maximum value of $sin^2\alpha\ cos\alpha$ is $\sim 0.38$, we infer from Eq.~\ref{e:m} that $M_{tot} \gtrsim 2.9 \times 10^{14} M_\odot$ and $0.54 \times 10^{14} M_\odot$ for the A1095 and A1926 pairs, respectively, which are smaller than their total gravitational masses ($M_{200}$), $5.0 \times 10^{14} M_\odot$ and $2.4 \times 10^{14} M_\odot$. In particular, the required value of $M_{tot}$  for A1926 may be substantially over-estimated here, because the  BCG in A1926S shows a large offset in both the position and velocity from its X-ray centroid (specz=0.136). Therefore, the two cluster pairs  could be bound and represent ongoing or future major mergers (with mass ratio $<$ 3:1).

The dynamic state of a cluster is intimately related to its substructure, which can be traced by various ICM and galaxy properties. First, we examine the temperature structures of our sample clusters.
In general, clusters can be categorized into cool core clusters (CCCs) and non-cool core clusters (NCCCs; e.g., Jones \& Forman 1984; Sanderson, O'Sullivan \& Ponman 2009).
A CCC tends to show a relaxed and symmetric morphology, whereas an NCCC often exhibits  disturbed overall shape and substructure. Cluster merging has been suspected to be the primary mechanism for transforming CCCs to NCCCs. Thus NCCCs are typically in dynamically young states (e.g., Sanderson, Ponman \& O'Sullivan 2006; Chen et al. 2007; Leccardi, Rossetti \& Molendi 2010). The bottom panels of Fig.~\ref{fig3} show that A1095W, A1095E and A1926N are most likely NCCCs, while the state of A1926S is uncertain because of the large measurement errors in the temperature distribution. 

\begin{figure*}
     \begin{center}
        \subfigure[A1095W]{%
            \includegraphics[width=0.495\textwidth]{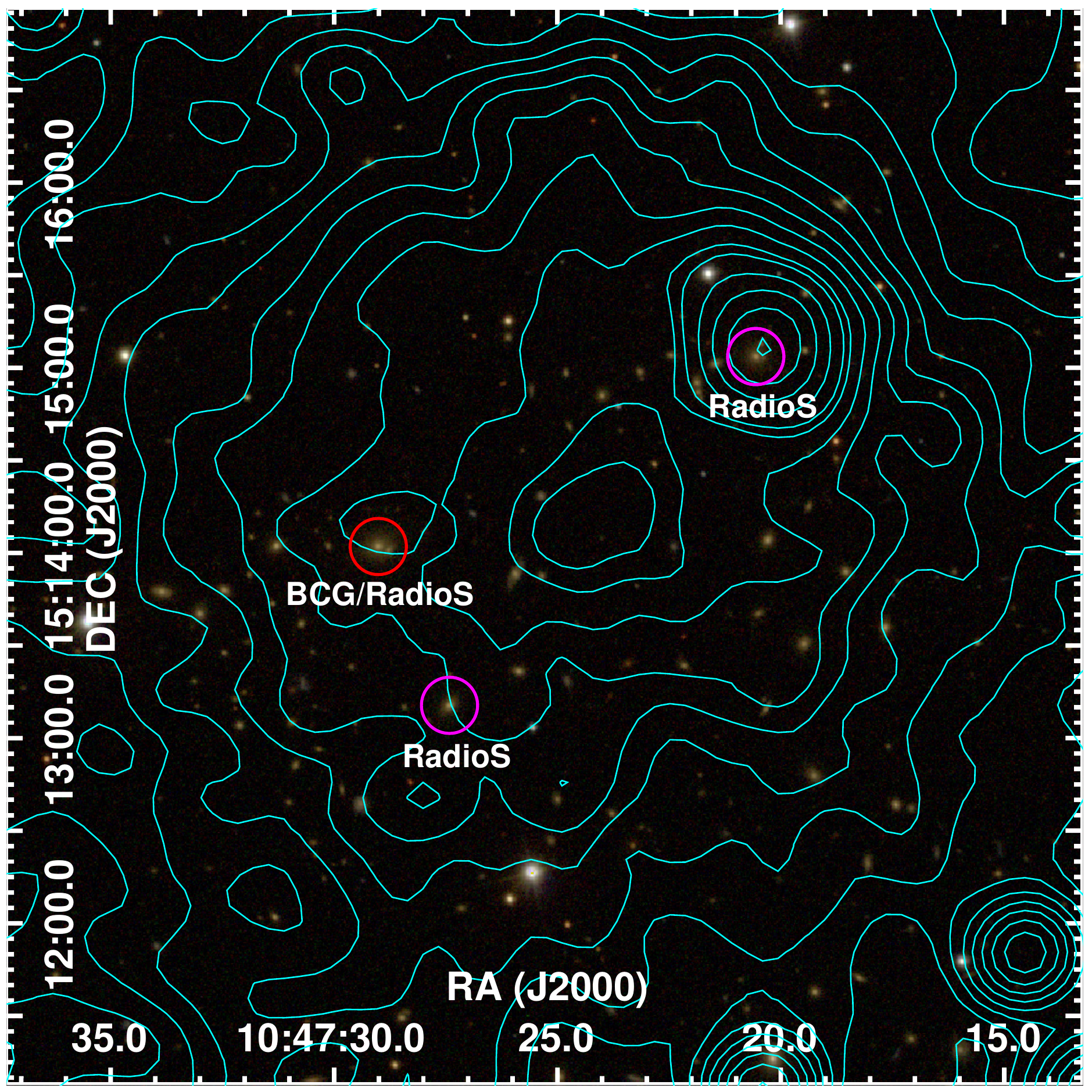}
        }%
        \subfigure[A1095E]{%
           \includegraphics[width=0.495\textwidth]{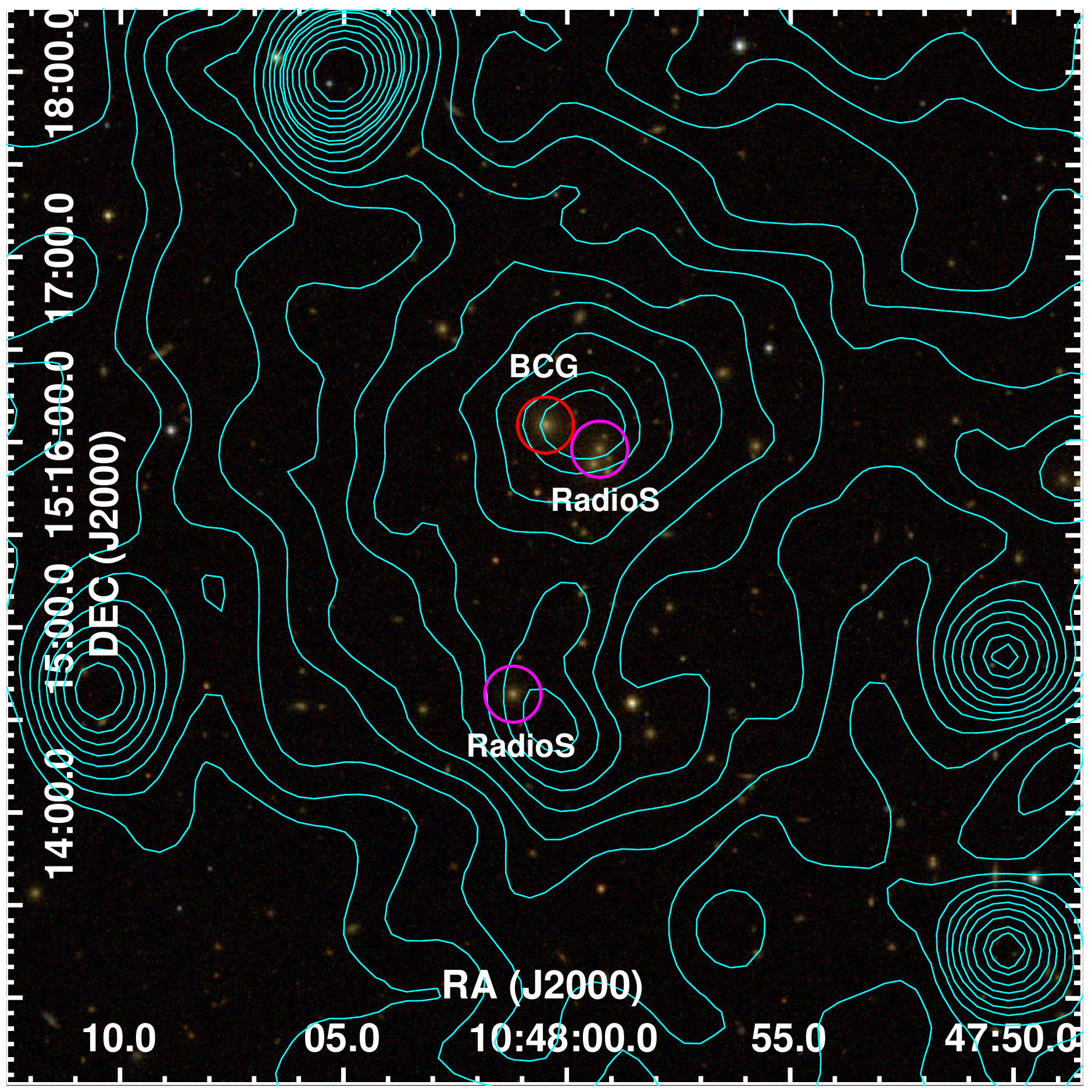}
        }\\ 
        \subfigure[A1926S]{%
            \includegraphics[width=0.495\textwidth]{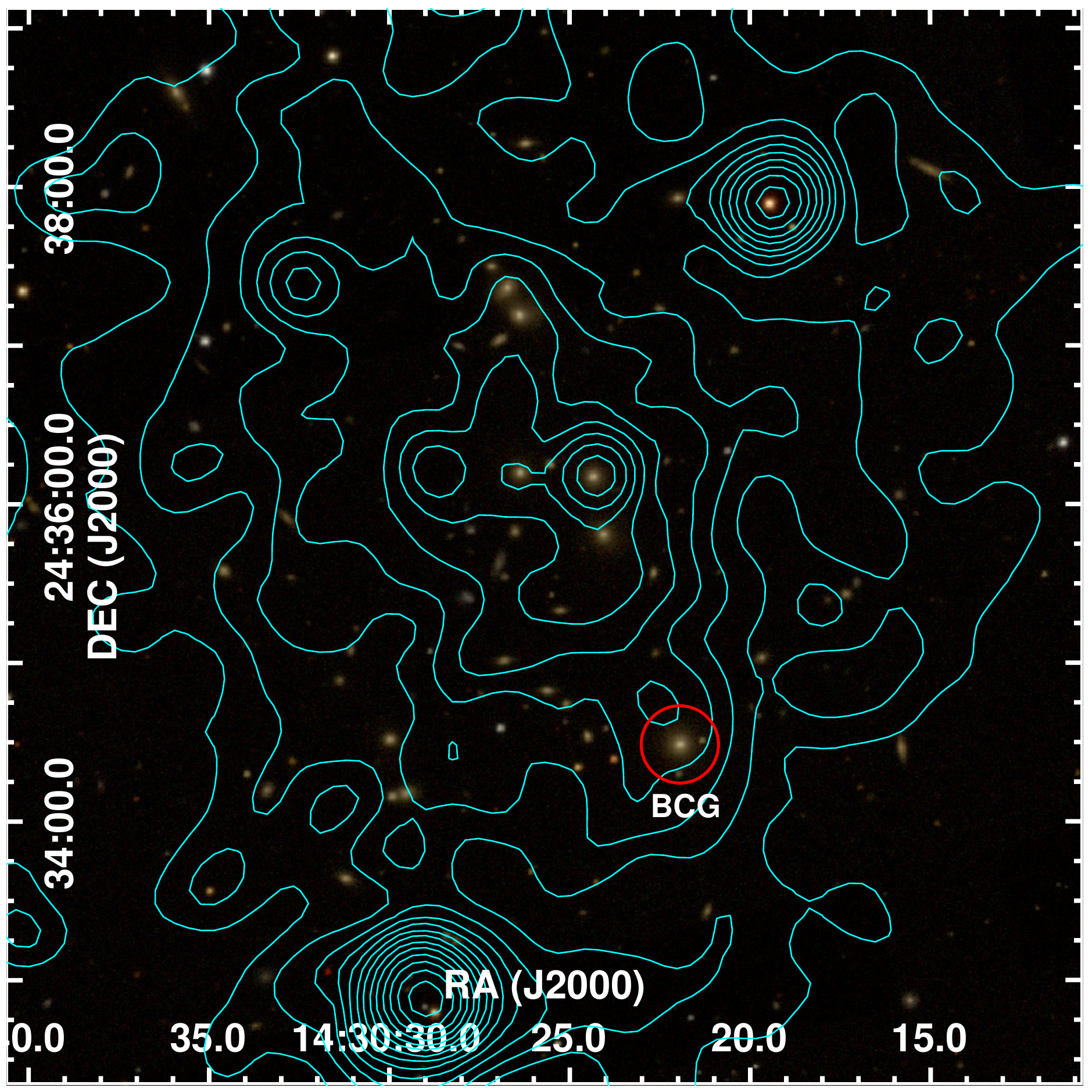}
        }%
        \subfigure[A1926N]{%
            \includegraphics[width=0.495\textwidth]{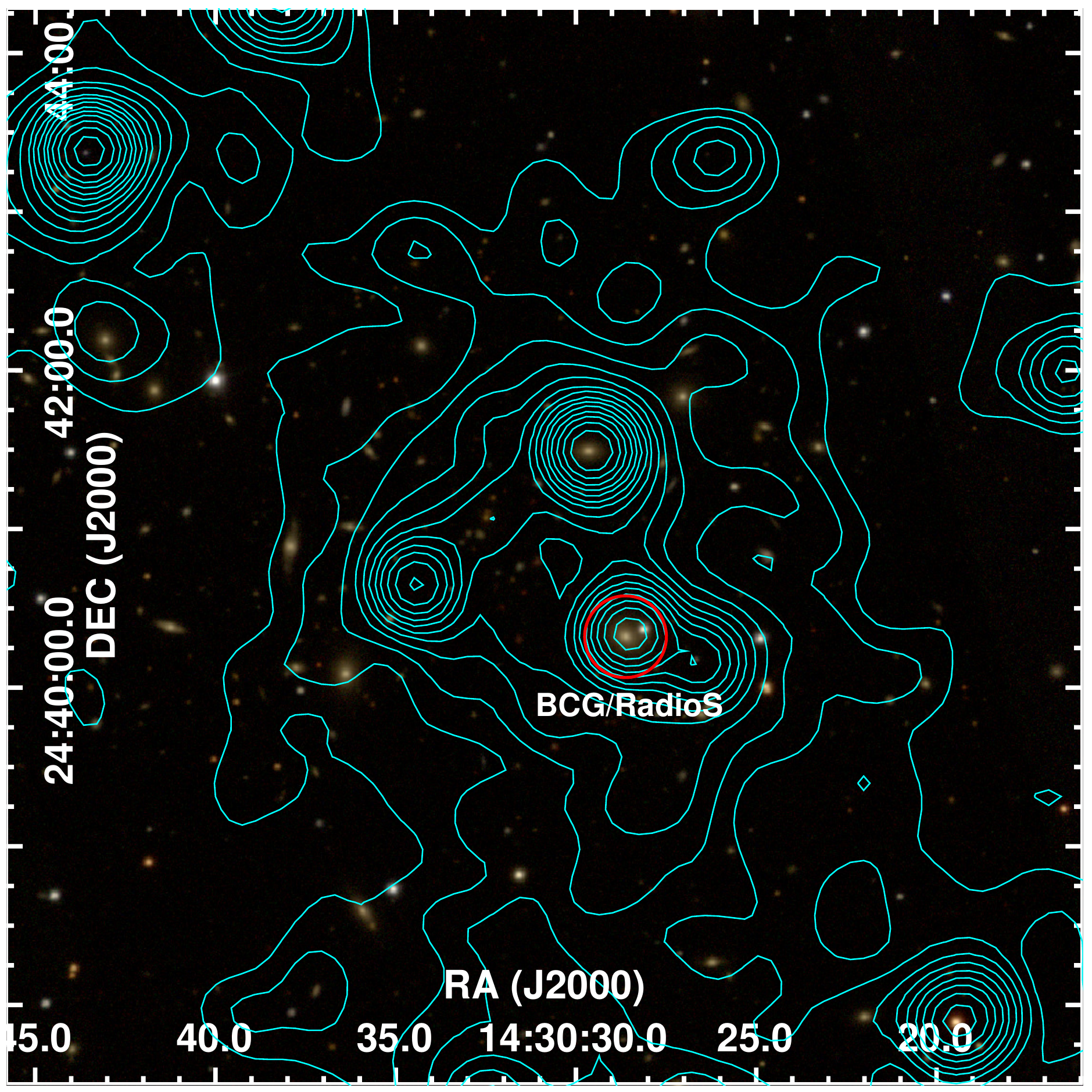}
        }%
    \end{center}
    \caption{%
SDSS tricolor (red = i-band, greed = r-band and blue = g-band) images, overlaid with the 0.5-7 keV X-ray intensity contours (in units of $\rm{counts\ s^{-1}\ deg^{-2}}$) at 16, 22, 29, 39, 53, 70, 94, 125, 166, 221, 254... for A1095W/A1095E, and 20, 28, 38, 51, 69, 93, 124, 166, 220, 293, 390... for A1926S/A1926N.
            }%
   \label{fig8}
\end{figure*}

Second, we check the X-ray morphologies
of the individual clusters, in comparison with the distributions of the 
galaxies. A close-up view of the clusters is shown in 
Fig.~\ref{fig8}. Particularly interesting are the 
morphologies of the X-ray peaks associated with BCGs, as well as their   
locations in the clusters. As the most luminous galaxy (which is almost always an elliptical), 
the BCG tends to be found  positionally at the gravitational center (Cui et al. 2016) and kinematically near the rest frame of a cluster (Schneider, Gunn \& Hoessel 1983; Hoessel \& Schneider 1985). Thus, the BCG of a relaxed cluster 
should generally lie at cluster X-ray centroids (Jones \& Forman 1984; Rhee \& Latour 1991; Lin \& Mohr 2004). 
However, most of the BCGs in our pair clusters are off their global X-ray
centroids, even in projection and are thus consistent with being in merging systems (e.g, Sanderson, Edge \& Smith 2009; Hudson et al. 2010; von der Linden et al. 2014; Rozo \& Rykoff 2014).

\begin{figure*}
     \begin{center}
        \subfigure[A1095W]{%
            \includegraphics[width=0.495\textwidth]{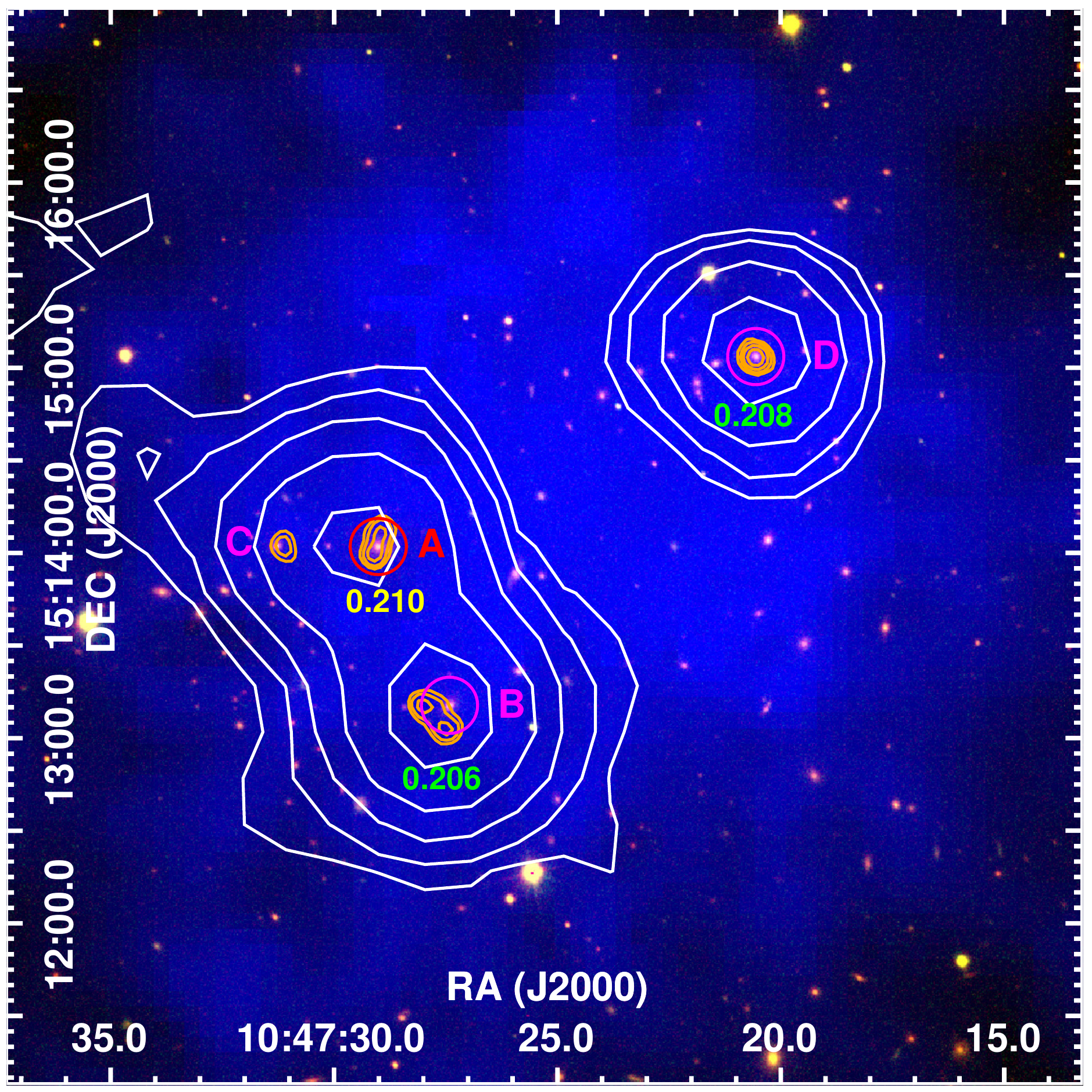}
        }%
        \subfigure[A1095E]{%
           \includegraphics[width=0.495\textwidth]{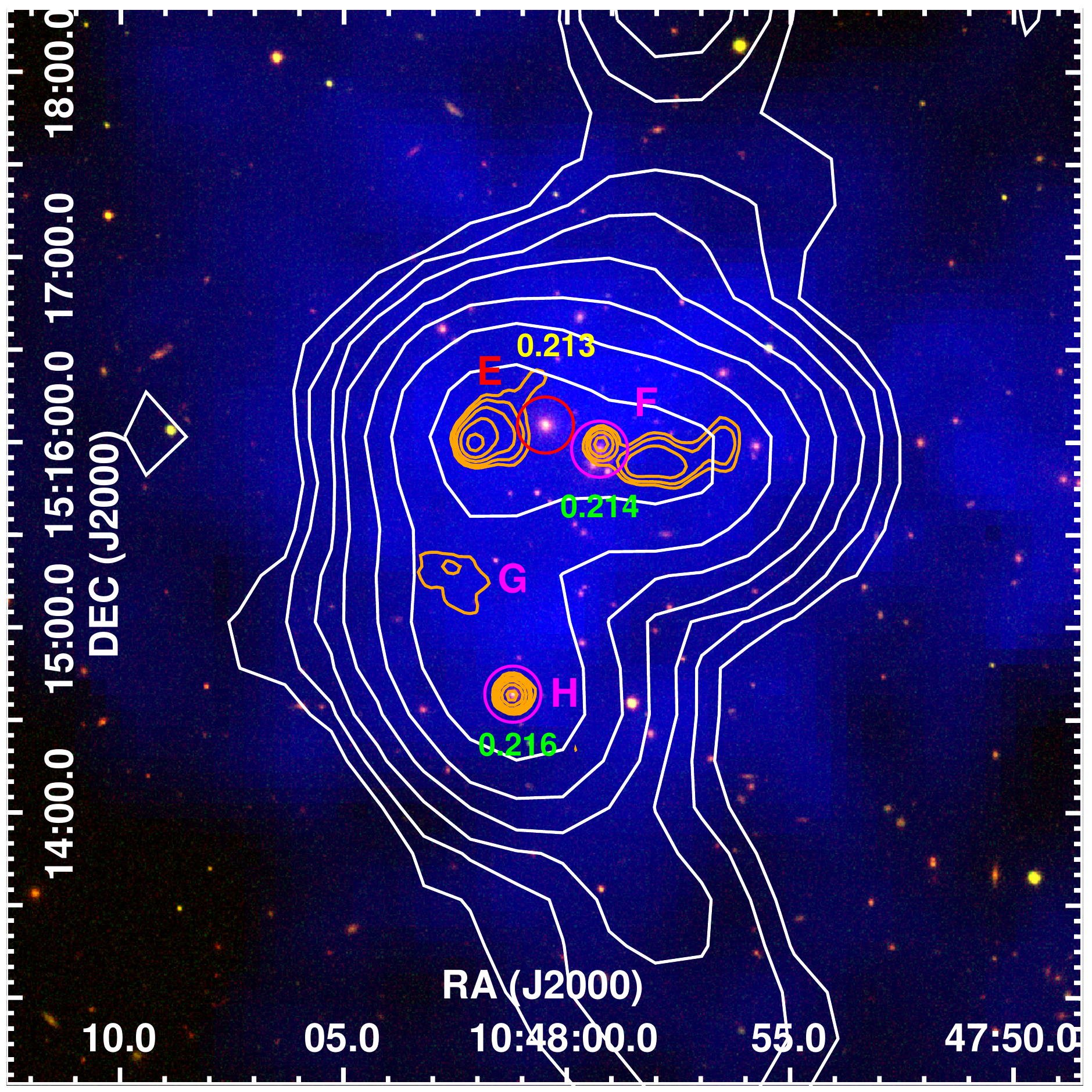}
        }\\ 
        \subfigure[A1926S]{%
            \includegraphics[width=0.495\textwidth]{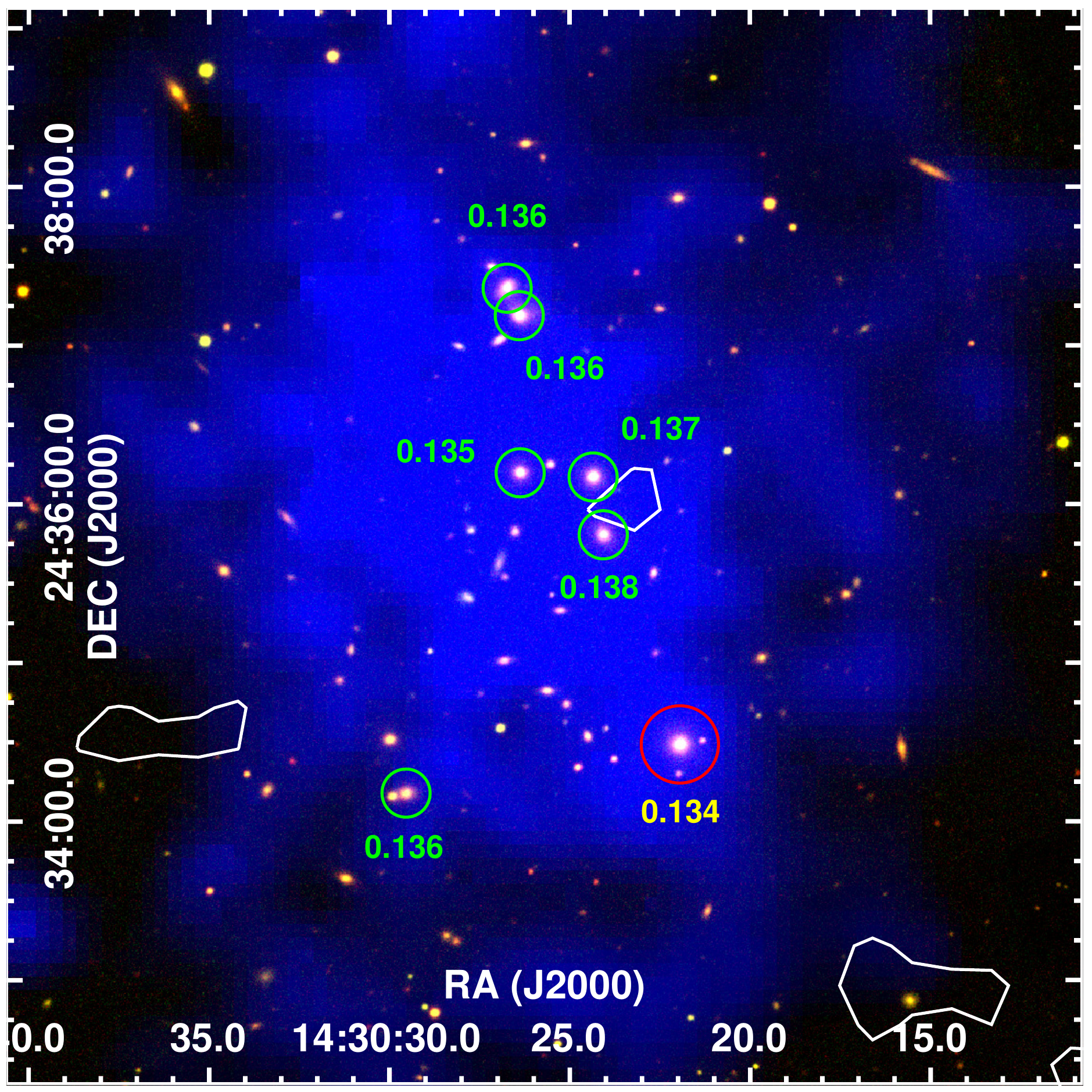}
        }%
        \subfigure[A1926N]{%
            \includegraphics[width=0.495\textwidth]{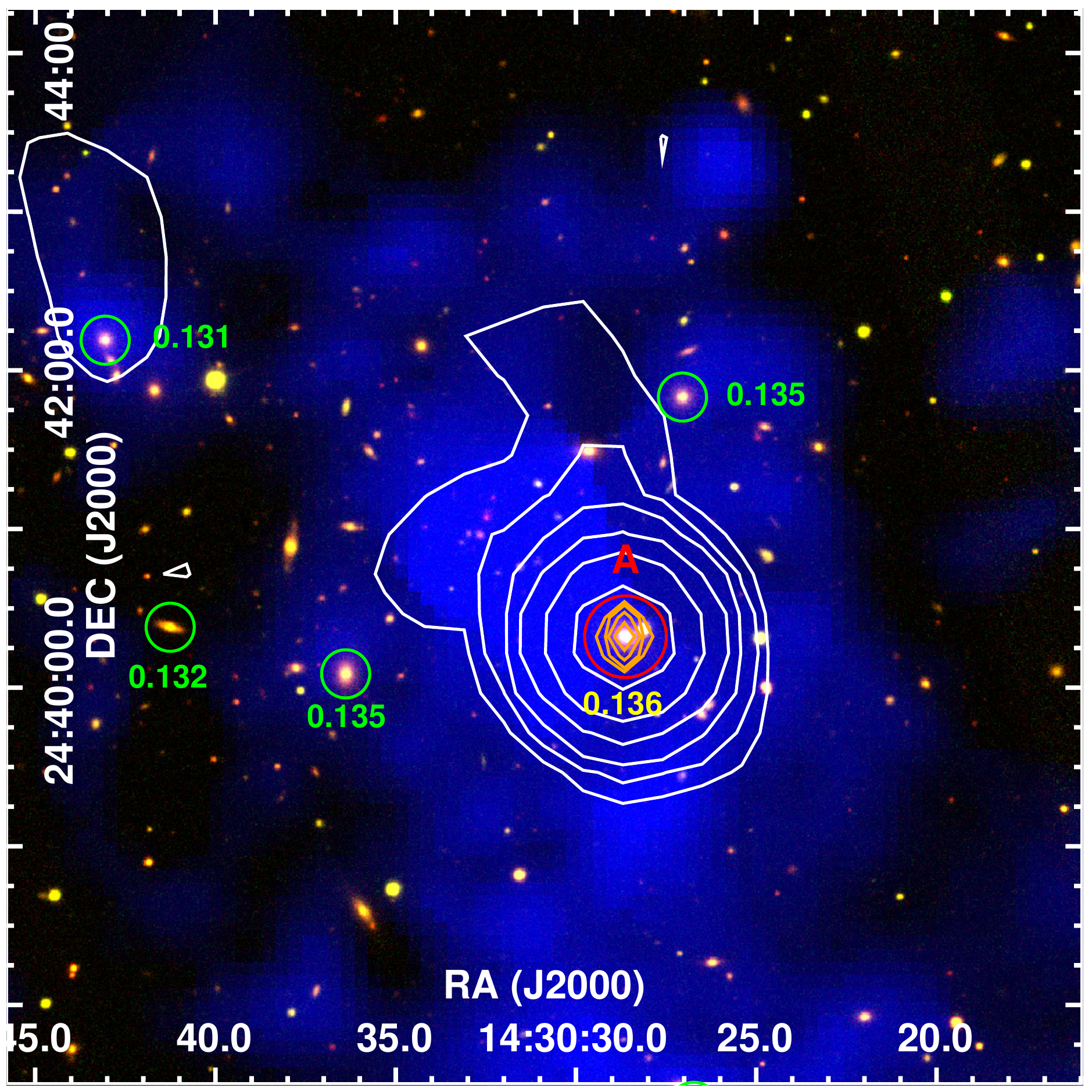}
        }%
    \end{center}
    \caption{%
Multi-wavelength images: SDSS r-band (red), g-band (green) and diffuse 0.5-2 keV X-ray emission
(blue). Overlaid with radio contours from NVSS (white contours: 1, 2, 4, 8, 16, 32, and 64 mJy beam$^{-1}$) and FIRST (orange contours: 1, 2, 4, 8, 16, and 32 mJy beam$^{-1}$).
            }%
   \label{fig9}
\end{figure*}

Third, we incorporate information from radio observations to further probe the states of the clusters. Radio emission, predominately representing synchrotron radiation
from relativistic electrons/positrons, can arise from diffuse halos, radio lobe relics and/or mini-halos (Feretti et al. 2012), as well as individual galaxies (e.g., jets from AGNs). Both the morphology and
intensity of the emission can be used to trace the dynamic state of a cluster: e.g., subcluster
mergers tend to increase the ram-pressures and lead to particle acceleration and amplification
of magnetic field in the ICM (Buote 2001; Feretti 2002).
Both A1095 and A1926 fields are covered by the NRAO VLA Sky Survey (NVSS; Condon et al. 1998) and the Faint Images of the Radio Sky at Twenty-Centimeters (FIRST; Becker, White \& Helfand 1995). 
The NVSS is the largest VLA radio survey at 1.4 GHz, with a limiting flux density of $\sim$ 2.5 mJy (5$\sigma_{rms}$) and an angular resolution of 45$^{\prime\prime}$ (FWHM). The NVSS radio intensity contours are included in Figs.~\ref{fig6}-\ref{fig7}, which give a global view of the cluster pairs.
The FIRST is also at 1.4 GHz and has a limiting flux density of $\sim$ 1 mJy (5$\sigma_{rms}$) and an angular resolution of 5\farcs4 (FWHM).
The  intensity contours from this survey, as well as those
from the NVSS, are shown in Fig.~\ref{fig9}, which presents a close-up view of the individual clusters. All the clusters, except for A1926S,
show significant radio emission. Accounting for the resolution difference
between the two surveys, the  NVSS intensity contours seem to trace well the distribution of the
FIRST sources, which mostly represent radio jets and lobes associated
with individual galaxies. A probable exception is the extended radio feature 
G in A1095E (Fig.~\ref{fig9}), which has no apparent
galaxy counterpart. Some of the lower surface brightness   
extensions demonstrated by the NVSS contours may be diffuse in origin. The 
examples are the large-scale 
extensions to the southwest and northwest of A1095E, as well as to the northeast of
A1926N. These extensions may represent the emission from materials stripped from individual
galaxies by the ram-pressure of the ICM. Such stripping may start in regions
far away from cluster cores, as indicated by the lopsided morphology of the
radio emission in the vicinity of a galaxy (or group) near the northeast edge
of the A1926N panel of Fig.~\ref{fig9}. More sensitive new
radio observations will help to further test this scenario. While the above discussion is focused on the common or comparative properties of the clusters, we describe their individual distinct signatures and corresponding implications for the dynamic states in the appendix.

The above discussion strongly suggests that the four clusters in the two pairs
all show lines of evidence for substantial substructures, which are probably best
explained by recent cluster assembling, ongoing subcluster mergers, and/or strong interplays. 

\subsection{Large-scale environment of the clusters}

It is widely believed that clusters are formed hierarchically, by the merging of smaller groups and clusters, typically at intersections of cosmic webs  (Jones \& Forman 1984; Jeltema et al. 2005; Andrade-Santos et al. 2013).  A1095 and A1926 fields may thus represent snapshots of this hierarchical formation process of massive bound structures. We have shown the presence of multiple clusters, as well as their substructures, which most likely represent ongoing subcluster mergers.

\begin{figure*}
 	\begin{center}
        \subfigure[A1095]{%
           \includegraphics[width=0.45\textwidth]{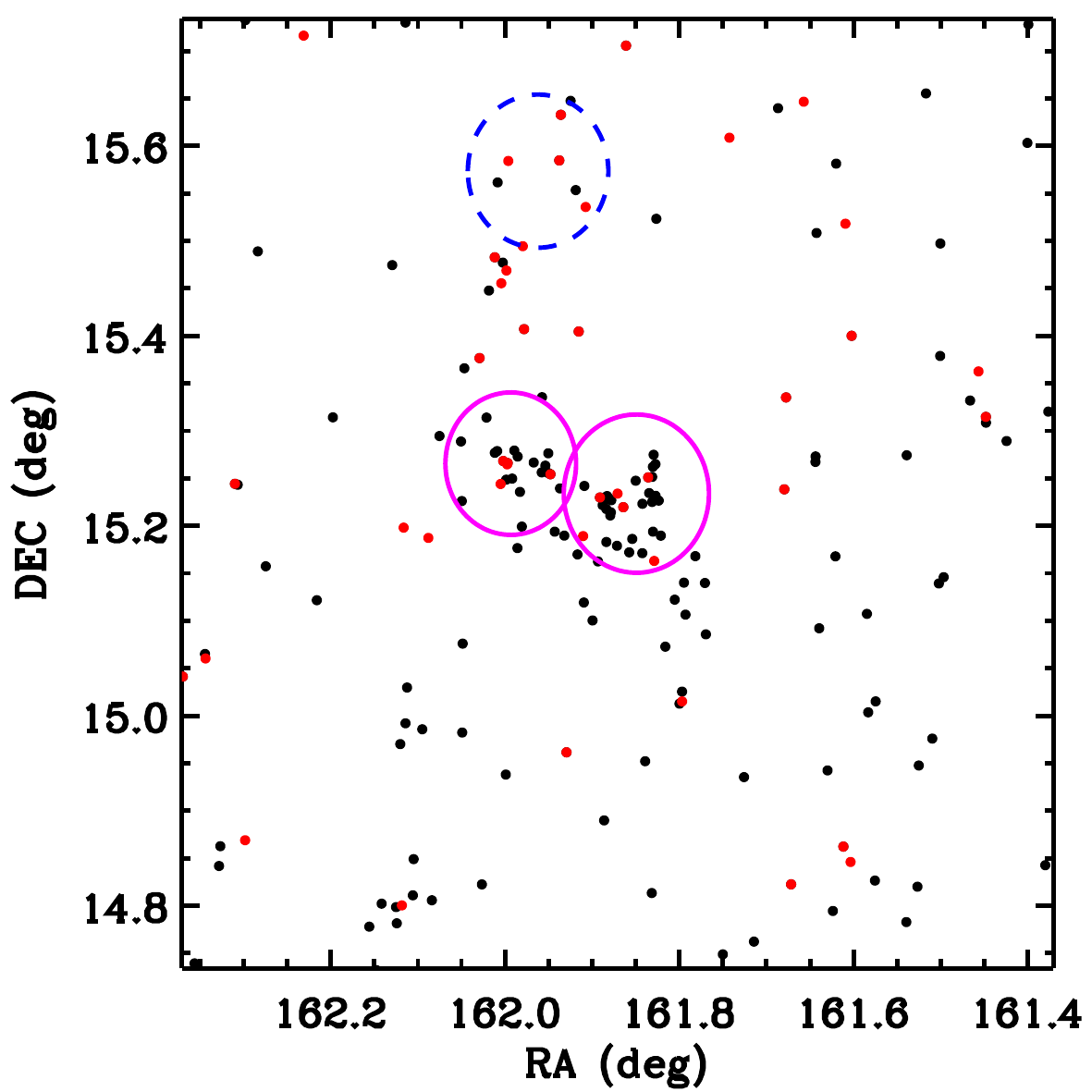}
           }
        \subfigure[A1926]{%
            \includegraphics[width=0.45\textwidth]{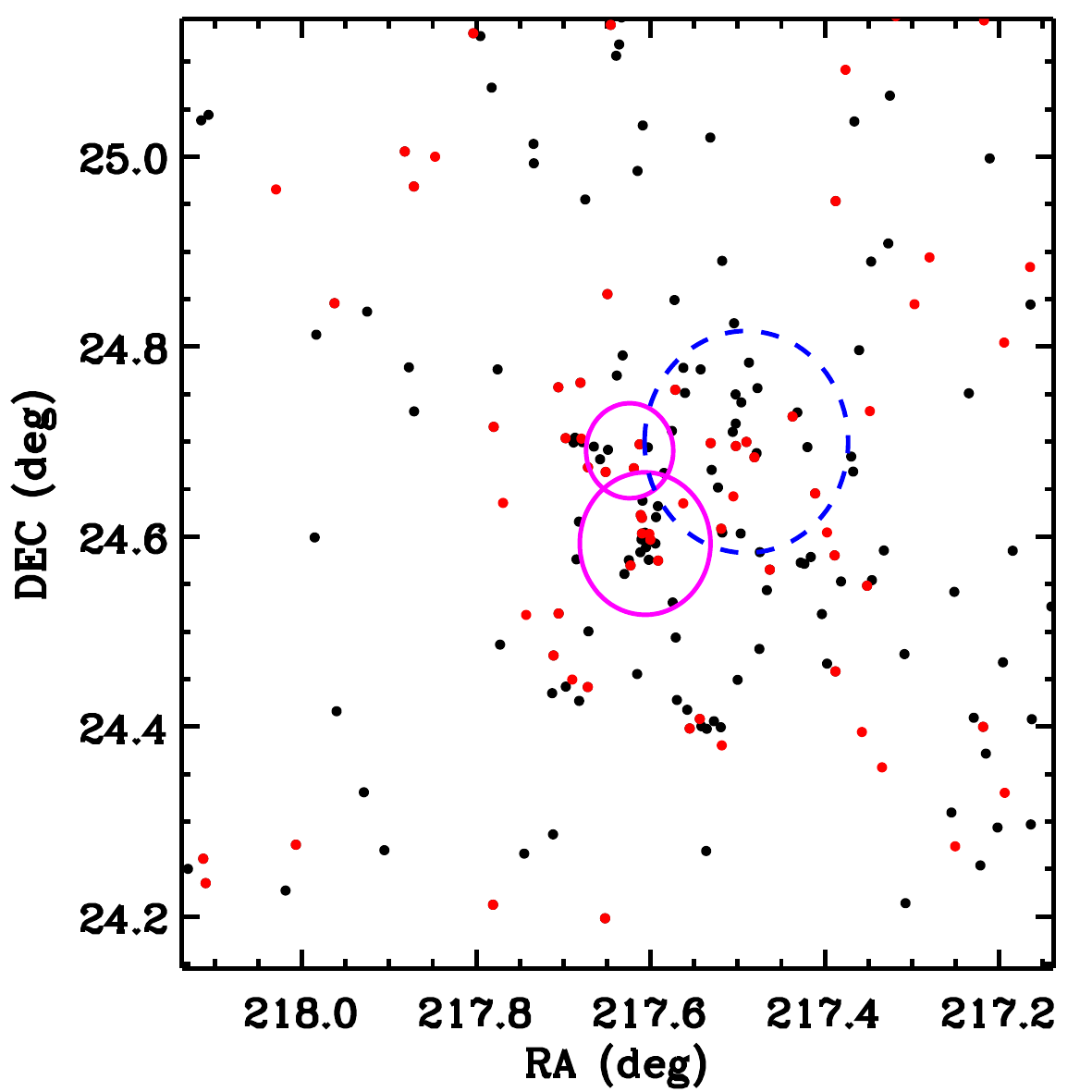}
        }%
 	\end{center}
    \caption{%
Distribution of the galaxies from SDSS with available redshifts over a large-scale field, including the surroundings of each cluster pair. Black points: all galaxies with SDSS measured photometric redshifts in the $0.210 \pm 0.02$ interval (i.e. about $\rm{\pm 6000 km\ s^{-1}}$ around the mean cluster velocity) for A1095 and $0.136 \pm 0.02$ interval for A1926; red points: galaxies with measured spectroscopic redshifts in the $\pm 0.01$ (about $\rm{\pm 3000 km\ s^{-1}}$) interval; solid magenta circles: the same as the larger white circles in Fig.~\ref{fig1}b for A1095 and Fig.~\ref{fig2}b for A1926; dashed blue circles: locations of additional clusters identified from NED with redshifts in the 0.02 redshift interval around A1095 and A1926. The diameter of each blue circle is 2 Mpc (i.e., the typical size of a cluster).
            }%
   \label{fig10}
\end{figure*}

Fig.~\ref{fig10} presents the large-scale environment in the A1095 and A1926 fields.
Each has a FOV of  $1^\circ \times 1^\circ$, corresponding to $\rm{12.5 \times 12.5\ Mpc^2}$ at the redshift of A1095 or $\rm{8.5 \times 8.5\ Mpc^2}$ at A1926. We mark the positions and redshifts of galaxies from SDSS and clusters identified in NASA/IPAC Extragalactic Database (NED\footnote{http://ned.ipac.caltech.edu/forms/nnd.html}). The projected separation is $\sim$ 1.8 Mpc between A1095W and A1095E. The nearest other cluster is WHL J104756.9+153431 (photo z=0.197), projected $\sim$ 3.8 Mpc north of A1095E. They may be connected by a filament of galaxies (cosmic web) and thus belong to the same large-scale structure. A1926S and A1926N, with a projected separation of $\sim$0.67 Mpc, probably have a companion cluster, GMBCG J217.49013+24.69973, which (though too faint to be firmly detected in X-ray) is located $\sim$1.1 Mpc west of the A1926S/A1926N pair center and at a comparable photo z=0.135. Therefore, the A1095 and A1926 pairs just represent the most outstanding peaks of the underlying evolving larger scale structures of galaxies. 

\subsection{Comparing optically selected and X-ray selected galaxy clusters}
\label{ss:dis-com}
Various methods have been used to discover clusters of galaxies.
Although massive clusters can now readily be detected via their SZ effects and sometimes their gravitational lensing effects, intermediate-mass ones, as dealt with here, are still mostly identified in optical and/or X-ray.
X-ray detection is sensitive to the emission measure of the ICM, tracing the deep potential wells of individual clusters, and is thus relatively free from projection contamination  (e.g., Rosati, Borgani \& Norman 2002). Optical surveys (e.g., Gal et al. 2003; Koester et al. 2007; Hao et al. 2010; Wen, Han \& Liu 2012) have been used to detect candidate clusters on all mass scales. The detection is based on identifying concentrations of galaxies, spatially and often kinematically. Because the detection signal depends on the concentration linearly,  projection effects can be serious. 

\begin{figure*}
 	\begin{center}
\includegraphics[width=0.495\textwidth]{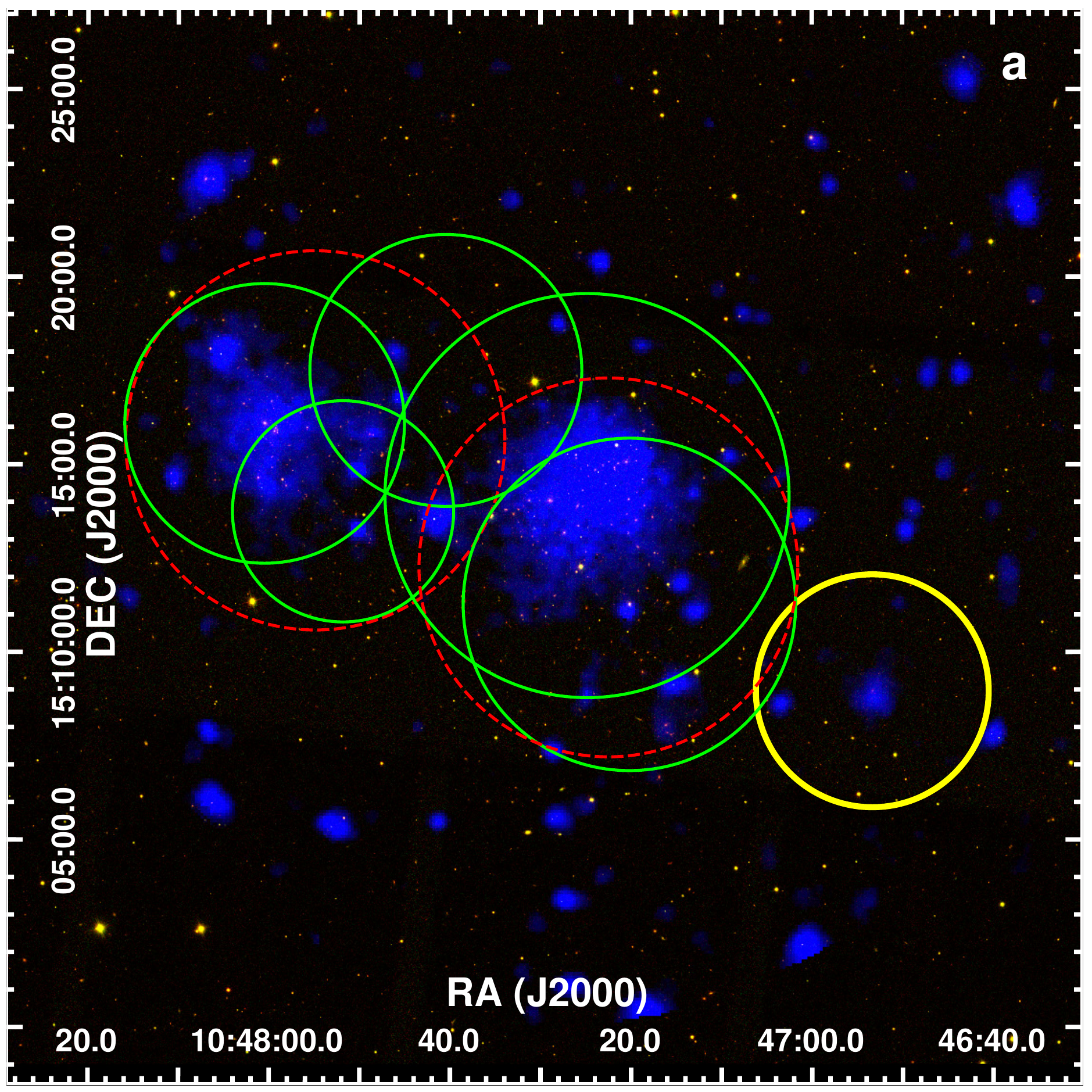}
\includegraphics[width=0.495\textwidth]{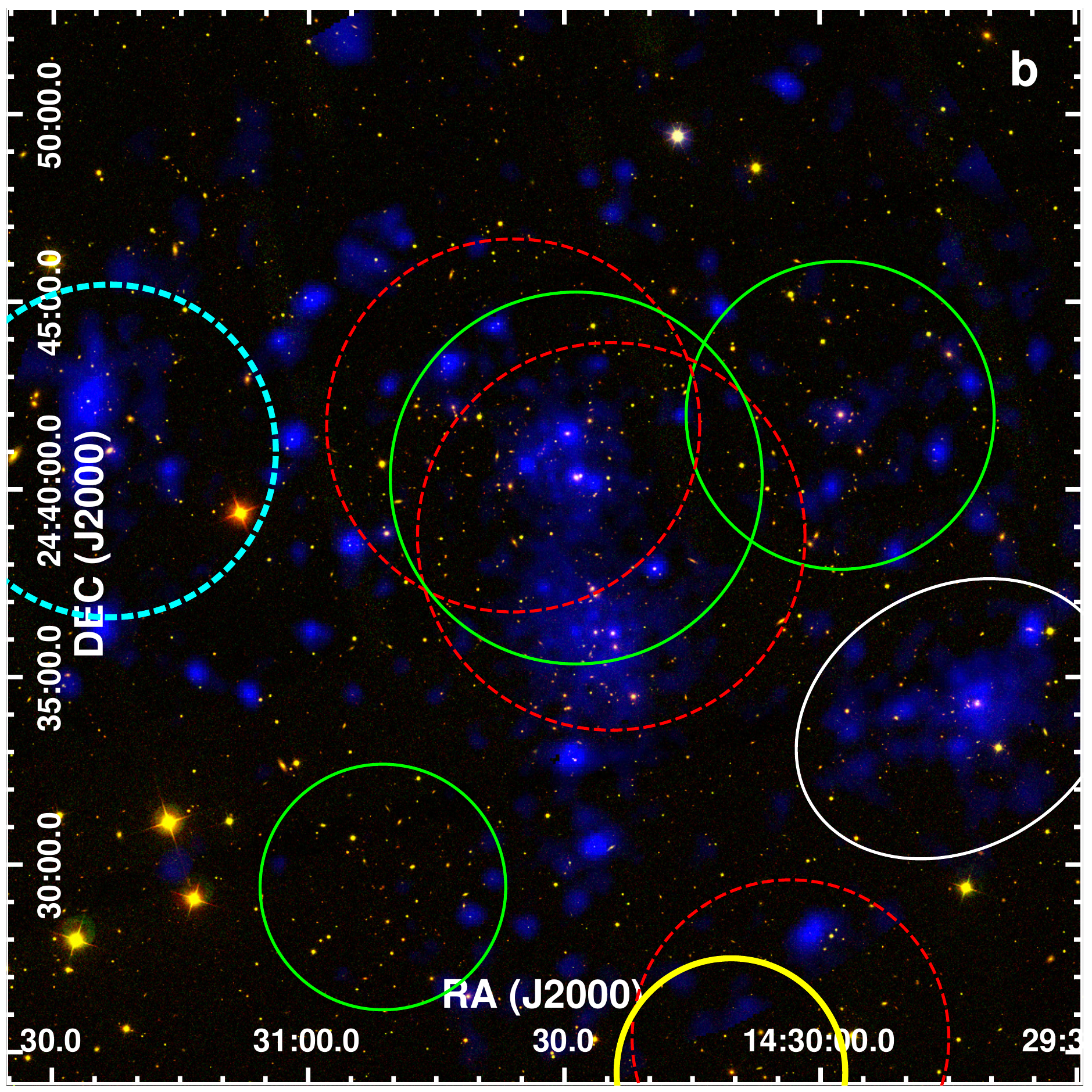}
 	\end{center}
    \caption{%
Cluster findings in the A1095 (a) and A1926 (b) fields. The three-color montages are composed of the images from SDSS r-band (red), g-band (green) and 0.5-7 keV X-ray emission (blue), while the overlaid circles represent the position and richness of optical detected galaxy clusters from various catalogues: the Gaussian Mixture Brightest Cluster Galaxy catalog (GMBCG; Hao et al. 2010; solid green circle), the Northern Sky optical Cluster catalog (NSC; Gal et al. 2003; dashed red circle), Wen+Han+Liu catalog (WHL; Wen et al. 2012; bold yellow circle), and the Maximum likelihood redshift Brightest Cluster Galaxy catalog (MaxBCG; Koester et al. 2007; dashed bold cyan circle).
In a theoretical model where the total optical light or number/richness of galaxies traces cluster mass,
the X-ray luminosity is proportional to the richness (N) square (e.g., Section 6 of Popesso et al. 2004),
while the distance is proportional to the redshift (z) in the local universe. Therefore, the radius (in unit of arcmin) of a cluster, represented by a circle here, is $R=log(N^2/z^2)$, roughly scaled with its expected X-ray flux.
            }%
   \label{fig11}
\end{figure*}

Let us use our cases to briefly examine the limitations of both optical and X-ray surveys, as well as how they may be used in a complementary fashion.
Fig.~\ref{fig11} illustrates the positions and richnesses of various versions of optically detected clusters. We see good correspondences between optically rich clusters with apparent diffuse X-ray structures,
which leads to the discovery of another X-ray-bright cluster MaxBCG J217.84740+24.68382 (z=0.097; bold cyan circle in Fig.~\ref{fig11}b).
These correspondences represent the firm identifications of these clusters.  
But the multiplicity, as well as the position and richness of the optical detections, vary from one version to another, presumably due to the differences in the adopted methodologies  (e.g., with or without spectroscopic redshifts, redshift range, apertures, etc.). This problem becomes especially serious for such a field of closely paired clusters, because of severe confusion. On the other hand, optical detections seem to be sensitive to potentially poor clusters (e.g., green circles in Fig.~\ref{fig11}b), which show little sign of any diffuse X-ray emission. Conversely, one X-ray cluster candidate (marked by the white ellipse in Fig.~\ref{fig11}b), not found in any optically selected cluster catalog, seems to be reasonably X-ray-bright and has an ICM temperature of  $1.9^{+1.0}_{-0.9}$ (corresponding to M$_{200} \simeq 0.9^{+0.9}_{-0.7} \times 10^{14} M_\odot$; R$_{200} \simeq 0.9^{+0.3}_{-0.2}$ Mpc) and a possible BCG SDSS J142941.65+243417.1 (z=0.147).  Although a quantitative comparison with the consideration of the richness uncertainties in both optical and X-ray is beyond the scope of the present work, this brief discussion demonstrates the importance of the imaging X-ray data with sufficient spatial resolution and sensitivity in the confirmation and characterization of galaxy clusters.

\subsection{Baryon content}
We first estimate the distribution and total mass of the hot ICM.
This estimate utilizes the results from the radial X-ray intensity
fit with the $\beta$-model (Eq.~1) and from the spectral fit  for each cluster (Table~\ref{t:clusters}). 
Assuming a spherical symmetry of the distribution,
the deprojected hydrogen density profile of the $\beta$-model as a function of the physical off-center radius ($r$) of a cluster is (e.g., Sarazin 1988)
\begin{equation}
n_{H}=n_0 \Big(1+\frac{r^2}{r_{c}^{2}}\Big)^{-\frac{3}{2}{\beta}},
\end{equation}
where we need to drive the central density $n_0$ in the following.

The model count rate of an optically-thin thermal plasma can be expressed as
\begin{equation}
CR=10^{14}A\Lambda\eta,
\end{equation}
where $A$ and $\Lambda$ are the instrument effective area and the photon emissivity function
 in our used band for the spectral fit, while $\eta$ is the normalization of the {\it apec} model,
\begin{equation}
\eta \equiv \frac{10^{-14}}{4 \pi [D_A(1+z)]^2}\int{n_e}{n_H}dV,
\end{equation}
where $n_e$ and $n_H$ are the electron and hydrogen number densities, $D_A$ and $z$ are the angular diameter distance and redshift of the cluster. We only need to use the ratio of  the
$CR$ and $\eta$ values, both of which can be readily read out from the spectral fit of a cluster and are
given in Table~\ref{t:clusters}.

The intensity of the thermal emission from the plasma in the same band  (in units of counts s$^{-1}$ deg$^{-2}$)  is 
\begin{equation}
I=\Big(\frac{\pi}{180}\Big)^2\frac{A\Lambda}{4\pi}EM,
\end{equation}
where the emission measure ($EM = \int n_{e}n_{H}dl$) along a line of sight at the  off-center projected distance ($R$) can be 
expressed in the $\beta$-model as [e.g., Eq. (5.68) of Sarazin (1988)]:
\begin{equation}
EM=\sqrt{\pi}\Big(\frac{n_e}{n_H}\Big)n_0^2 r_c \frac{\Gamma(3\beta-1/2)}{\Gamma(3\beta)}(1+x^{2})^{1/2-3\beta},
\end{equation}
where $\Gamma$ is the gamma function.
Using Eqs.~(6), (8) and (9), we can express $n_0$ as:
\begin{equation}
n_0=\frac{180}{\pi}\sqrt{\frac{10^{14}4\sqrt{\pi}I_0\Gamma(3\beta)}{(\frac{n_e}{n_H})\frac{CR}{\eta}r_c\Gamma(3\beta-1/2)}}
\end{equation}
Using the parameter values of $\frac{CR}{\eta}$, $I_{0}$, $r_{c}$, and $\beta$  listed in Table~\ref{t:clusters} for each cluster, as well as $n_e \simeq 1.2n_H$, we estimate $n_0$ and include it in the table.

We can now estimate the hot ICM mass ($M_{gas}$)
within $r_{500}$ of each cluster, using Eq.~(5).  The estimate is listed in Table~\ref{t:clusters}, which also includes the hot ICM mass fraction (relative to the gravitational mass).
 
For future comparison, we here estimate the column density of the hot ICM at the projected distances of the QSOs for individual clusters. Using the $\beta$-model, we can express the column density of hot protons through a cluster as
\begin{equation}
N_p=\sqrt{\pi}n_0 r_c\frac{\Gamma(3\beta/2-1/2)}{\Gamma(3\beta/2)}(1+x^{2})^{1/2-3\beta/2}.
\end{equation}
This expression is for $\beta > 1/3$, which is satisfied by the measured values (Table~\ref{t:clusters}). We include the $N_p$ values estimated  for the individual clusters in Table~\ref{t:clusters}.

\begin{figure}
     \begin{center}
\includegraphics[width=0.495\textwidth,keepaspectratio=true,clip=true]{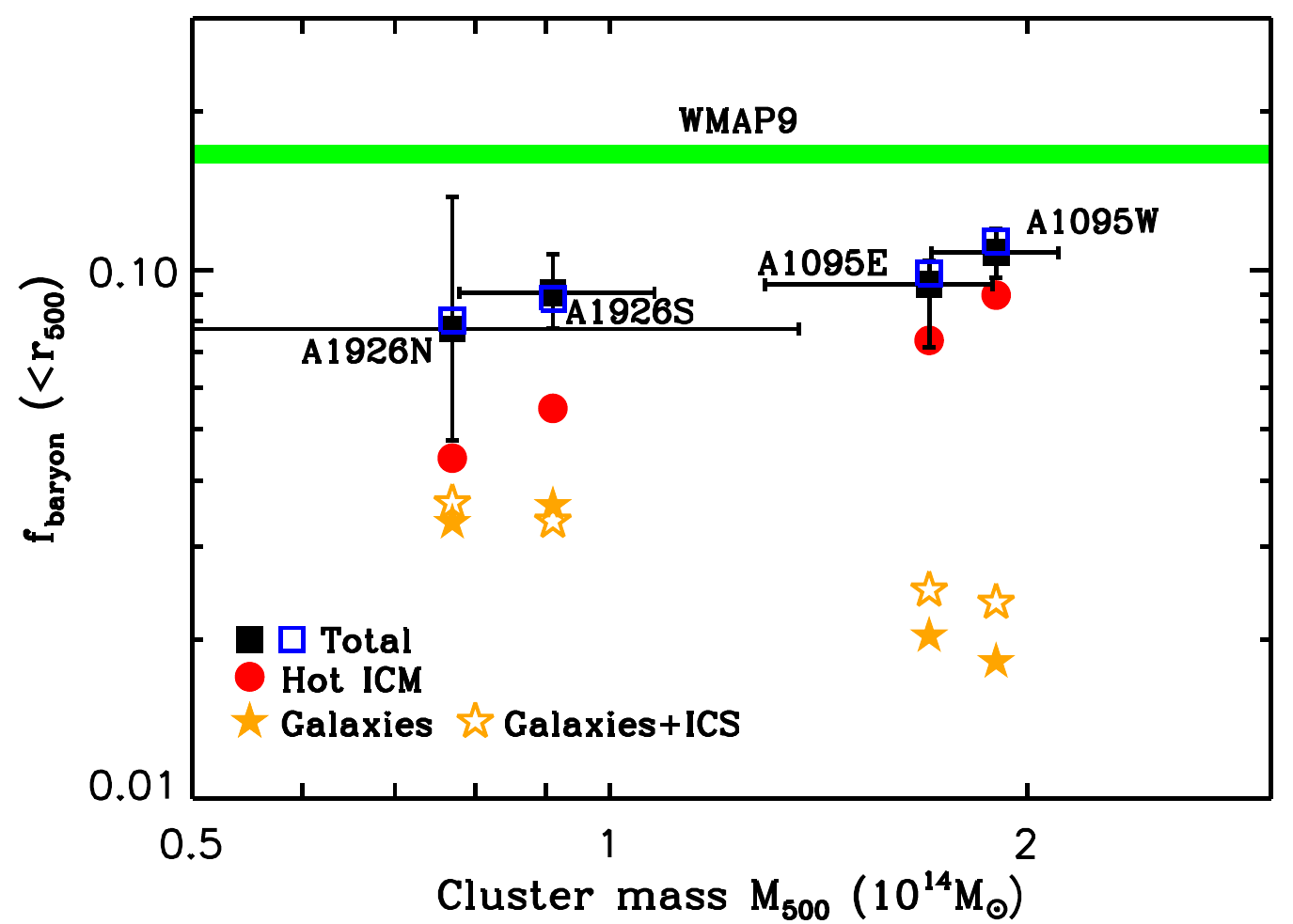}
    \end{center}
    \caption{
Baryon fraction within $r_{500}$ as a function of $M_{500}$: the red dots are the hot ICM fractions from the X-ray observations; the filled stars represent the stellar mass fractions estimated with the SDSS galaxy data, while the open stars present the stellar mass fractions based on the scaling relation of Gonzalez et al. (2013), which includes the contribution from ICS and applies the projection correction; the filled boxes are the total baryon fractions of hot ICM plus stellar mass fractions from the SDSS galaxy data, while the open boxes are the total baryon fractions of the hot ICM plus stellar mass fractions from Gonzalez et al. (2013); the error is dominated by the uncertainty in the gravitational mass estimate.
The green band marks the uncertain range of the WMAP 9-year baryon fraction (Bennett et al. 2013).
     }%
   \label{fig12}
\end{figure}

We now turn to estimate the stellar masses of cluster galaxies observed by the SDSS. We fit the photometric redshift distribution of galaxies within $r_{500}$ of each cluster with a Gaussian function. To minimize the confusion with background and foreground galaxies. Those with redshifts deviating more than $\pm 1 \sigma$ from the Gaussian peak (mean) are excluded (Fig.~\ref{fig5}); this exclusion may lead to slight underestimate of the mass (which, however, should not change our later conclusion in any significant way). For each of the remaining galaxies, we convert the SDSS r-band magnitude (after both extinction correction and k-correction) to the stellar mass by adopting the color-dependent (g-r color) mass-to-light ratio (Bell et al. 2003) and the photometric redshift. The results are plotted as filled star in Fig.~\ref{fig12}.

The above method is straightforward yet biased toward the low redshift because of increasing incompleteness of the galaxy sample with increasing redshift, moreover misses the fainter galaxies below the detection limit of the SDSS and the intracluster stars (ICS), while contaminates by galaxies outside $r_{500}$ sphere in the $r_{500}$ cylinder along the line-of-sight direction due to projection effect and galaxies in the overlap region in such close pairs. Thus we also estimate the stellar mass fraction based on the scaling relation of $M_{\star,3D}=3.2\times10^{-2}(M_{500}/10^{14}\ M_{\odot})^{0.52}$ from Gonzalez et al. (2013) to double check the results. They perform a statistical background subtraction, include a completeness correction for the fainter galaxies, take account of the contribution from the ICS, and apply the deprojection correction. The results are plotted as open stars in Fig.~\ref{fig12} and listed in Table~\ref{t:clusters}, the ratio between these stellar mass fractions and the previous ones is $\sim0.9-1.3$. However, the difference between these two estimates has minimal impact on the total baryon fractions.

Fig.~\ref{fig12} shows that the hot ICM and total observed baryonic mass fractions increase with the cluster mass, while the stellar mass fraction and stellar-to-gas mass ratio (star-formation efficiency) decreases with the cluster mass, these results are consistent with previous findings (e.g., Mohr, Mathiesen \& Evrard 1999; Lin, Mohr \& Stanford 2003; Andreon 2010; Dai et al. 2010; Gonzalez et al. 2013; Chiu et al. 2016). The observed baryon fraction of the most massive cluster in our sample is $\sim$11\%, compared with the cosmological fraction determined from the Wilkinson Microwave Anisotropy Probe (WMAP) 9-year data ($\sim$17\%; Bennett et al.2013), there is many missing baryons.
On one hand, the missing baryons may be ejected out of central $r_{500}$ region due to energetic process such as AGN feedback (e.g. McCarthy et al. 2010, 2011). However, the baryon budget outside of $r_{500}$ is hard to make because one need to rely on the extrapolation of the $\beta-$model, which is hardly justified, to get close to the virial radius of the cluster. On the other hand, the missing baryons may be in the form of warm ICM, which is undetected by both X-ray and optical observations. Therefore, the results of upcoming QSO absorption line observations would help to test this assumption.

\section{Summary}

We have carefully investigated the diffuse emission from the hot ICM and X-ray sources in two optically-selected sample cluster fields, based on {\it XMM-Newton} observations. We have further incorporated observations from SDSS and NVSS/FIRST to probe the relationship among the clusters in each field, their dynamic states, their large-scale environments,  the complementarity of the X-ray, UV, optical and radio observations of the clusters. Our main results
and conclusions are as follows:

\begin{itemize}
\item We find that each of our sample fields actually contains a pair of X-ray luminous clusters at similar redshifts ($z = 0.210$ and $0.213$ for A1095W and A1095E; $z = 0.136$ for both A1926S and A1926N). Member clusters in each pair have comparable ICM mean temperatures ($\sim 3.6$~keV vs. 3.3~keV for A1095W/A1095E and $\sim 2.2$~keV vs. 2.0~keV for A1926S/A1926N).
The minimum mass required to bind each cluster pair, is smaller than its total gravitational mass.
The clusters in each pair  also morphologically appear to be mutually perturbed.
Therefore, these paired clusters are most likely bound systems and would eventually merge into two massive clusters. 

\item The paired clusters  are characterized by asymmetric diffuse X-ray morphologies with direct evidence for subclustering, by the large offsets of the BCGs from the global centroids of diffuse X-ray emission, and/or by significant radio emission on various scales. Therefore, the individual clusters themselves are probably the results of recent subcluster mergers or still in early formation stages, and they may represent the most outstanding snapshots of the underlying larger scale structures, consistent with the hierarchical formation paradigm of galaxy clusters.

\item We have demonstrated the power of the modern X-ray observations in detecting clusters and in characterizing their ICM properties, as well as the complementary values of the optical and radio surveys in studying the relationship among the apparently paired clusters and their dynamic states.
In addition to the two pairs, we have also identified two X-ray-bright poor clusters (z$\simeq$0.097 and 0.147) in the A1926 field. Moreover, one needs to
be extremely cautious in using optically selected clusters (especially for paired ones, because of the large uncertainties in their centroids, masses, and structures) for cosmological studies.

\item We have estimated the hot ICM and stellar masses in each cluster. The hot ICM fraction increases with the cluster mass, while the stellar fraction decreases. The total observed baryon fraction, which is smaller than the expected cosmological baryonic mass fraction, leaves ample room for a substantial mass presence of warm ICM.

\end{itemize}

The results from this study will aid any further investigation of the clusters, especially the interpretation of the QSO absorption line observations. We hope to characterize how the heating/cooling may depend on the dynamic states of the clusters, as well as their mass and projected distances (e.g. Wang \& Walker 2014), advancing our understanding of the structure formation in the universe.

\section*{ACKNOWLEDGEMENTS}
We thank the anonymous referee for helpful suggestions that 
improved the presentation of this work, which was partially supported by 
the National Natural
Science Foundation of China under grants 11273015 and
11133001, and the National Basic Research Program
No.2013CB834905. C.G. acknowledges support from the
program of China Scholarships Council No.201206190041
during his visit to the University of Massachusetts. Q.D.W. is grateful to the hospitality that he received in the School of Astronomy and Space Science, Nanjing University, during his visit as a Yixing Visiting Chair Professor. Z.L. acknowledges support
from the Recruitment Program of Global Youth Experts.
L.J. acknowledges support from the Joint Funds of the National Natural Science Foundation of China 
under grant U1531248.

\newpage

\begin{landscape}
\begin{deluxetable}{lccccccccccccc}
\tablecolumns{14}
\tabletypesize{\scriptsize}
  \tablecaption{A1095 field source list}
  \tablewidth{0pt}
  \tablehead{
  \colhead{Source} &
  \colhead{Name } &
  \colhead{RA} &
  \colhead{DEC} &
  \colhead{$\delta$} &
  \colhead{DET$\_$ML} &
  \colhead{CR} &
  \colhead{Flux} &
  \colhead{HR} &
  \colhead{Counterpart name} &
  \colhead{Type} &
 \colhead{Sep} &
 \colhead{Redshift} &
 \colhead{Luminosity} \\
	&		&	(J2000)	&	(J2000)	&	(1$^{\prime\prime}$)	&		&	(counts$/$ks)	&	($\rm{10^{-14}\ erg\ cm^{-2} s^{-1}}$)	&		&		&		&	(1$^{\prime\prime}$)	&		&	($\rm{erg\ s^{-1}}$)	\\
  \noalign{\smallskip}
  \colhead{(1)} &
 \colhead{(2)} &
 \colhead{(3)} &
 \colhead{(4)} &
 \colhead{(5)} &
 \colhead{(6)} &
 \colhead{(7)} & 
 \colhead{(8)} &  
 \colhead{(9)} &                  
 \colhead{(10)} &                  
 \colhead{(11)} &                  
 \colhead{(12)} &                  
 \colhead{(13)} &                  
 \colhead{(14)} \\
   }
\startdata
1	&	XMMU J104627.61+151325.5	&	161.615051	&	15.223760	&	3.75	&	17.6	&	$ 13.54\pm3.88 $	&	$ 3.76\pm1.08 $	&	$ -0.81\pm0.26 $	&	-	&	-	&	-	&	-	&	-	\\
2	&	XMMU J104636.86+152150.2	&	161.653595	&	15.363952	&	1.67	&	73.5	&	$ 40.01\pm6.21 $	&	$ 11.11\pm1.72 $	&	$ -0.57\pm0.14 $	&	SDSS J104637.21+152152.3	&	G	&	5.5	&	-	&	-	\\
3	&	XMMU J104640.08+150747.4	&	161.666988	&	15.129833	&	1.73	&	22.2	&	$ 13.39\pm3.88 $	&	$ 2.96\pm0.70 $	&	$ -0.67\pm0.24 $	&	SDSS J104640.22+150748.8	&	$\star$	&	2.5	&	1.155$^p$	&	$2.20\pm0.52\times10^{44}$	\\
4	&	XMMU J104643.80+151723.2	&	161.682491	&	15.289785	&	1.81	&	11.7	&	$ 11.64\pm4.02 $	&	$ 1.80\pm0.55 $	&	$ -0.60\pm0.31 $	&	-	&	-	&	-	&	-	&	-	\\
5	&	XMMU J104649.90+151316.3	&	161.707902	&	15.221182	&	2.40	&	16.2	&	$ 12.68\pm3.05 $	&	$ 2.02\pm0.54 $	&	$ -0.66\pm0.19 $	&	NVSS J104649+151312	&	RadioS	&	4.0	&	-	&	-	\\
6	&	XMMU J104653.41+150846.4	&	161.722552	&	15.146230	&	2.92	&	14.3	&	$ 9.98\pm2.73 $	&	$ 2.00\pm0.54 $	&	$ -0.47\pm0.22 $	&	SDSS J104653.42+150845.3	&	G	&	1.1	&	-	&	-	\\
7	&	XMMU J104659.84+152334.3	&	161.749341	&	15.392857	&	1.54	&	17.1	&	$ 8.80\pm2.56 $	&	$ 1.89\pm0.50 $	&	$ -0.60\pm0.24 $	&	SDSS J104659.79+152331.7	&	G	&	2.7	&	-	&	-	\\
8	&	XMMU J104700.95+150213.5	&	161.753951	&	15.037085	&	1.80	&	32.6	&	$ 44.87\pm14.23 $	&	$ 8.41\pm2.12 $	&	$ -0.64\pm0.28 $	&	SDSS J104701.13+150212.1	&	G	&	3.0	&	-	&	-	\\
9	&	XMMU J104701.31+151332.3	&	161.755464	&	15.225645	&	0.99	&	60.8	&	$ 16.67\pm2.68 $	&	$ 2.93\pm0.48 $	&	$ -0.05\pm0.16 $	&	-	&	-	&	-	&	-	&	-	\\
10	&	XMMU J104703.59+150835.3	&	161.764945	&	15.143133	&	1.41	&	26.9	&	$ 8.02\pm2.09 $	&	$ 2.10\pm0.50 $	&	$ -0.21\pm0.23 $	&	SDSS J104703.61+150834.3	&	G	&	1.0	&	-	&	-	\\
11	&	XMMU J104709.26+151512.7	&	161.788602	&	15.253522	&	1.43	&	14.7	&	$ 5.98\pm1.70 $	&	$ 1.21\pm0.31 $	&	$ -0.49\pm0.23 $	&	-	&	-	&	-	&	-	&	-	\\
12	&	$^6$XMMU J104712.94+151103.9	&	161.803915	&	15.184421	&	1.13	&	37.4	&	$ 5.60\pm1.15 $	&	$ 2.57\pm0.53 $	&	$ -0.28\pm0.05 $	&	SDSS J104712.82+151057.8	&	G	&	6.3	&	-	&	-	\\
13	&	XMMU J104713.36+151412.8	&	161.805661	&	15.236898	&	1.98	&	13.9	&	$ 7.18\pm1.78 $	&	$ 1.07\pm0.29 $	&	$ -0.01\pm0.24 $	&	SDSS J104713.33+151412.6	&	G	&	0.4	&	-	&	-	\\
14	&	$^5$XMMU J104714.53+151150.5	&	161.810543	&	15.197351	&	1.25	&	31.5	&	$ 5.48\pm1.13 $	&	$ 2.51\pm0.52 $	&	$ -0.28\pm0.20 $	&	-	&	-	&	-	&	-	&	-	\\
15	&	$^7$XMMU J104715.08+150908.9	 &	161.812845	&	15.152466	&	0.84	&	154.8	&	$ 26.64\pm3.16 $	&	$ 5.57\pm0.67 $	&	$ -0.45\pm0.10 $	&	SDSS J104715.34+150907.5	&	$\star$	&	4.0	&	1.795$^p$	&	$1.23\pm0.15\times10^{45}$	\\
16	&	XMMU J104719.08+150030.7	&	161.829483	&	15.008514	&	1.76	&	22.4	&	$ 20.50\pm7.09 $	&	$ 4.66\pm1.23 $	&	$ -0.96\pm0.25 $	&	SDSS J104718.98+150028.8	&	G	&	2.2	&	-	&	-	\\
17	&	$^3$XMMU J104720.36+151509.7	&	161.834839	&	15.252690	&	0.51	&	580.1	&	$ 79.76\pm4.81 $	&	$ 12.92\pm0.79 $	&	$ -0.83\pm0.04 $	&	NVSS J14720+151505 	&	RadioS	&	6.7	&	0.208	&	$1.62\pm0.10\times10^{43}$	\\
18	&	$^4$XMMU J104720.43+151103.9	&	161.835121	&	15.184425	&	0.94	&	89.9	&	$ 17.97\pm2.40 $	&	$ 3.17\pm0.42 $	&	$ -0.20\pm0.12 $	&	\begin{tiny}
GALEXASC J104720.62+151105.8
\end{tiny}	&	UvS	&	3.5	&	-	&	-	\\
19	&	XMMU J104720.73+151340.7	&	161.836394	&	15.227971	&	1.43	&	19.6	&	$ 8.63\pm2.04 $	&	$ 1.31\pm0.30 $	&	$ -0.73\pm0.16 $	&	SDSS J104720.50+151340.2	&	G	&	3.4	&	-	&	-	\\
20	&	XMMU J104723.13+151414.5	&	161.846360	&	15.237359	&	1.70	&	12.8	&	$ 8.75\pm2.10 $	&	$ 1.64\pm0.37 $	&	$ -0.69\pm0.17 $	&	-	&	-	&	-	&	-	&	-	\\
21	&	XMMU J104723.44+152022.6	&	161.847668	&	15.339611	&	0.83	&	77.5	&	$ 21.29\pm3.46 $	&	$ 3.79\pm0.66 $	&	$ -0.43\pm0.15 $	&	SDSS J104723.37+152020.8	&	$\star$	&	2.0	&	-	&	-	\\
22	&	XMMU J104724.00+151511.1	&	161.850001	&	15.253074	&	1.59	&	43.4	&	$ 15.31\pm2.42 $	&	$ 2.36\pm0.39 $	&	$ -0.70\pm0.12 $	&	-	&	-	&	-	&	-	&	-	\\
23	&	XMMU J104727.07+150105.2	&	161.862782	&	15.018099	&	1.65	&	33.3	&	$ 14.68\pm3.34 $	&	$ 3.12\pm0.71 $	&	$ -0.69\pm0.19 $	&	SDSS J104727.19+150105.2	&	$\star$	&	1.9	&	2.625$^p$	&	$1.74\pm0.40\times10^{45}$	\\
24	&	XMMU J104727.28+150322.3	&	161.863678	&	15.056194	&	1.31	&	46.3	&	$ 15.19\pm2.91 $	&	$ 3.44\pm0.68 $	&	$ -0.17\pm0.15 $	&	SDSS J104727.69+150321.6	&	$\star$	&	6.0	&	0.065$^p$	&	$3.51\pm0.69\times10^{41}$	\\
25	&	XMMU J104728.16+151842.6	&	161.867340	&	15.311824	&	1.37	&	32.5	&	$ 7.28\pm1.80 $	&	$ 1.65\pm0.34 $	&	$ -0.50\pm0.20 $	&	SDSS J104728.42+151840.8	&	$\star$	&	4.2	&	2.255$^p$	&	$6.40\pm1.31\times10^{44}$	\\
26	&	XMMU J104728.53+150533.4	&	161.868865	&	15.092601	&	1.65	&	21.7	&	$ 8.66\pm2.09 $	&	$ 1.91\pm0.47 $	&	$ -0.42\pm0.22 $	&	SDSS J104728.71+150533.0	&	$\star$	&	2.7	&	-	&	-	\\
27	&	XMMU J104728.60+150722.5	&	161.869175	&	15.122909	&	1.35	&	26.7	&	$ 9.27\pm2.18 $	&	$ 1.97\pm0.30 $	&	$ -0.38\pm0.21 $	&	SDSS J104728.53+150719.9	&	G	&	2.7	&	-	&	-	\\
28	&	$^2$XMMU J104729.28+151408.3	&	161.872019	&	15.235652	&	1.72	&	34.1	&	$ 12.22\pm2.11 $	&	$ 2.25\pm0.38 $	&	$ -0.61\pm0.14 $	&	NVSS J104729+151404	&	RadioS$^b$	&	5.9	&	-	&	-	\\
29	&	XMMU J104733.45+152158.9	&	161.889368	&	15.366356	&	2.02	&	12.3	&	$ 5.24\pm1.90 $	&	$ 1.23\pm0.39 $	&	$ -0.27\pm0.32 $	&	-	&	-	&	-	&	-	&	-	\\
30	&	XMMU J104738.30+152144.7	&	161.909590	&	15.362430	&	2.35	&	8.5	&	$ 4.72\pm1.57 $	&	$ 0.74\pm0.27 $	&	$ -0.07\pm0.33 $	&	SDSS J104738.63+152142.0	&	G	&	5.5	&	-	&	-	\\
31	&	XMMU J104741.38+150528.7	&	161.922412	&	15.091312	&	1.26	&	31.4	&	$ 11.80\pm2.41 $	&	$ 1.80\pm0.37 $	&	$ -0.44\pm0.18 $	&	SDSS J104741.47+150528.0	&	G	&	1.5	&	-	&	-	\\
32	&	$^1$XMMU J104741.54+151334.6	&	161.923103	&	15.226291	&	0.22	&	2364.2	&	$ 193.17\pm8.02 $	&	$ 33.34\pm1.26 $	&	$ -0.63\pm0.03 $	&	SDSS J104741.75+151332.2	&	QSO$^a$	&	3.8	&	0.386	&	$1.73\pm0.07\times10^{44}$	\\
33	&	$^{12}$XMMU J104745.66+151644.6 &	161.940257	&	15.279063	&	1.28	&	26.6	&	$ 4.08\pm0.95 $	&	$ 1.87\pm0.44 $	&	$ -0.04\pm0.23 $	&	SDSS J104745.88+151642.8	&	$\star$	&	3.7	&	3.045$^p$	&	$1.50\pm0.35\times10^{45}$	\\
34	&	XMMU J104746.10+151758.6	&	161.942083	&	15.299610	&	1.38	&	16.4	&	$ 3.17\pm0.89 $	&	$ 1.46\pm0.41 $	&	$ -0.68\pm0.21 $	&	SDSS J104746.36+151757.4	&	G	&	3.9	&	-	&	-	\\
35	&	$^{13}$XMMU J104748.48+151550.7	&	161.952008	&	15.264084	&	2.16	&	17.9	&	$ 7.69\pm1.88 $	&	$ 1.41\pm0.32 $	&	$ -0.87\pm0.13 $	&	SDSS J104748.54+151546.8	&	G	&	4.0	&	-	&	-	\\
36	&	$^{14}$XMMU J104750.24+151450.1	&	161.959353	&	15.247242	&	0.78	&	83.8	&	$ 17.07\pm2.41 $	&	$ 3.20\pm0.43 $	&	$ -0.20\pm0.13 $	&	SDSS J104750.48+151447.9	&	$\star$	&	4.1	&	-	&	-	\\
37	&	$^{15}$XMMU J104750.25+151314.2	&	161.959355	&	15.220614	&	0.88	&	114.7	&	$ 19.41\pm2.46 $	&	$ 3.48\pm0.43 $	&	$ -0.30\pm0.11 $	&	SDSS J104750.43+151312.8	&	$\star$	&	3.1	&	-	&	-	\\
38	&	XMMU J104752.90+150521.5	&	161.970401	&	15.089299	&	0.55	&	426.0	&	$ 77.66\pm6.70 $	&	$ 13.71\pm1.07 $	&	$ -0.58\pm0.07 $	&	SDSS J104753.10+150520.2	&	$\star$	&	3.2	&	1.305$^p$	&	$1.38\pm0.11\times10^{45}$	\\
39	&	XMMU J104754.93+152404.5	&	161.978856	&	15.401249	&	1.48	&	12.9	&	$ 4.26\pm2.03 $	&	$ 1.47\pm0.49 $	&	$ -0.47\pm0.32 $	&	SDSS J104755.07+152401.5	&	G	&	3.6	&	-	&	-	\\
40	&	XMMU J104758.79+150409.6	&	161.994960	&	15.069329	&	1.73	&	15.0	&	$ 4.83\pm1.60 $	&	$ 1.38\pm0.42 $	&	$ 0.50\pm0.28 $	&	-	&	-	&	-	&	-	&	-	\\
41	&	$^8$XMMU J104800.00+151607.9	&	161.999985	&	15.268854	&	1.55	&	13.6	&	$ 5.21\pm1.38 $	&	$ 2.39\pm0.63 $	&	$ -0.79\pm0.18 $	&	SDSS J104759.83+151609.7	&	BCG$^c$	&	3.0	&	-	&	-	\\
42	&	$^{11}$XMMU J104800.64+151426.3	&	162.002665	&	15.240643	&	2.40	&	10.9	&	$ 4.14\pm1.16 $	&	$ 1.90\pm0.53 $	&	$ -0.11\pm0.27 $	&	SDSS J104800.82+151422.5 	&	$\star$	&	4.6	&	2.595$^p$	&	$1.03\pm0.29\times10^{45}$	\\
43	&	$^9$XMMU J104805.17+151758.1	&	162.021546	&	15.299467	&	0.41	&	764.2	&	$ 119.80\pm7.62 $	&	$ 19.46\pm1.17 $	&	$ -0.57\pm0.05 $	&	SDSS J104805.33+151755.9	&	$\star$	&	3.1	&	1.065$^p$	&	$1.19\pm0.07\times10^{45}$	\\
44	&	XMMU J104806.17+150600.6	&	162.025692	&	15.100174	&	0.48	&	383.2	&	$ 68.97\pm5.78 $	&	$ 12.73\pm1.04 $	&	$ -0.55\pm0.07 $	&	SDSS J104806.30+150558.4	&	G	&	2.9	&	-	&	-	\\
45	&	XMMU J104806.73+150752.3	&	162.028044	&	15.131198	&	1.01	&	51.4	&	$ 15.90\pm2.81 $	&	$ 2.94\pm0.51 $	&	$ -0.52\pm0.15 $	&	-	&	-	&	-	&	-	&	-	\\
46	&	XMMU J104806.97+152237.6	&	162.029034	&	15.377112	&	1.59	&	113.8	&	$ 37.99\pm4.80 $	&	$ 8.34\pm1.03 $	&	$ -0.96\pm0.05 $	&	2MASX J10480700+1522356	&	G	&	2.0	&	0.213	&	$1.10\pm0.14\times10^{43}$	\\
47	&	$^{10}$XMMU J104810.53+151439.7	&	162.043866	&	15.244365	&	1.83	&	28.0	&	$ 12.77\pm2.80 $	&	$ 2.25\pm0.46 $	&	$ -0.69\pm0.16 $	&	-	&	-	&	-	&	-	&	-	\\
\hline 
\enddata
\begin{tablenotes}
      \scriptsize
      \item $Note.$ 
      (1): Generic source number. 
      (2): The International Astronomical Union-style J2000.0 position-based names for the X-ray sources, the prefix ``XMMU" means ``{\it XMM-Newton}, Unregistered", the front superscript notes the source number in Table~\ref{t:spec}.
      (3)(4): Source position, RA, DEC in J2000. 
      (5): Position uncertainty. 
      (6): Detection likelihood, defined as $DET\_ML=-ln(P)$, where P is the probability that a Poissonian fluctuation of the background is detected as a spurious source.
      (7): Source count rate in the 0.5-7 keV, which is the total
from PN, MOS1, and MOS2, except for locations where the damaged CCD3/6 of 
MOS1 or CCD gaps are encountered.
      (8): Source flux in the 0.5-7 keV, the energy conversion factors (ECF) was computed by assuming a power-law spectrum with spectral index $\Gamma = 1.7$ and Galactic column density $N_H = 2.76 \times 10^{20}\ cm^{-2}$ (Dickey \& Lockman 1990), accounting for missing counts from the damaged CCDs and CCD gaps, as well as other location-dependent instrumental effects.
      (9): The hardness ratios defined as HR=(H-S)/(H+S), where S and H are the source count rates in the 0.5-2, 2-7 keV, respectively.
      (10): X-ray source counterpart by cross matching with NED within separations less than 7$^{\prime\prime}$ (Watson et al. 2009).
      (11): Type of X-ray source counterpart, with G=galaxy; BCG=Brightest Cluster Galaxy; $\star$=star (the stars with photometric redshifts are QSO candidates); QSO=quasar; RadioS=Radio Source and  UvS=UV Source.
      (12): Separation between X-ray source and its counterpart.
      (13): Redshift of counterpart, with superscript of $^{p}$ representing photometric redshift.
      (14): Luminosity of X-ray source estimated from X-ray flux and redshift of its counterpart.

$^a$: The UV bright QSO.
$^b$: This radio source is near the BCG (SDSS J104729.01+151402.0, Sep=1.4$^{\prime\prime}$) of A1095W.
$^c$: The BCG of A1095E.
    \end{tablenotes} 
\label{t:s-a1095}
\end{deluxetable}
\end{landscape}

\newpage
\begin{landscape}
\begin{deluxetable}{lccccccccccccc}
\tablecolumns{14}
\tabletypesize{\scriptsize}
  \tablecaption{A1926 field source list}
  \tablewidth{0pt}
  \tablehead{
  \colhead{Source} &
  \colhead{Name } &
  \colhead{RA} &
  \colhead{DEC} &
  \colhead{$\delta$} &
  \colhead{DET$\_$ML} &
  \colhead{CR} &
  \colhead{Flux} &
  \colhead{HR} &
  \colhead{Counterpart name} &
  \colhead{Type} &
 \colhead{Sep} &
 \colhead{Redshift} &
 \colhead{Luminosity} \\
	&		&	(J2000)	&	(J2000)	&	(1$^{\prime\prime}$)	&		&	(counts$/$ks)	&	($\rm{10^{-14}\ erg\ cm^{-2} s^{-1}}$)	&		&		&		&	(1$^{\prime\prime}$)	&		&	($\rm{erg\ s^{-1}}$)	\\
  \noalign{\smallskip}
  \colhead{(1)} &
 \colhead{(2)} &
 \colhead{(3)} &
 \colhead{(4)} &
 \colhead{(5)} &
 \colhead{(6)} &
 \colhead{(7)} & 
 \colhead{(8)} &  
 \colhead{(9)} &                  
 \colhead{(10)} &                  
 \colhead{(11)} &                  
 \colhead{(12)} &                  
 \colhead{(13)} &                  
 \colhead{(14)} \\
   }
\startdata
1	&	XMMU J217.397067+24.604126	&	217.397067	&	24.604126	&	0.60	&	238.1	&	$ 44.09\pm3.72 $	&	$ 7.86\pm0.71 $	&	$ 0.12\pm0.08 $	&	2MASX J14293545+2436154	&	G	&	1.9	&	0.128	&	$3.36\pm0.30\times10^{42}$	\\
2	&	XMMU J217.420615+24.577670	&	217.420615	&	24.577670	&	2.21	&	11.5	&	$ 11.00\pm2.53 $	&	$ 2.01\pm0.50 $	&	$ -0.66\pm0.19 $	&	-	&	-	&	-	&	-	&	-	\\
3	&	XMMU J217.422412+24.562411	&	217.422412	&	24.562411	&	2.69	&	10.9	&	$ 9.11\pm2.18 $	&	$ 1.46\pm0.39 $	&	$ -0.32\pm0.22 $	&	-	&	-	&	-	&	-	&	-	\\
4	&	XMMU J217.422811+24.571962	&	217.422811	&	24.571962	&	1.18	&	126.8	&	$ 36.11\pm3.59 $	&	$ 6.00\pm0.67 $	&	$ -0.11\pm0.10 $	&	SDSS J142941.65+243417.1	&	G	&	3.1	&	0.147	&	$3.47\pm0.39\times10^{42}$	\\
5	&	XMMU J217.427498+24.714159	&	217.427498	&	24.714159	&	1.13	&	48.3	&	$ 13.08\pm2.12 $	&	$ 2.70\pm0.44 $	&	$ -0.58\pm0.14 $	&	SDSS J142942.72+244252.0	&	$\star$	&	2.0	&	2.025$^p$	&	$8.04\pm1.31\times10^{44}$	\\
6	&	XMMU J217.433129+24.552466	&	217.433129	&	24.552466	&	1.30	&	21.2	&	$ 10.16\pm2.08 $	&	$ 1.58\pm0.37 $	&	$ -0.41\pm0.20 $	&	SDSS J142944.06+243308.7	&	G	&	1.5	&	-	&	-	\\
7	&	XMMU J217.442860+24.682915	&	217.442860	&	24.682915	&	1.08	&	88.1	&	$ 17.78\pm2.19 $	&	$ 2.83\pm0.39 $	&	$ -0.64\pm0.10 $	&	SDSS J142946.33+244058.3	&	G	&	0.7	&	-	&	-	\\
8	&	XMMU J217.460527+24.567928	&	217.460527	&	24.567928	&	1.21	&	25.8	&	$ 8.92\pm1.69 $	&	$ 1.76\pm0.36 $	&	$ -0.30\pm0.18 $	&	SDSS J142950.56+243406.4	&	G	&	2.0	&	-	&	-	\\
9	&	XMMU J217.463546+24.637864	&	217.463546	&	24.637864	&	1.97	&	42.5	&	$ 10.58\pm1.70 $	&	$ 1.81\pm0.33 $	&	$ -0.37\pm0.16 $	&	SDSS J142951.24+243817.4	&	$\star$	&	1.1	&	-	&	-	\\
10	&	XMMU J217.483206+24.594784	&	217.483206	&	24.594784	&	1.45	&	17.0	&	$ 5.50\pm1.21 $	&	$ 1.10\pm0.26 $	&	$ -0.46\pm0.20 $	&	SDSS J142956.18+243541.4	&	G	&	2.9	&	-	&	-	\\
11	&	XMMU J217.486443+24.559335	&	217.486443	&	24.559335	&	2.64	&	20.7	&	$ 5.89\pm1.33 $	&	$ 0.99\pm0.29 $	&	$ -0.93\pm0.13 $	&	SDSS J142956.92+243337.3	&	$\star$	&	4.4	&	0.995$^p$	&	$5.10\pm1.50\times10^{43}$	\\
12	&	XMMU J217.490144+24.700731	&	217.490144	&	24.700731	&	1.56	&	21.8	&	$ 6.46\pm1.36 $	&	$ 1.11\pm0.25 $	&	$ -0.86\pm0.14 $	&	2MASX J14295759+2441585	&	G	&	3.6	&	0.133	&	$5.19\pm1.17\times10^{41}$	\\
13	&	XMMU J217.492865+24.565309	&	217.492865	&	24.565309	&	1.71	&	17.8	&	$ 4.68\pm1.18 $	&	$ 1.43\pm0.37 $	&	$ -0.41\pm0.22 $	&	SDSS J142958.03+243350.6	&	G	&	5.6	&	-	&	-	\\
14	&	XMMU J217.495035+24.667446	&	217.495035	&	24.667446	&	0.97	&	55.8	&	$ 10.47\pm1.55 $	&	$ 2.02\pm0.34 $	&	$ -0.18\pm0.15 $	&	SDSS J142959.20+244001.0	&	G	&	5.7	&	-	&	-	\\
15	&	XMMU J217.504884+24.469533	&	217.504884	&	24.469533	&	0.57	&	335.4	&	$ 44.46\pm3.45 $	&	$ 9.67\pm0.82 $	&	$ -0.40\pm0.07 $	&	SDSS J143001.25+242811.3	&	G	&	1.6	&	-	&	-	\\
16	&	XMMU J217.517846+24.846114	&	217.517846	&	24.846114	&	1.15	&	72.8	&	$ 17.26\pm2.41 $	&	$ 3.87\pm0.58 $	&	$ -0.55\pm0.13 $	&	SDSS J143004.36+245045.6	&	G	&	1.2	&	-	&	-	\\
17	&	XMMU J217.550165+24.746981	&	217.550165	&	24.746981	&	0.91	&	132.7	&	$ 15.02\pm1.61 $	&	$ 2.65\pm0.31 $	&	$ -0.43\pm0.10 $	&	SDSS J143012.15+244450.0	&	$\star$	&	1.8	&	0.375$^p$	&	$1.28\pm0.15\times10^{43}$	\\
18	&	XMMU J217.568480+24.699673	&	217.568480	&	24.699673	&	0.88	&	54.9	&	$ 6.85\pm1.03 $	&	$ 1.06\pm0.18 $	&	$ -0.48\pm0.13 $	&	SDSS J143016.41+244159.6	&	G	&	0.9	&	-	&	-	\\
19	&	$^5$XMMU J217.580858+24.631305	&	217.580858	&	24.631305	&	0.77	&	145.5	&	$ 7.67\pm0.87 $	&	$ 3.52\pm0.40 $	&	$ -0.93\pm0.06 $	&	SDSS J143019.46+243753.4	&	$\star$	&	1.1	&	-	&	-	\\
20	&	XMMU J217.596146+24.808146	&	217.596146	&	24.808146	&	1.15	&	38.3	&	$ 6.60\pm1.26 $	&	$ 1.08\pm0.22 $	&	$ 0.58\pm0.18 $	&	SDSS J143023.14+244830.3	&	G	&	1.5	&	-	&	-	\\
21	&	$^6$XMMU J217.601277+24.602510	&	217.601277	&	24.602510	&	0.58	&	226.8	&	$ 20.90\pm1.80 $	&	$ 3.76\pm0.38 $	&	$ -0.54\pm0.08 $	&	2MASXi J1430243+243610	&	G	&	1.4	&	-	&	-	\\
22	&	XMMU J217.601295+24.596945	&	217.601295	&	24.596945	&	2.77	&	13.3	&	$ 4.94\pm1.06 $	&	$ 0.83\pm0.20 $	&	$ -0.79\pm0.14 $	&	2MASX J14302407+2435486	&	G	&	3.3	&	0.138	&	$4.23\pm1.02\times10^{41}$	\\
23	&	XMMU J217.602192+24.788778	&	217.602192	&	24.788778	&	1.09	&	39.9	&	$ 7.03\pm1.21 $	&	$ 1.20\pm0.23 $	&	$ 0.04\pm0.17 $	&	SDSS J143024.59+244719.9	&	$\star$	&	1.0	&	-	&	-	\\
24	&	XMMU J217.609198+24.722463	&	217.609198	&	24.722463	&	1.29	&	25.3	&	$ 4.04\pm0.81 $	&	$ 0.69\pm0.15 $	&	$ -0.44\pm0.18 $	&	SDSS J143026.36+244321.6	&	G	&	2.3	&	-	&	-	\\
25	&	XMMU J217.610152+24.507977	&	217.610152	&	24.507977	&	0.49	&	513.8	&	$ 46.53\pm3.33 $	&	$ 10.23\pm0.73 $	&	$ -0.60\pm0.06 $	&	SDSS J143026.47+243029.6	&	$\star$	&	1.1	&	2.315$^p$	&	$4.22\pm0.30\times10^{45}$	\\
26	&	XMMU J217.610747+24.602797	&	217.610747	&	24.602797	&	1.58	&	31.6	&	$ 7.60\pm1.40 $	&	$ 1.20\pm0.31 $	&	$ -0.77\pm0.13 $	&	2MASX J14302634+2436116	&	G	&	3.3	&	0.135	&	$5.77\pm1.49\times10^{41}$	\\
27	&	$^4$XMMU J217.610843+24.669255	&	217.610843	&	24.669255	&	0.62	&	164.4	&	$ 9.52\pm0.95 $	&	$ 4.37\pm0.44 $	&	$ -0.33\pm0.09 $	&	SDSS J143026.65+244010.7	&	$\star$	&	1.6	&	2.065$^p$	&	$1.36\pm0.14\times10^{45}$	\\
28	&	XMMU J217.611431+24.610636	&	217.611431	&	24.610636	&	1.14	&	36.3	&	$ 7.46\pm1.14 $	&	$ 1.22\pm0.20 $	&	$ -0.49\pm0.13 $	&	-	&	-	&	-	&	-	&	-	\\
29	&	XMMU J217.613827+24.804217	&	217.613827	&	24.804217	&	1.88	&	24.5	&	$ 5.19\pm1.10 $	&	$ 1.01\pm0.24 $	&	$ 0.88\pm0.15 $	&	-	&	-	&	-	&	-	&	-	\\
30	&	XMMU J217.618981+24.483501	&	217.618981	&	24.483501	&	1.77	&	20.1	&	$ 5.47\pm1.22 $	&	$ 0.86\pm0.23 $	&	$ -0.16\pm0.22 $	&	SDSS J143028.41+242900.4	&	$\star$	&	1.9	&	-	&	-	\\
31	&	$^3$XMMU J217.619170+24.671927	&	217.619170	&	24.671927	&	0.34	&	1259.5	&	$ 57.91\pm2.58 $	&	$ 9.83\pm0.46 $	&	$ -0.68\pm0.03 $	&	2MASX J14302862+2440196	&	BCG$^b$	&	0.5	&	0.136	&	$4.82\pm0.23\times10^{42}$	\\
32	&	XMMU J217.619889+24.603738	&	217.619889	&	24.603738	&	0.94	&	37.3	&	$ 11.12\pm2.97 $	&	$ 1.79\pm0.34 $	&	$ -0.73\pm0.19 $	&	-	&	-	&	-	&	-	&	-	\\
33	&	$^7$XMMU J217.620972+24.547558	&	217.620972	&	24.547558	&	0.49	&	822.3	&	$ 54.88\pm2.98 $	&	$ 12.64\pm0.89 $	&	$ -0.54\pm0.05 $	&	SDSS J143029.02+243250.0	&	$\star$	&	1.1	&	2.205$^p$	&	$4.63\pm0.33\times10^{45}$	\\
34	&	XMMU J217.622311+24.454404	&	217.622311	&	24.454404	&	1.94	&	15.1	&	$ 5.16\pm1.27 $	&	$ 1.13\pm0.32 $	&	$ -0.55\pm0.22 $	&	SDSS J143029.01+242716.1	&	$\star$	&	4.6	&	1.925$^p$	&	$2.97\pm0.84\times10^{44}$	\\
35	&	$^1$XMMU J217.623346+24.691364	&	217.623346	&	24.691364	&	0.27	&	1515.9	&	$ 79.70\pm4.27 $	&	$ 15.24\pm0.65 $	&	$ -0.09\pm0.05 $	&	2MASX J14302965+2441296	&	G	&	1.0	&	-	&	-	\\
36	&	XMMU J217.623980+24.499409	&	217.623980	&	24.499409	&	2.91	&	8.6	&	$ 3.53\pm0.99 $	&	$ 0.84\pm0.27 $	&	$ -0.53\pm0.25 $	&	SDSS J143029.70+242956.6	&	G	&	1.4	&	-	&	-	\\
37	&	XMMU J217.631514+24.671674	&	217.631514	&	24.671674	&	1.11	&	20.5	&	$ 3.83\pm0.81 $	&	$ 0.72\pm0.16 $	&	$ -0.58\pm0.16 $	&	SDSS J143031.58+244018.1	&	G	&	0.3	&	-	&	-	\\
38	&	XMMU J217.631873+24.420660	&	217.631873	&	24.420660	&	2.36	&	10.7	&	$ 5.70\pm1.54 $	&	$ 1.11\pm0.38 $	&	$ -0.60\pm0.24 $	&	-	&	-	&	-	&	-	&	-	\\
39	&	XMMU J217.633844+24.684826	&	217.633844	&	24.684826	&	2.39	&	19.0	&	$ 5.49\pm1.05 $	&	$ 0.83\pm0.18 $	&	$ -0.84\pm0.12 $	&	SDSS J143032.48+244104.3	&	G	&	5.0	&	-	&	-	\\
40	&	XMMU J217.635647+24.623048	&	217.635647	&	24.623048	&	1.46	&	14.2	&	$ 5.62\pm1.45 $	&	$ 0.79\pm0.26 $	&	$ -0.69\pm0.20 $	&	-	&	-	&	-	&	-	&	-	\\
41	&	XMMU J217.643173+24.472308	&	217.643173	&	24.472308	&	1.67	&	19.7	&	$ 5.53\pm1.23 $	&	$ 1.01\pm0.28 $	&	$ -0.43\pm0.21 $	&	SDSS J143034.25+242820.7	&	$\star$	&	1.5	&	1.965$^p$	&	$2.79\pm0.77\times10^{44}$	\\
42	&	$^2$XMMU J217.643825+24.677384	&	217.643825	&	24.677384	&	0.46	&	479.4	&	$ 26.73\pm1.70 $	&	$ 5.07\pm0.34 $	&	$ -0.53\pm0.05 $	&	-	&	-	&	-	&	-	&	-	\\
43	&	XMMU J217.652843+24.806537	&	217.652843	&	24.806537	&	2.61	&	8.9	&	$ 2.80\pm0.87 $	&	$ 0.55\pm0.19 $	&	$ 0.50\pm0.31 $	&	SDSS J143036.66+244823.1	&	G	&	0.5	&	-	&	-	\\
44	&	XMMU J217.659965+24.493268	&	217.659965	&	24.493268	&	1.43	&	27.7	&	$ 5.29\pm1.09 $	&	$ 1.40\pm0.30 $	&	$ -0.67\pm0.16 $	&	SDSS J143038.48+242937.5	&	$\star$	&	2.1	&	2.505$^p$	&	$7.00\pm1.50\times10^{44}$	\\
45	&	XMMU J217.660017+24.738530	&	217.660017	&	24.738530	&	0.80	&	132.9	&	$ 12.88\pm1.44 $	&	$ 2.25\pm0.27 $	&	$ -0.54\pm0.09 $	&	SDSS J143038.40+244418.5	&	$\star$	&	0.2	&	1.325$^p$	&	$2.36\pm0.28\times10^{44}$	\\
46	&	XMMU J217.667892+24.553894	&	217.667892	&	24.553894	&	1.43	&	13.4	&	$ 2.64\pm0.70 $	&	$ 0.60\pm0.15 $	&	$ -0.82\pm0.17 $	&	-	&	-	&	-	&	-	&	-	\\
47	&	XMMU J217.668669+24.774961	&	217.668669	&	24.774961	&	1.45	&	30.0	&	$ 5.20\pm1.01 $	&	$ 0.89\pm0.19 $	&	$ -0.33\pm0.18 $	&	-	&	-	&	-	&	-	&	-	\\
48	&	XMMU J217.670962+24.477033	&	217.670962	&	24.477033	&	1.31	&	55.5	&	$ 11.73\pm1.74 $	&	$ 2.04\pm0.35 $	&	$ -0.51\pm0.13 $	&	SDSS J143041.07+242837.3	&	$\star$	&	0.6	&	2.255$^p$	&	$7.90\pm1.35\times10^{44}$	\\
49	&	XMMU J217.671460+24.441456	&	217.671460	&	24.441456	&	1.48	&	28.0	&	$ 7.84\pm1.58 $	&	$ 1.44\pm0.34 $	&	$ -0.85\pm0.15 $	&	2MASX J14304135+2426303	&	G	&	2.2	&	0.136	&	$7.08\pm1.67\times10^{41}$	\\
50	&	XMMU J217.681630+24.722818	&	217.681630	&	24.722818	&	0.35	&	1097.6	&	$ 54.95\pm2.68 $	&	$ 9.78\pm0.56 $	&	$ -0.65\pm0.04 $	&	SDSS J143043.60+244322.1	&	$\star$	&	0.1	&	0.595$^p$	&	$1.43\pm0.08\times10^{44}$	\\
51	&	XMMU J217.687178+24.574344	&	217.687178	&	24.574344	&	2.36	&	8.2	&	$ 2.55\pm0.75 $	&	$ 0.55\pm0.17 $	&	$ -0.47\pm0.27 $	&	-	&	-	&	-	&	-	&	-	\\
52	&	XMMU J217.704484+24.778175	&	217.704484	&	24.778175	&	1.38	&	36.0	&	$ 6.49\pm1.16 $	&	$ 1.37\pm0.27 $	&	$ -0.57\pm0.15 $	&	SDSS J143048.96+244636.9	&	G	&	4.7	&	-	&	-	\\
53	&	XMMU J217.704969+24.587968	&	217.704969	&	24.587968	&	1.32	&	17.0	&	$ 2.64\pm0.68 $	&	$ 0.53\pm0.14 $	&	$ 0.48\pm0.26 $	&	-	&	-	&	-	&	-	&	-	\\
54	&	XMMU J217.710450+24.784631	&	217.710450	&	24.784631	&	1.84	&	17.7	&	$ 4.79\pm1.10 $	&	$ 1.03\pm0.26 $	&	$ -0.78\pm0.18 $	&	-	&	-	&	-	&	-	&	-	\\
55	&	XMMU J217.711438+24.711956	&	217.711438	&	24.711956	&	1.03	&	53.9	&	$ 6.35\pm0.98 $	&	$ 1.09\pm0.19 $	&	$ 0.10\pm0.15 $	&	SDSS J143050.80+244243.5	&	$\star$	&	0.9	&	-	&	-	\\
56	&	XMMU J217.711914+24.647066	&	217.711914	&	24.647066	&	1.08	&	25.2	&	$ 4.33\pm0.84 $	&	$ 0.75\pm0.15 $	&	$ -0.71\pm0.15 $	&	2MASX J14305086+2438493	&	G	&	0.3	&	0.098	&	$1.84\pm0.37\times10^{41}$	\\
57	&	XMMU J217.718106+24.768559	&	217.718106	&	24.768559	&	2.04	&	8.8	&	$ 3.39\pm0.97 $	&	$ 0.52\pm0.18 $	&	$ -0.79\pm0.20 $	&	SDSS J143052.13+244607.2	&	$\star$	&	3.0	&	-	&	-	\\
58	&	XMMU J217.727425+24.771655	&	217.727425	&	24.771655	&	4.69	&	13.3	&	$ 4.28\pm1.06 $	&	$ 0.70\pm0.20 $	&	$ -0.78\pm0.18 $	&	SDSS J143054.97+244614.6	&	G	&	6.3	&	-	&	-	\\
59	&	XMMU J217.728702+24.575579	&	217.728702	&	24.575579	&	1.22	&	29.3	&	$ 4.24\pm0.83 $	&	$ 0.71\pm0.16 $	&	$ -0.60\pm0.16 $	&	SDSS J143054.86+243432.1	&	G	&	0.3	&	-	&	-	\\
60	&	XMMU J217.729768+24.641952	&	217.729768	&	24.641952	&	0.29	&	1756.6	&	$ 66.31\pm2.69 $	&	$ 11.59\pm0.50 $	&	$ -0.57\pm0.03 $	&	SDSS J143055.18+243830.7	&	QSO	&	0.6	&	1.534	&	$1.74\pm0.08\times10^{45}$	\\
61	&	XMMU J217.733217+24.865270	&	217.733217	&	24.865270	&	1.58	&	26.3	&	$ 3.98\pm1.00 $	&	$ 3.65\pm0.92 $	&	$ -0.60\pm0.23 $	&	SDSS J143055.90+245154.0	&	$\star$	&	1.3	&	1.685$^p$	&	$6.92\pm1.74\times10^{44}$	\\
62	&	XMMU J217.747892+24.477957	&	217.747892	&	24.477957	&	1.32	&	29.1	&	$ 5.79\pm1.18 $	&	$ 1.27\pm0.27 $	&	$ -0.68\pm0.15 $	&	-	&	-	&	-	&	-	&	-	\\
63	&	XMMU J217.748642+24.750468	&	217.748642	&	24.750468	&	0.50	&	589.7	&	$ 38.82\pm2.44 $	&	$ 7.82\pm0.53 $	&	$ -0.49\pm0.05 $	&	SDSS J143059.63+244502.5	&	$\star$	&	1.1	&	1.285$^p$	&	$7.60\pm0.51\times10^{44}$	\\
64	&	XMMU J217.749314+24.603408	&	217.749314	&	24.603408	&	0.85	&	97.1	&	$ 10.27\pm1.27 $	&	$ 1.90\pm0.26 $	&	$ -0.40\pm0.11 $	&	SDSS J143059.77+243612.7	&	$\star$	&	1.0	&	-	&	-	\\
65	&	XMMU J217.755977+24.638639	&	217.755977	&	24.638639	&	1.57	&	23.1	&	$ 3.33\pm0.82 $	&	$ 0.59\pm0.15 $	&	$ 0.72\pm0.26 $	&	-	&	-	&	-	&	-	&	-	\\
66	&	XMMU J217.757946+24.688726	&	217.757946	&	24.688726	&	0.39	&	874.4	&	$ 43.78\pm2.36 $	&	$ 7.93\pm0.47 $	&	$ -0.52\pm0.05 $	&	SDSS J143101.95+244119.8	&	QSO	&	0.9	&	1.506	&	$1.14\pm0.07\times10^{45}$	\\
67	&	XMMU J217.759705+24.727007	&	217.759705	&	24.727007	&	2.10	&	12.0	&	$ 3.43\pm0.88 $	&	$ 0.62\pm0.18 $	&	$ -0.41\pm0.23 $	&	SDSS J143102.34+244337.7	&	$\star$	&	0.6	&	0.685$^p$	&	$1.28\pm0.37\times10^{43}$	\\
68	&	XMMU J217.764628+24.732372	&	217.764628	&	24.732372	&	1.01	&	87.4	&	$ 10.00\pm1.31 $	&	$ 2.05\pm0.30 $	&	$ -0.41\pm0.12 $	&	SDSS J143103.52+244357.0	&	$\star$	&	0.6	&	2.015$^p$	&	$6.03\pm0.88\times10^{44}$	\\
69	&	XMMU J217.767838+24.672625	&	217.767838	&	24.672625	&	1.25	&	34.4	&	$ 4.91\pm1.09 $	&	$ 0.99\pm0.27 $	&	$ 0.36\pm0.23 $	&	-	&	-	&	-	&	-	&	-	\\
70	&	XMMU J217.779265+24.575482	&	217.779265	&	24.575482	&	0.81	&	175.0	&	$ 18.31\pm1.87 $	&	$ 2.92\pm0.38 $	&	$ -0.47\pm0.09 $	&	SDSS J143107.03+243431.6	&	$\star$	&	0.2	&	-	&	-	\\
71	&	XMMU J217.805305+24.584931	&	217.805305	&	24.584931	&	0.93	&	104.9	&	$ 13.49\pm1.61 $	&	$ 2.32\pm0.31 $	&	$ -0.57\pm0.10 $	&	-	&	-	&	-	&	-	&	-	\\
72	&	XMMU J217.815673+24.499656	&	217.815673	&	24.499656	&	2.33	&	10.5	&	$ 2.95\pm0.85 $	&	$ 1.36\pm0.39 $	&	$ -0.21\pm0.29 $	&	-	&	-	&	-	&	-	&	-	\\
73	&	XMMU J217.818822+24.675090	&	217.818822	&	24.675090	&	1.60	&	48.7	&	$ 9.74\pm1.47 $	&	$ 1.45\pm0.25 $	&	$ -0.45\pm0.14 $	&	SDSS J143116.53+244028.3	&	$\star$	&	1.9	&	-	&	-	\\
74	&	XMMU J217.846068+24.682884	&	217.846068	&	24.682884	&	1.58	&	17.6	&	$ 6.84\pm1.44 $	&	$ 1.29\pm0.29 $	&	$ -0.60\pm0.17 $	&	2MASX J14312340+2441013	&	BCG$^c$	&	5.5	&	0.097	&	$3.09\pm0.69\times10^{41}$	\\
75	&	XMMU J217.851676+24.607276	&	217.851676	&	24.607276	&	0.44	&	880.8	&	$ 70.20\pm3.71 $	&	$ 12.86\pm0.72 $	&	$ -0.58\pm0.04 $	&	SDSS J143124.44+243626.5	&	$\star$	&	0.6	&	1.365$^p$	&	$1.45\pm0.08\times10^{45}$	\\
76	&	$^{8}$XMMU J217.857867+24.705732	&	217.857867	&	24.705732	&	0.26	&	7077.2	&	$ 413.46\pm9.19 $	&	$ 73.21\pm1.86 $	&	$ -0.60\pm0.02 $	&	SDSS J143125.88+244220.6	&	QSO$^a$	&	0.1	&	0.407	&	$4.30\pm0.11\times10^{44}$	\\
77	&	XMMU J217.860248+24.675353	&	217.860248	&	24.675353	&	0.49	&	391.9	&	$ 48.95\pm3.36 $	&	$ 8.49\pm0.64 $	&	$ -0.50\pm0.06 $	&	SDSS J143126.42+244031.5	&	G	&	0.5	&	-	&	-	\\
78	&	XMMU J217.868882+24.578807	&	217.868882	&	24.578807	&	2.00	&	20.8	&	$ 4.14\pm0.97 $	&	$ 1.90\pm0.44 $	&	$ -0.42\pm0.22 $	&	SDSS J143128.55+243444.2	&	$\star$	&	0.6	&	0.315$^p$	&	$6.13\pm1.42\times10^{42}$	\\
\enddata
\begin{tablenotes}
      \scriptsize
      \item $Note.$ 
The same as Table. 4 except (8), when calculating the ECF, the Galactic column density of A1926 field is $N_H = 2.59 \times 10^{20}\ cm^{-2}$.

$^a$: The UV bright QSO.
$^b$: The BCG of A1926N.
$^c$: The BCG of MaxBCG J217.84740+24.68382.
    \end{tablenotes} 
\label{t:s-a1926}
\end{deluxetable}
\end{landscape}

\begin{table}
\caption{X-ray spectral properties of luminous compact sources}
\tabcolsep=0.10cm
\begin{tabular}{lcccc}
\hline\hline
Source \#& Model & $\chi^{2}/$d.o.f. & Parameter & $^a$Flux\\
A1095\\
1 & \emph{wabs*power} & 218/159 & $\Gamma=2.4^{+0.1}_{-0.1}$ & $20.1^{+0.9}_{-0.9}$\\
2 & \emph{wabs*apec} & 27/30 & $^b$kT=$5.8^{+3.3}_{-1.6}$  & $4.8^{+0.6}_{-0.6}$\\
3 & \emph{wabs*apec} & 139/145 & kT$=2.0^{+0.1}_{-0.1}$  & $18.9^{+0.8}_{-0.8}$\\
4 & \emph{wabs*power} & 20/17 & $\Gamma=1.6^{+0.2}_{-0.2}$ & $3.6^{+0.6}_{-0.5}$\\
5 & \emph{wabs*power} & 2/3 & $\Gamma=1.9^{+0.8}_{-0.7}$ & $3.0^{+0.7}_{-0.7}$\\
6 & \emph{wabs*power} & 1/2 & $\Gamma=1.4^{+0.9}_{-0.8}$ & $2.2^{+1.0}_{-0.7}$\\
7 & \emph{wabs*power} & 19/14 & $\Gamma=1.8^{+0.2}_{-0.2}$ & $3.5^{+0.6}_{-0.6}$\\
8 & \emph{wabs*apec} & 29/32 & kT=$3.2^{+0.9}_{-0.6}$ & $5.8^{+0.7}_{-0.7}$\\
9 & \emph{wabs*power} & 81/74 & $\Gamma=1.9^{+0.1}_{-0.1}$ & $16.9^{+1.2}_{-1.2}$\\
10 & \emph{wabs*power} & 4/3 & $\Gamma=1.6^{+0.7}_{-0.6}$ & $2.5^{+0.9}_{-0.7}$\\
11 & \emph{wabs*apec} & 8/11 & kT=$4.1^{+2.7}_{-1.4}$ & $3.8^{+0.7}_{-0.7}$\\
12 & \emph{wabs*power} & 3/3 & $\Gamma=2.0^{+0.7}_{-0.5}$ & $1.5^{+0.4}_{-0.4}$\\
13 & \emph{wabs*apec} & 14/11 & kT$=2.4^{+5.1}_{-0.9}$ & $1.9^{+1.1}_{-0.6}$\\
14 & \emph{wabs*power} & 21/22 & $\Gamma=1.8^{+0.2}_{-0.2}$ & $4.1^{+0.5}_{-0.5}$\\
15 & \emph{wabs*power} & 20/21 & $\Gamma=1.5^{+0.1}_{-0.1}$ & $4.5^{+0.6}_{-0.6}$\\
\hline
A1926\\
1 & \emph{wabs*power} & 62/70 & $\Gamma=1.2^{+0.1}_{-0.1}$ & $23.2^{+1.6}_{-1.6}$\\
2 & \emph{wabs*power} & 22/26 & $\Gamma=1.8^{+0.2}_{-0.1}$  & $5.0^{+0.6}_{-0.6}$\\
3 & \emph{wabs*power} & 63/57 & $\Gamma=2.2^{+0.1}_{-0.1}$  & $9.1^{+0.7}_{-0.7}$\\
4 & \emph{wabs*power} & 9/10 & $\Gamma=1.4^{+0.3}_{-0.3}$ & $4.5^{+1.0}_{-0.9}$\\
5 & \emph{wabs*apec} & 4/6 & $^c$kT$=0.6^{+0.2}_{-0.2}$  & $1.8^{+0.3}_{-0.3}$\\
6 & \emph{wabs*power} & 14/14 & $\Gamma=1.8^{+0.2}_{-0.2}$ & $4.0^{+0.7}_{-0.7}$\\
7 & \emph{wabs*power} & 32/40 & $\Gamma=2.0^{+0.1}_{-0.1}$ & $11.5^{+1.2}_{-1.2}$\\
8 & \emph{wabs*power} & 184/167 & $\Gamma=2.1^{+0.1}_{-0.1}$ & $60.4^{+2.8}_{-2.8}$\\
\hline 
\end{tabular}
\begin{tablenotes}
      \small
      \item
Sources numbers are the same as those marked in Figs.~\ref{fig1}a 
and \ref{fig2}a and have counterparts
of the detected X-ray sources in Tables~\ref{t:s-a1095} or \ref{t:s-a1926} 
 (marked as the front superscripts of the corresponding entries in Column 2).
Errors are quoted at the 90\% confidence level.\\ 
$^a$: fluxes in the 0.5-7 keV band and in units of $\rm{10^{-14}\ erg\ cm^{-2}\ s^{-1}}$. \\
$^b$: in units of keV; the metal abundances are fixed at 0.3 solar unless stated otherwise.\\      
$^c$: from a thermal plasma fit, which also gives a fitted metal abundance of $0.05^{+0.07}_{-0.03}$ solar.
    \end{tablenotes} 
\label{t:spec}    
\end{table}

\appendix
\section{Features of individual cluster}

\noindent {\bf A1095W}: This cluster itself seems to be undergoing a subcluster merger. 
As shown in \S~\ref{ss:x-morph}, 
the cluster stands out in terms of its radial X-ray profile. The $\beta$ value 
of the cluster ($\sim$1.1) is larger than those of the other three clusters 
($\sim$0.5) or the average value ($\sim$0.65) of 46 clusters studied 
by Jones \& Forman (1984). This is a strong indication for the presence of 
substantial substructure in the ICM distribution of A1095W. The presence 
of such substructure leads to an effective increase  
of the core radius and a steep drop in the outer part of the profile,
which is reflected in a large fitted $\beta$ value (Jones 
\& Forman 1999). Indeed, the core radius of A1095W is almost three times 
larger than that of A1095E, although they have similar temperatures 
as listed in Table~\ref{t:clusters}. 

Strong evidence for the substructure is also present in the galaxy
distribution of A1095W. The distribution shows two concentrations 
of galaxies, morphologically resembling the northwest-southeast 
X-ray elongation as shown in Fig.~\ref{fig8}a. The northwest concentration is associated with the most 
prominent X-ray substructure, which is detected as a bright and extended X-ray peak (marked as Source 3 in Fig.~\ref{fig1}a; its surface brightness profile is much extended than the PSF of {\it XMM-Newton}, e.g., Ghizzardi 2002) and has a radio counterpart 
(D in Fig.~\ref{fig9}a). The
concentration most likely represents a subcluster, which still retains 
its own ICM with a temperature lower than its surroundings 
(Fig.~\ref{fig4}c; A1095 Source 3 in Table~\ref{t:spec}; 
\S~\ref{ss:x-spec}). The other galaxy concentration is around the BCG (SDSS J104729.00+151402.0), which is $\sim$ 210 kpc east of the X-ray centroid of the cluster. The radio
emission predominantly arises from multiple compact structures 
resolved in the FIRST map (Fig.~\ref{fig9}a). Both A and B components likely represent double lobes. The separation between the lobes is 7$^{\prime\prime}$/23 kpc for A or 10$^{\prime\prime}$/32 kpc for B; the optical counterparts of A (SDSS J104729.00+151402.0; z=0.210) and B (SDSS J104727.41+151310.6; z=0.206) appear to be normal early-type galaxies (ETGs), according to their SDSS spectra. The counterpart of C (SDSS J104731.29+151402.2; photo z=$0.227$) is also an ETG.  In fact, none of the bright radio-emitting ETGs in A1095W show strong optical emission lines (according to their SDSS spectra). Only the radio galaxy D, the central galaxy (SDSS J104720.55+151503.6; though not classified as a BCG) of the northwest 
galaxy concentration, shows weak
line emission and is classified as a low-ionization nuclear emission-line region, according to the emission line ratios in the BPT diagram (Kauffmann et al 2003).
Therefore, much of the radio emission from the galaxies may 
represent relics of their past AGN activities, probably related to an
early stage of the subcluster merger.

\noindent {\bf A1095E}: In addition to its very asymmetric X-ray  morphology and 
the apparently large-scale elongated diffuse radio features, as already described above, this
cluster shows a strikingly rich set of small-scale radio structures (Fig.~\ref{fig9}b).
Interestingly, the BCG (SDSS J104800.48+151605.6; z=0.213) itself is not a radio source and is 
a normal ETG according to its SDSS spectrum. To the east of the BCG appears one-sided prominent radio lobe E. To the west of the BCG
is a more-elongated one-sided radio trail F. Its point-like eastern head seems to coincide with another small group of galaxies, one of which (SDSS J104759.27+151557.8; z=0.214) is  spectroscopically a normal ETG. The compact radio source H, far away from the BCG and in the southern part of the cluster, has an optical counterpart (SDSS J104801.21+151438.4; z=0.216), which is an ETG with weak emission line. The locations of these radio sources trace well the triangular X-ray morphology of the cluster. Therefore, the radio, optical, and X-ray properties of the cluster all suggest that it is a very unrelaxed system.

\noindent {\bf A1926S}: The chain appearance of bright galaxies in this cluster resembles 
 its northeast-southwest elongated X-ray morphology as shown in Fig.~\ref{fig8}c. The large
offset ($\sim$ 290 kpc in projection) of the BCG (SDSS J143021.94+243429.1; z=0.134) from the X-ray centroid of the cluster is particularly striking. But no significant radio emission is found in the NVSS and FIRST map.
These facts may indicate that the cluster is being assembled gently (i.e., not via a major merger), which is consistent with its flat radial temperature structure (Fig.~\ref{fig3}). The location of the z=0.136 pair of 
galaxies (Fig.~\ref{fig9}c) to the north coincides with the hard X-ray peak in Fig.~\ref{fig2}d, which may be part of the ICM enhancement.

\noindent {\bf A1926N}: 
The trailing northeastern low surface brightness radio emission is oriented
in a fashion similar to the extension of the diffuse X-ray emission of this
cluster (Fig.~\ref{fig9}d). However, the X-ray emission is
 weak and is strongly contaminated by relatively bright point-like X-ray
sources, the spectra of which can be well characterized by power law spectra (Table \ref{t:spec}). 
A high-resolution sensitive X-ray observation
is  needed to further the investigation of the cluster.
Although classified as a normal ETG according to its SDSS spectrum,
the BCG  (SDSS J143028.62+244019.2; z=0.136) positionally coincides with a bright X-ray 
source (marked as Source 3 in Fig.~\ref{fig2}a) and with a radio source (Fig.~\ref{fig9}d).
The BCG  appears  $\sim$ 90~kpc off from the global centroid of the large-scale X-ray emission of the cluster. Thus we may still tentatively conclude that the cluster is a dynamically 
young system.

\end{document}